\documentclass[reprint,
 amsmath,amssymb,
 aps,prd,
 superscriptaddress,
]{revtex4-1}

\usepackage{graphicx}

\usepackage{dcolumn}
\usepackage{bm}
\usepackage{comment}
\usepackage{xspace}
\usepackage{units}
\usepackage{float}
\usepackage{multirow}
\usepackage{xcolor}
\usepackage[caption=false]{subfig}
\usepackage{amssymb}
\usepackage{amsmath}
\usepackage{commath}
\usepackage{verbatim}

\usepackage[colorlinks = true,
            linkcolor = red,
            urlcolor  = blue,
            citecolor = red,
            anchorcolor = blue]{hyperref}

\newcommand{\nue}{$\nu_e$\xspace}

\newcommand{\numu}{$\nu_{\mu}$\xspace}

\newcommand{\npsel}{1$e$N$p$0$\pi$\xspace}
\newcommand{\zpsel}{1$e$0$p$0$\pi$\xspace}

\newcommand{\dedx}{d$E$/d$x$\xspace}

\newcommand{\Bern}{Universit{\"a}t Bern, Bern CH-3012, Switzerland}
\newcommand{\BNL}{Brookhaven National Laboratory (BNL), Upton, NY, 11973, USA}
\newcommand{\UCSB}{University of California, Santa Barbara, CA, 93106, USA}
\newcommand{\Cambridge}{University of Cambridge, Cambridge CB3 0HE, United Kingdom}
\newcommand{\CIEMAT}{Centro de Investigaciones Energ\'{e}ticas, Medioambientales y Tecnol\'{o}gicas (CIEMAT), Madrid E-28040, Spain}
\newcommand{\Chicago}{University of Chicago, Chicago, IL, 60637, USA}
\newcommand{\Cincinnati}{University of Cincinnati, Cincinnati, OH, 45221, USA}
\newcommand{\CSU}{Colorado State University, Fort Collins, CO, 80523, USA}
\newcommand{\Columbia}{Columbia University, New York, NY, 10027, USA}
\newcommand{\Edinburgh}{University of Edinburgh, Edinburgh EH9 3FD, United Kingdom}
\newcommand{\FNAL}{Fermi National Accelerator Laboratory (FNAL), Batavia, IL 60510, USA}
\newcommand{\Granada}{Universidad de Granada, Granada E-18071, Spain}
\newcommand{\Harvard}{Harvard University, Cambridge, MA 02138, USA}
\newcommand{\IIT}{Illinois Institute of Technology (IIT), Chicago, IL 60616, USA}
\newcommand{\KSU}{Kansas State University (KSU), Manhattan, KS, 66506, USA}
\newcommand{\Lancaster}{Lancaster University, Lancaster LA1 4YW, United Kingdom}
\newcommand{\LANL}{Los Alamos National Laboratory (LANL), Los Alamos, NM, 87545, USA}
\newcommand{\Manchester}{The University of Manchester, Manchester M13 9PL, United Kingdom}
\newcommand{\MIT}{Massachusetts Institute of Technology (MIT), Cambridge, MA, 02139, USA}
\newcommand{\Michigan}{University of Michigan, Ann Arbor, MI, 48109, USA}
\newcommand{\Minnesota}{University of Minnesota, Minneapolis, MN, 55455, USA}
\newcommand{\NMSU}{New Mexico State University (NMSU), Las Cruces, NM, 88003, USA}
\newcommand{\Oxford}{University of Oxford, Oxford OX1 3RH, United Kingdom}
\newcommand{\Pitt}{University of Pittsburgh, Pittsburgh, PA, 15260, USA}
\newcommand{\Rutgers}{Rutgers University, Piscataway, NJ, 08854, USA}
\newcommand{\SLAC}{SLAC National Accelerator Laboratory, Menlo Park, CA, 94025, USA}
\newcommand{\SDSMT}{South Dakota School of Mines and Technology (SDSMT), Rapid City, SD, 57701, USA}
\newcommand{\Maine}{University of Southern Maine, Portland, ME, 04104, USA}
\newcommand{\Syracuse}{Syracuse University, Syracuse, NY, 13244, USA}
\newcommand{\TelAviv}{Tel Aviv University, Tel Aviv, Israel, 69978}
\newcommand{\Tennessee}{University of Tennessee, Knoxville, TN, 37996, USA}
\newcommand{\UTA}{University of Texas, Arlington, TX, 76019, USA}
\newcommand{\Tufts}{Tufts University, Medford, MA, 02155, USA}
\newcommand{\VTech}{Center for Neutrino Physics, Virginia Tech, Blacksburg, VA, 24061, USA}
\newcommand{\Warwick}{University of Warwick, Coventry CV4 7AL, United Kingdom}
\newcommand{\Yale}{Wright Laboratory, Department of Physics, Yale University, New Haven, CT, 06520, USA}

\begin{document}

\title{Search for an anomalous excess of charged-current $\nu_e$ interactions without pions in the final state with the MicroBooNE experiment}

\affiliation{\Bern}
\affiliation{\BNL}
\affiliation{\UCSB}
\affiliation{\Cambridge}
\affiliation{\CIEMAT}
\affiliation{\Chicago}
\affiliation{\Cincinnati}
\affiliation{\CSU}
\affiliation{\Columbia}
\affiliation{\Edinburgh}
\affiliation{\FNAL}
\affiliation{\Granada}
\affiliation{\Harvard}
\affiliation{\IIT}
\affiliation{\KSU}
\affiliation{\Lancaster}
\affiliation{\LANL}
\affiliation{\Manchester}
\affiliation{\MIT}
\affiliation{\Michigan}
\affiliation{\Minnesota}
\affiliation{\NMSU}
\affiliation{\Oxford}
\affiliation{\Pitt}
\affiliation{\Rutgers}
\affiliation{\SLAC}
\affiliation{\SDSMT}
\affiliation{\Maine}
\affiliation{\Syracuse}
\affiliation{\TelAviv}
\affiliation{\Tennessee}
\affiliation{\UTA}
\affiliation{\Tufts}
\affiliation{\VTech}
\affiliation{\Warwick}
\affiliation{\Yale}

\author{P.~Abratenko} \affiliation{\Tufts} 
\author{R.~An} \affiliation{\IIT}
\author{J.~Anthony} \affiliation{\Cambridge}
\author{L.~Arellano} \affiliation{\Manchester}
\author{J.~Asaadi} \affiliation{\UTA}
\author{A.~Ashkenazi}\affiliation{\TelAviv}
\author{S.~Balasubramanian}\affiliation{\FNAL}
\author{B.~Baller} \affiliation{\FNAL}
\author{C.~Barnes} \affiliation{\Michigan}
\author{G.~Barr} \affiliation{\Oxford}
% ...

\collaboration{The MicroBooNE Collaboration}
\thanks{microboone\_info@fnal.gov}\noaffiliation

\date{\today}

\begin{abstract}
This article presents a measurement of \nue interactions without pions in the final state using the MicroBooNE experiment and an investigation into the excess of low-energy electromagnetic events observed by the MiniBooNE collaboration.
The measurement is performed in exclusive channels with (\npsel) and without (\zpsel) visible final-state protons using 6.86$\times 10^{20}$ protons on target of data collected from the Booster Neutrino Beam at Fermilab.
Events are reconstructed with the Pandora pattern recognition toolkit and selected using additional topological information from the MicroBooNE liquid argon time projection chamber.
Using a goodness-of-fit test the data are found to be consistent with the predicted number of events with nominal flux and interaction models with a $p$-value of 0.098 in the two channels combined.
A model based on the low-energy excess observed in MiniBooNE is introduced to quantify the strength of a possible \nue excess.
The analysis suggests that, if an excess is present, it is not consistent with a scaling of the \nue contribution to the flux as predicted by the signal model used in the analysis.
Combined, the \npsel and \zpsel channels do not give a conclusive indication about the tested model, but separately they both disfavor the low-energy excess model at $>$90\% CL.
The observation in the most sensitive \npsel channel is below the prediction and consistent with no excess. In the less sensitive \zpsel channel the observation at low energy is above the prediction, while overall there is agreement over the full energy spectrum.
\end{abstract}

\addtolength{\tabcolsep}{5pt}

\maketitle

\section{Introduction}
\label{sec:introduction}
Neutrino physics has entered an era of precision measurements of the parameters that describe three-flavor oscillations~\cite{PDG}. 
At the same time, a broad set of experimental results, collectively referred to as short-baseline anomalies~\cite{gallium,reactor,LSND,MiniBooNELEE,BEST}, is in tension with the three-neutrino paradigm and remains without resolution. 
These short-baseline anomalies have often been linked to the physics signature of $\mathcal{O}$(eV) sterile neutrinos~\cite{sterileWP}. Recent long-baseline sterile-neutrino oscillation searches~\cite{Aartsen:2020iky,disappearance}, however, show tension with this interpretation, and other explanations for these anomalies may need to be considered. 
The observation of an excess of low-energy electromagnetic activity by the MiniBooNE experiment~\cite{MiniBooNELEE} is one example of these anomalies.
Many scenarios have been suggested to explain the origin of the MiniBooNE excess of low-energy electromagnetic showers, including new physics such as sterile neutrino oscillations and decay~\cite{Vergani2021,Fischer2021}, dark-sector portals~\cite{Alvarez-Ruso:2017hdm,Abdullahi2020,Bertuzzo2018}, heavy neutral leptons~\cite{Ballett2019,Gninenko2011}, non-standard Higgs models~\cite{AbdallahEtAl,AsaadiEtAl,DuttaEtAl}, or Standard Model processes such as an enhancement of photon backgrounds~\cite{deltarad}.

The MicroBooNE experiment~\cite{ubdetector} was built to explore the nature of the low-energy excess of events observed by MiniBooNE. 
Operating in the same Fermilab Booster Neutrino Beamline (BNB), it is in a position to examine the nature of low-energy electromagnetic activity with the capabilities of the liquid argon time projection chamber (LArTPC) detector technology. A set of analyses, including the one presented here, have been designed to measure both electron neutrino interactions in multiple topologies~\cite{eLEEPRL,DLPRD,WCPRD}, as well as single photon events~\cite{gLEE}. This article presents a measurement of the rate of charged current (CC) electron neutrino interactions without pions in the final state, and investigates the possibility of  low-energy \nue interactions as an explanation for the MiniBooNE observation of an anomalous excess.

\begin{figure*}
\subfloat[\npsel candidate event\label{sfig:evd_np}]{%
  \includegraphics[height=7cm,width=.49\linewidth]{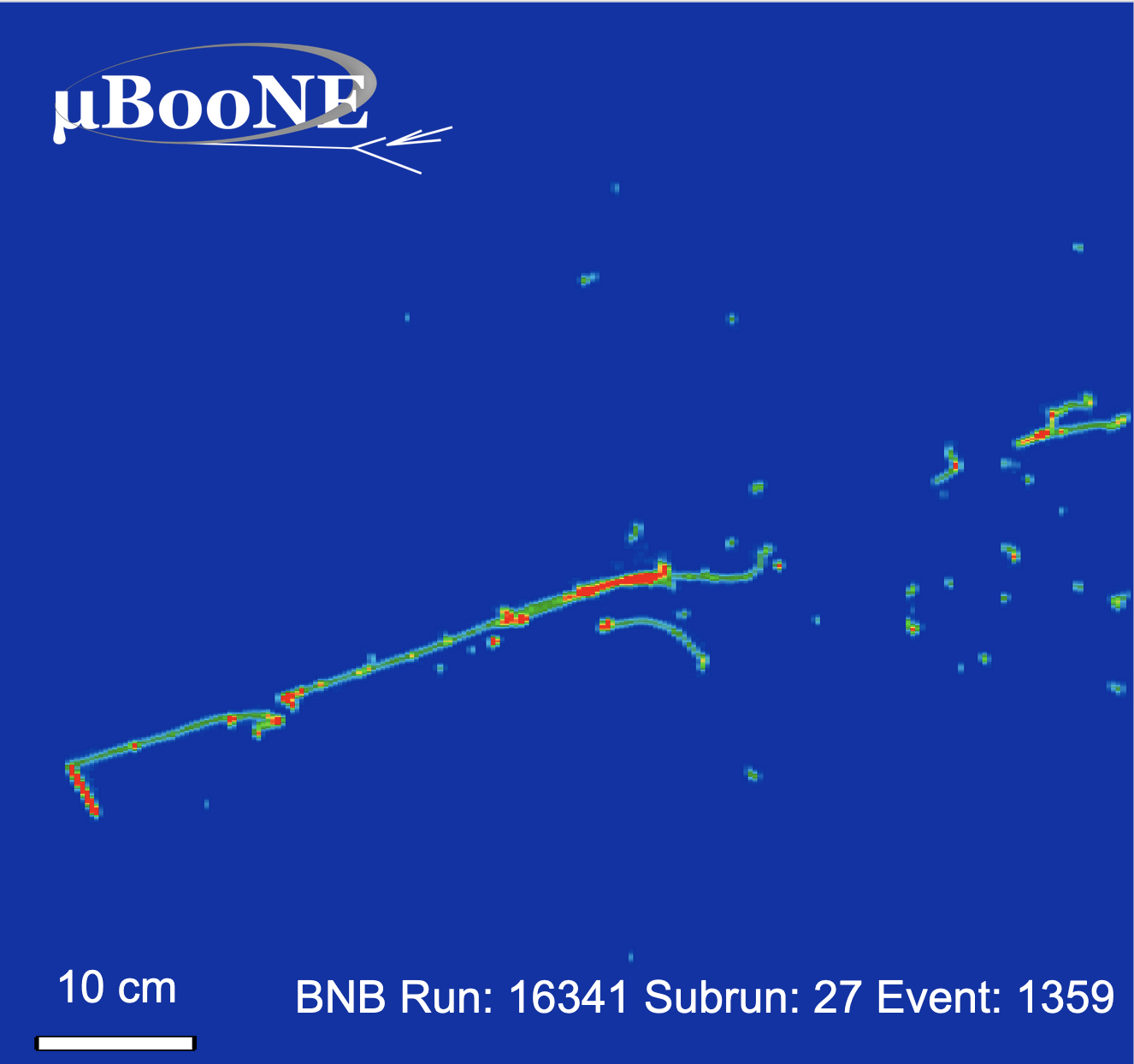}
}\hfill
\subfloat[\zpsel candidate event\label{sfig:evd_zp}]{%
  \includegraphics[height=7cm,width=.49\linewidth]{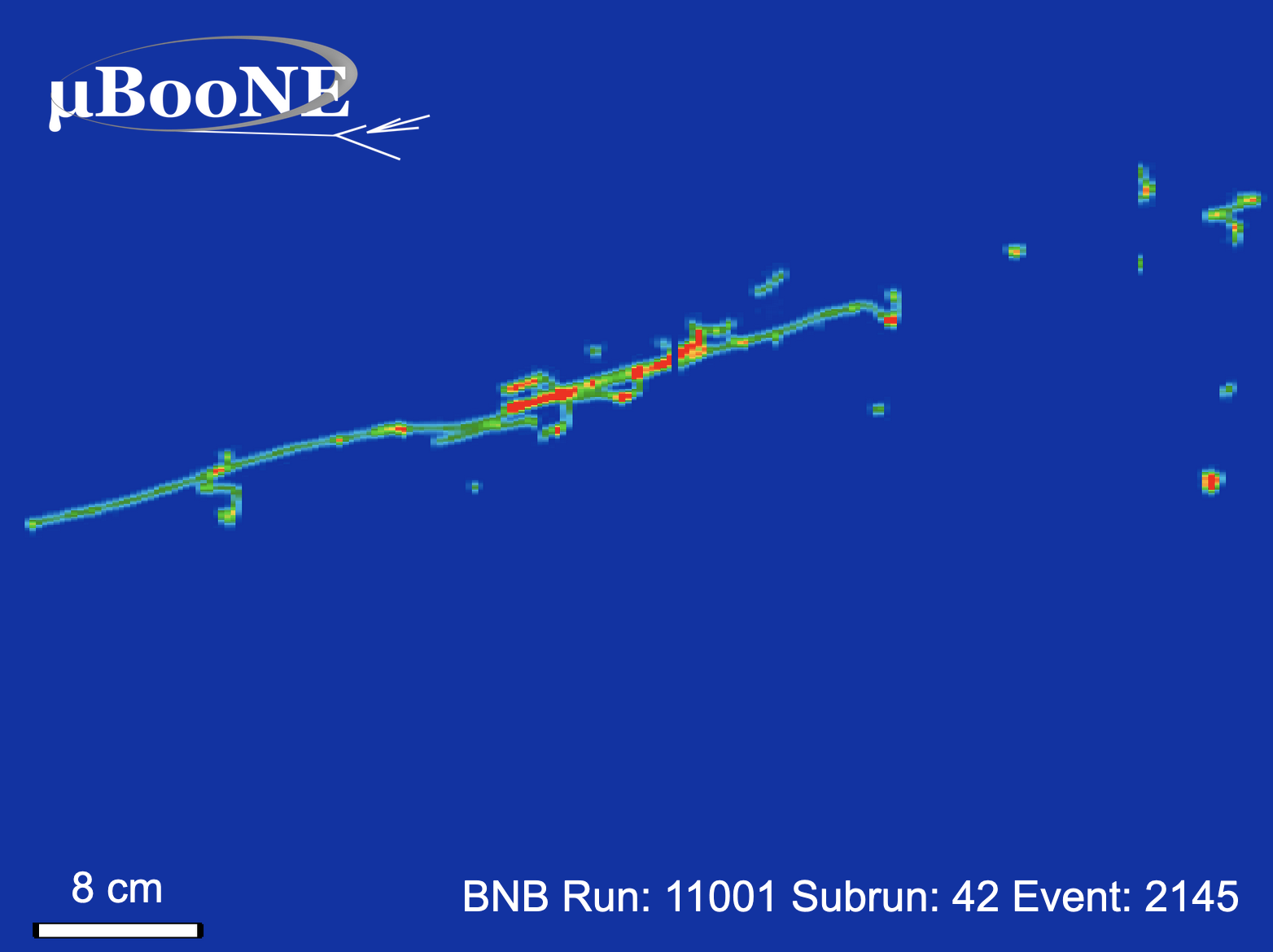}
}
\caption{\label{fig:evd} Event displays of selected electron neutrino candidate data events. The horizontal axis corresponds to the wire number, which is converted into a distance based on the wire spacing, the vertical axis corresponds to the time of the recorded charge, which is converted to a distance along the TPC drift direction using the drift velocity in the TPC drift direction, and the color scale corresponds to the deposited charge. The \npsel event shown~\protect\subref{sfig:evd_np} has a long electron shower and a short proton track attached at the vertex with a large amount of deposited energy.  The \zpsel event shown~\protect\subref{sfig:evd_zp} consists of a single electron shower.}
\end{figure*}

Electron neutrinos that undergo CC interactions will produce a visible electron in the detector, which develops into an electromagnetic (EM) shower, along with possible hadronic activity such as protons and pions.
This analysis performs a measurement of \nue interactions with any number of protons and without visible pions in the final state (1$e$X$p$0$\pi$, X $\geq 0$). This is designed to match MiniBooNE's single electron-like Cherenkov ring signal topology~\cite{MiniBooNE2007}. The presence of visible protons (\npsel, N $\geq 1$) provides additional handles for background rejection.  Furthermore, the  final state with no protons (\zpsel) may be sensitive to a broad range of models~\cite{Chang:2021myh,Alvarez-Ruso:2017hdm,Bertuzzo2018,Abdullahi2020,AsaadiEtAl,Ballett2019} that predict single-shower events and go beyond the electron neutrino interpretation of the MiniBooNE anomaly.
Together these motivate the choice to develop separate, orthogonal selections for events with and without protons in the final state. We focus on the use of calorimetric and topological information provided by the LArTPC technology to perform particle identification (PID) and measure electron neutrino interactions in a broad phase space. Example candidate events in these topologies from data are shown in Fig.~\ref{fig:evd}.
The neutrino flux and interaction systematic uncertainties associated with the selected electron neutrinos are constrained using a high-statistics inclusive measurement of CC muon neutrino interactions in the BNB.  Analysis results are obtained through a series of statistical tests with the introduction of an empirical model which interprets the MiniBooNE anomaly as an enhancement of the flux of low energy electron neutrinos.

This article is organized as follows. Section~\ref{sec:beamdetector} describes the neutrino beamline and MicroBooNE detector. Section~\ref{sec:simulations} provides details of the tools used to simulate neutrino events. Section~\ref{sec:reconstruction} presents the reconstruction methods used to identify neutrino interactions.
Section~\ref{sec:selections} presents the PID methods as well as the \numu and \nue event selections.
Section~\ref{sec:sidebands} describes the blinding procedure and studies on data sidebands.  
Section~\ref{sec:constraint} details the formalism of the procedure used to reduce uncertainties based on the \numu observation, referred to as the \numu constraint. Section~\ref{sec:results} presents the analysis results.

\section{Beamline and Detector Overview}
\label{sec:beamdetector}
This section provides a brief overview of the Booster Neutrino Beamline, the MicroBooNE detector, and the dataset used for the analysis.
The MicroBooNE detector sits at a distance of \unit[468.5]{m} from the BNB target, on-axis with respect to the neutrino beam. 
The neutrino beam begins with 8 GeV protons extracted from the Fermilab Booster synchrotron. 
These protons interact with a beryllium target and produce pions and kaons, which then decay to produce neutrinos.  
The resulting neutrino beam is composed predominantly of muon neutrinos with a small ($< 1\%$) electron neutrino component. This electron neutrino component produced by meson decay chains in the BNB is referred to as ``intrinsic \nue'' in this article. The BNB is structured in spills, each with a duration of \unit[1.6]{$\mu$s} and an intensity of up to $5\times 10^{12}$ protons, with an average repetition rate of up to 5 Hz. Additional details on the BNB are found in Ref.~\cite{BNB,mbflux}.

The MicroBooNE detector~\cite{ubdetector} consists of a time projection chamber (TPC) and a photon detection system.
The TPC measures \unit[2.56]{m} (drift coordinate, $x$) $\times$  \unit[2.32]{m} (vertical, $y$) $\times$ \unit[10.36]{m} (beam direction, $z$) and contains 85 tonnes of  liquid argon in its active volume. Charged particles traversing the detector ionize the argon leaving trails of ionization electrons which drift under the \unit[273]{V/cm} electric field towards the anode where three planes of wires record induced currents and collect the ionization electrons. The three planes of wires, spaced \unit[3]{mm} apart and oriented at 0 degrees (vertical) and at $\pm$60 degrees, produce three different two-dimensional views of the neutrino interaction and allow for three-dimensional reconstruction with $\mathcal{O}(\unit[]{mm})$ spatial resolution. The low-noise TPC electronics allow for measurement of the charge with few percent resolution~\cite{ubsig1}. Combined, these features enable the MicroBooNE detector to record the final-state particles produced by neutrino interactions with the detail required to perform particle identification and accurately measure particle kinematics. 
The light detection system, composed of 32 photomultiplier tubes (PMTs), has a timing resolution of $\mathcal{O}$($\mu$s), which allows us to select events in the BNB time window and to remove a large fraction of the cosmic-ray background.
In addition, a cosmic-ray tagger (CRT)~\cite{ubCRT} is used to reject cosmogenic interactions with precise spatial and timing information.

 MicroBooNE has collected approximately $12 \times 10^{20}$ protons on target (POT) of BNB data since October 2015.
This analysis uses data corresponding to a total of $6.86 \times 10^{20}$ POT collected between February 2016 and July 2018, which corresponds to the first three operation run periods.

\section{Simulation and Modelling}
\label{sec:simulations}
This section describes the tools used in this analysis to model the neutrino flux, neutrino interactions on argon, and the detector response, as well as the low-energy excess signal.  
The simulation packages referenced below are used in the LArSoft software framework~\cite{LArSoft}.

\subsection{Flux model}
The neutrino flux prediction from the BNB at MicroBooNE is made using the flux simulation developed by the MiniBooNE collaboration~\cite{mbflux} and takes into account the different  detector positions. 
The MicroBooNE and MiniBooNE experiments operated simultaneously between 2015 and 2019. No significant variations in the flux were observed by the MiniBooNE collaboration during this time~\cite{MiniBooNE2020}.
The flux prediction is therefore assumed to be invariant over time.

There are two primary sources of uncertainty in the beam flux modeling. The first is hadron production cross sections, which includes the production of $\pi^{\pm}$, $K^{\pm}$, and $K^{0}_L$ particles in the beamline.  The second is related to to the beamline, in particular the modeling of the horn configuration and current, as well as of secondary interactions on the beryllium target and aluminum horn. Flux uncertainties in this analysis are treated analogously to those in MicroBooNE cross section measurements~\cite{ubnuminue,ubnuminueMCC9,ubnumunp,ubccqexsec,ubccincl,ubccpi0} and follow the implementation of the MiniBooNE collaboration as described in Ref.~\cite{mbflux} including the improvements described in Ref.~\cite{mbflux2}. 
Flux uncertainties have a 5\textrm{--}10\% impact on the event rate after selection and, for neutrinos with energy below 0.8 GeV, are dominated by hadron production uncertainties.

\subsection{Neutrino interaction model}
MicroBooNE relies on the GENIE~\cite{GENIE} event generator to simulate neutrino interactions in the detector and model the outgoing final-state particles produced. This analysis uses GENIE v3.0.6 G18\_10a\_02\_11a, which incorporates theoretical models and experimental results relevant for 0.1\textrm{--}1 GeV neutrino interactions. A tune of this model was developed by fitting parameters of particular importance to the modeling of sub-GeV CC interactions to external data~\cite{ubtune}.

Modeling uncertainties on neutrino interactions in GENIE are obtained in three different ways. For model parameters that are estimated using the tune, we use the parameter uncertainty from the fit. 
For all other parameters in the model we use uncertainties as provided by the GENIE collaboration.
Finally, for parameters for which an uncertainty was not provided, we estimate the uncertainty in other ways, such as by choosing the full range between different available models that cover the world data. The treatment of systematic uncertainties in the neutrino interaction model is further detailed in Ref.~\cite{ubtune}, including an inconsistency in the treatment of the final state interaction uncertainty which has a negligible impact on the analysis.
The effect of cross section uncertainties on the predicted neutrino event rate after selection is about 20\%, and constitutes the leading source of systematic uncertainty for the analysis.

\subsection{Detector model}
MicroBooNE's detector response is modeled with multiple simulation tools. Geant4~\cite{geant4} v10\_3\_03c is used to simulate the propagation of particles through the detector. The propagation of light and charge in the detector is done within LArSoft.

The MicroBooNE TPC readout electronics and wire response are determined using a simulation of the induced charge from drifting electrons~\cite{ubsig1,ubsig2}. The production of scintillation light is simulated through a voxelized look-up library created from a detailed Geant4 simulation to model photon propagation.
Several detector simulation components are implemented using a data-driven approach. These involve effects that lead to a non-uniform detector response in space and time. Electric field distortions due to space-charge buildup in the active volume are accounted for through MicroBooNE's data-driven electric field maps~\cite{ubefield,efieldcosmics}.  Non-uniformities across the detector due to electron lifetime or wire response are simulated in a time-dependent way when appropriate. Ion recombination is simulated using a modified box model~\cite{argoneutrecomb}.

As MicroBooNE is a surface detector, cosmic rays are the largest background to neutrino interactions.
A data-driven method is used to eliminate the need to simulate the high rate of cosmic rays passing through the detector as well as intrinsic noise in the TPC and PMT electronics.
This starts with a dedicated data stream which is collected in periods when there is no beam and provides a sample of detector activity from both cosmic rays and electronics noise. Then, to form the beam simulation, TPC and PMT waveforms from simulated neutrino interactions are merged with this beam-off data-stream, ensuring faithful modelling of cosmic-ray backgrounds and noise. We use this approach for the simulation of neutrino interactions taking place both inside and outside the TPC fiducial volume; the latter are referred to as “dirt” background events in this analysis and include interactions in the LAr outside the TPC fiducial and in the walls of the cryostat as well as the rock around the detector cavern.

Detector response systematic uncertainties include the propagation of final-state particles as well as the formation of light and charge signals.
Uncertainties associated with the TPC dominate over the ones associated with the light collection system, with a total uncertainty of 10$\textrm{--}$15\% in the $\nu_e$ event rate. Mis-modeling of the wire response, electric field map, and ion recombination each contribute at a smaller level but in similar magnitude. The treatment of wire response systematics are discussed in~\cite{DetSyst}.
The impact of final state particle propagation is assessed by varying charged pion and proton reinteraction cross sections available from external data, using the \texttt{Geant4Reweight}~\cite{G4Reweight} framework. This leads to $\mathcal{O}$(1\%) uncertainties on the event rate.
Systematic uncertainties associated with light production impact only the first stages of the analysis related to cosmic-ray rejection and lead to uncertainties of 3$\textrm{--}$5\% on the event rate.
Uncertainties due to the limited sample size are also included in the analysis and in terms of \nue event rate they vary from O(1\%) for the most populated bins to O(100\%) for the bins with very small prediction at high energy.

\subsection{Unfolded median MiniBooNE \nue excess model} 
\label{sec:eleemodel}

This analysis searches for an excess of electron neutrino events over the predicted intrinsic interaction rate. To benchmark the analysis performance and calculate sensitivity to potential new physics, we adopt a model constructed using the MiniBooNE dataset to obtain a prediction of a $\nu_e$-like excess in the MicroBooNE detector.  

\begin{figure}
\includegraphics[width=0.45\textwidth]{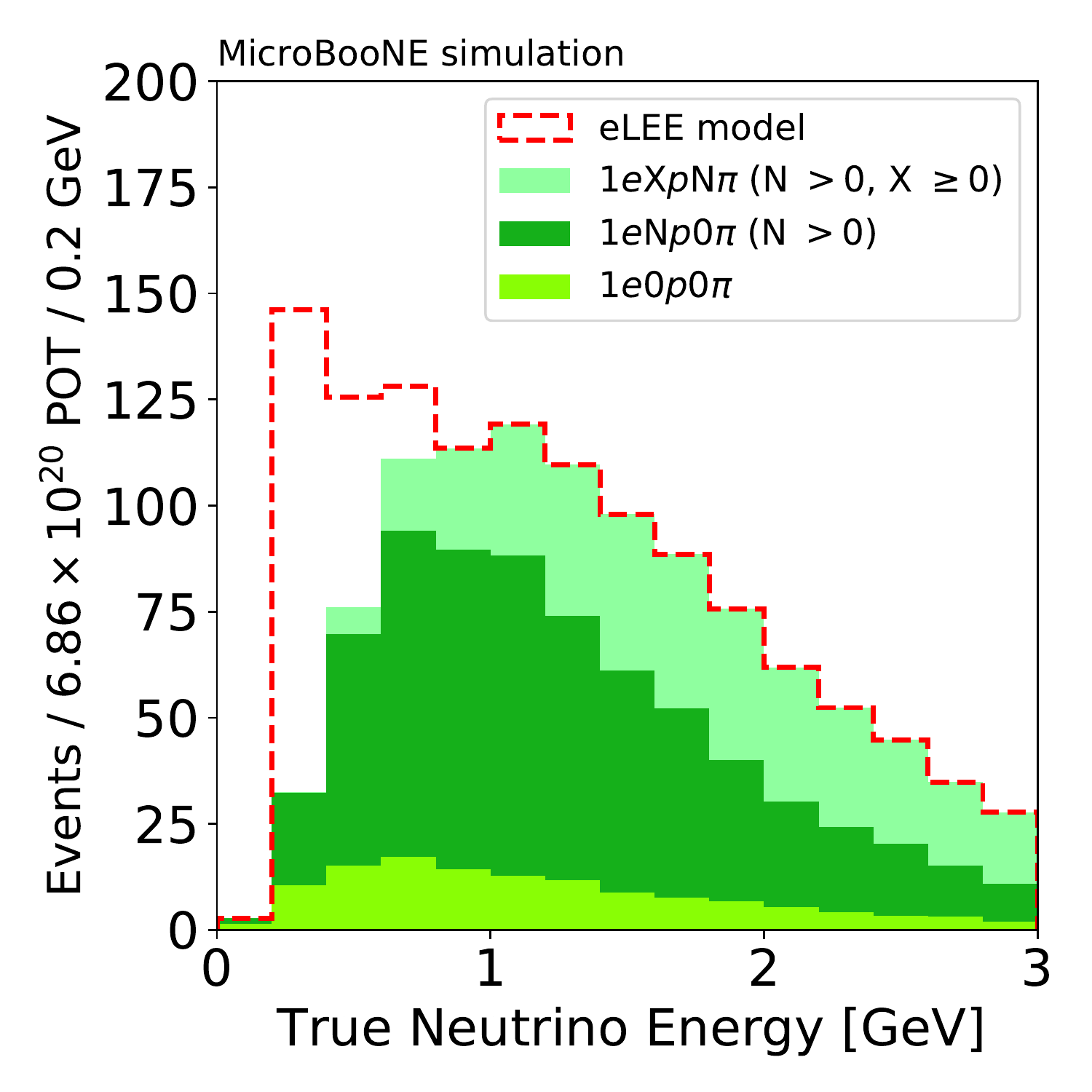}
\caption{\label{fig:eLEE}Predicted electron neutrino events broken down by true number of visible protons and pions. The final states selected in this analysis (\npsel and \zpsel) are shown along with all other \nue interactions (1$e$X$p$N$\pi$). Pion categorization refers to both charged and neutral pions.}
\end{figure}

To construct the model, the background-subtracted excess of data events from MiniBooNE's 2018 result~\cite{MiniBooNELEE} is unfolded using MiniBooNE's electron neutrino energy reconstruction smearing matrix, constructed with the NUANCE~\cite{nuance} neutrino interaction simulation using a CC quasi-elastic energy definition, and accounting for MiniBooNE's energy smearing and selection efficiency. This predicts the rate of electron neutrinos as a function of true neutrino energy above 200 MeV.
The ratio between the predicted rate from MiniBooNE and that of the intrinsic electron neutrino component in MiniBooNE's simulation is used to obtain an energy-dependent flux scaling factor for the excess under the electron neutrino hypothesis. These energy-dependent weights are applied to the rate of intrinsic electron neutrino events predicted by MicroBooNE's flux and cross section simulation to obtain a prediction for the MiniBooNE \nue-like excess in the MicroBooNE detector. Uncertainties from the MiniBooNE measurement are not propagated in our signal prediction as an accurate determination of correlations with MiniBooNE uncertainties is beyond the scope of this work and will require a joint analysis of the two experiments. We will refer to this model in the article as the eLEE model. Figure~\ref{fig:eLEE} shows the truth-level intrinsic $\nu_e$ spectrum, broken into final state particles, and the additional contribution of the  prediction of eLEE model events. In this plot, and throughout this work, protons and charged pions are considered ``visible" and counted if their true kinetic energy is above 40 MeV. A scaling factor $\mu$ is used to vary the normalization of the excess component of the flux. Systematic uncertainties on eLEE signal events are applied analogously to those for intrinsic \nue interactions, consistent with their implementation as an enhancement of the intrinsic \nue flux. Since it is constructed using an unfolding procedure based on the neutrino energy, this model is only used to predict the event rate as a function of reconstructed neutrino energy and not for predictions in other kinematic variables. 

Although it does not fully characterize the MiniBooNE excess, this empirical model provides a benchmark that allows the analysis to quantitatively relate to the \nue interpretation of the MiniBooNE excess and to provide a reproducible reference for further interpretations. As mentioned in Sec.~\ref{sec:introduction} however, different theoretical models, such as sterile neutrino oscillations or more exotic phenomena, can lead to different predictions in MicroBooNE. These are not directly explored here and are left for future work. 

\section{Neutrino Event Reconstruction}
\label{sec:reconstruction}

This section describes the methods used to reconstruct events, separate cosmic-rays from neutrino interactions, and calculate neutrino energy.

\subsection{Event reconstruction}

Event reconstruction in the MicroBooNE TPC starts with the processing of the electronic signal waveforms recorded on wires, which includes noise suppression~\cite{ubnoise} and signal processing~\cite{ubsig1,ubsig2}. Gaussian-shaped pulses on the waveforms, called ``hits'', are then identified, and the associated wire number, time, and integrated charge are inputs to later reconstruction steps.

This analysis uses the Pandora~\cite{pandora} event-reconstruction. 
Pandora is a multi-algorithm pattern recognition toolkit that performs particle tracking for LArTPC detectors. It has been extensively developed within the MicroBooNE collaboration and used for numerous published results~\cite{ubnuminue,ubnuminueMCC9,ubnumunp,ubccqexsec,ubccincl,ubccpi0,ubtrkmult,ubhportal,ubhsl}. It starts with the two-dimensional hit coordinates from each plane and outputs three-dimensional particles categorized in terms of their hierarchy in the neutrino interaction.
In this hierarchy primary particles are directly produced in the neutrino interaction, while those from their decay or interaction with the argon are secondaries.
Particles are classified as shower-like (electrons and photons) or track-like (muons, charged pions, and protons) using a score ranging from 0 to 1. The Pandora pattern recognition is further complemented by analysis-specific tools that  enhance the PID and track-shower separation capabilities with an  emphasis on ensuring powerful muon/proton and electron/photon  separation for particles from low energy neutrino interactions (see Sec.~\ref{sec:pid}). 

TPC detector calibrations implement position- and time-dependent corrections that provide a uniform detector response in addition to absolute gain calibrations necessary for calorimetric energy measurements and PID. These include a calibration of the position-dependent electric field~\cite{ubefield} using MicroBooNE's UV-laser as well as calibrations of the electron lifetime, wire-response, and absolute charge. MicroBooNE's overall calibration strategy is described in Ref.~\cite{ubcalib} and relies on  through-going and stopping cosmic muons.

\subsection{Cosmic-ray rejection}

The near-surface location and the low neutrino interaction rate in the detector lead to significant cosmic-ray contamination. For each beam spill, $\mathcal{O}$(10) cosmic rays cross the detector in the \unit[2.3]{ms} TPC drift window.  Conversely, approximately one in every 1000 spills lead to a neutrino interaction in the active volume for a $\nu$-to-cosmic ratio of $10^{-4}$. Scintillation light is used to suppress cosmic-ray backgrounds, first as part of an online trigger selection and subsequently through an offline analysis filter.  A requirement to observe prompt scintillation light in coincidence with the beam rejects 98\% of recorded beam-spills while accepting more than $99\%$ for \nue interactions with neutrino energy greater than 200 MeV.

At this stage in the analysis, selected events are still dominated by cosmic-ray interactions which  occur in time with the \unit[1.6]{$\mu$s} BNB spill. 
Through-going or out-of-time TPC interactions are rejected as obvious cosmic rays and removed from further analysis. The three-dimensional charge-pattern of the remaining interactions identified in the TPC  is compared to the pattern of scintillation light collected on the detector's PMTs.  Compatibility between the absolute charge and light, as well as their relative position in the TPC, is required. Including these tools in the selection leads to an additional suppression of cosmic-ray interactions by a factor of ten, with an integrated efficiency of 83\% for both \nue and \numu CC interactions.

The CRT~\cite{ubCRT} provides an additional tool for cosmic rejection. CRT information is available only for data taken after December 2017, when it was integrated in MicroBooNE's analysis chain. Its usage in this analysis is limited to the \numu selection (Sec.~\ref{sec:numu}).

\subsection{Energy reconstruction}
\label{sec:evtReco}

The MicroBooNE LArTPC can detect particles with a threshold of few to tens of MeV and measure the energy deposited in the neutrino interaction with high precision .
Energy reconstruction is performed calorimetrically for electromagnetic (EM) showers and based on measurements of particle range for track-like particles.  Selections in this analysis require particle containment in the detector (see Sec.~\ref{sec:selections}). Range-based track energy measurements  deliver very good energy resolution, which is estimated from simulation to be 3\% for muons and $<2\%$ ($<9\%$) for protons with kinetic energy $>100$ MeV ($>40$ MeV). The EM shower energy is measured by integrating the deposited energy ($E_{\textrm{calorimetric}}$) and relying on simulation of electron showers to obtain a correction factor which accounts for inefficiencies in collecting the full energy deposited~\cite{uBpi0reco}. This leads to a reconstructed energy definition of $E_{\textrm{corrected}} = E_{\textrm{calorimetric}} / 0.83$. The electron energy resolution is $\leq 12\%$ and is dominated by the charge clustering inefficiencies discussed above. The reconstructed neutrino energy for \nue and \numu interaction candidates is calculated using:

\begin{subequations}
\begin{align}
    \label{eq:ereconue}
    E^{\nu_e}_{\textrm{reco}} &= E^{\textrm{electron}}_{\textrm{corrected}} + \sum_{\textrm{tracks}} E^{\textrm{proton}}_{\textrm{range}} \\
    \label{eq:ereconumu}
    E^{\nu_{\mu}}_{\textrm{reco}} &= E^{\textrm{muon}}_{\textrm{range}} + \sum_{\textrm{other tracks}} E^{\textrm{proton}}_{\textrm{range}} + 0.105 \; \textrm{GeV}
\end{align}
\end{subequations}

where 0.105 GeV is the muon mass. 
In the energy definition we assume that all tracks other than the selected muon are protons, which matches the \nue selection without pions, but represents an approximation for the inclusive \numu selection.  This definition achieves 15\% energy resolution for both selected \nue and \numu events in the low energy region primarily targeted by this analysis.
For \nue events this definition measures the the energy deposited by charged final state particles above threshold and provides an accurate estimate, with an average bias at the percent level; when compared to the true neutrino energy, however, it typically underestimates by ~16\% (9\%) for selected \zpsel (\npsel) events.
More details on PID are described in Sec.~\ref{sec:selections}.

\section{Neutrino Event Selections}
\label{sec:selections}

Neutrino candidate events are initially identified using the reconstruction methods described in Sec.~\ref{sec:reconstruction}. 
The following section presents a description of several of the PID tools developed for this analysis as well as the \numu and \nue selections in which they are used.

\subsection{Particle identification}
\label{sec:pid}

The primary PID tasks required for this analysis are the separation of highly ionizing proton tracks from minimally ionizing muons and pions as well as the separation of photon and electron electromagnetic showers. 
To distinguish stopping muons from protons we leverage the difference in the energy loss profile at the Bragg peak through a measurement of the energy loss per unit length (\dedx) versus particle residual range. 
A probability density function for simulated protons and muons is used to construct a likelihood function that combines the measured \dedx at each point along a particle’s trajectory from the calorimetric information on all three planes~\cite{llrpid}. 
This tool provides a 90\% relative efficiency for proton selection with a 5\% mis-identification rate. Track PID is used to identify muon candidates produced by \numu CC interactions, isolate protons, and remove pion candidates.

\begin{figure}
\includegraphics[width=0.5\textwidth]{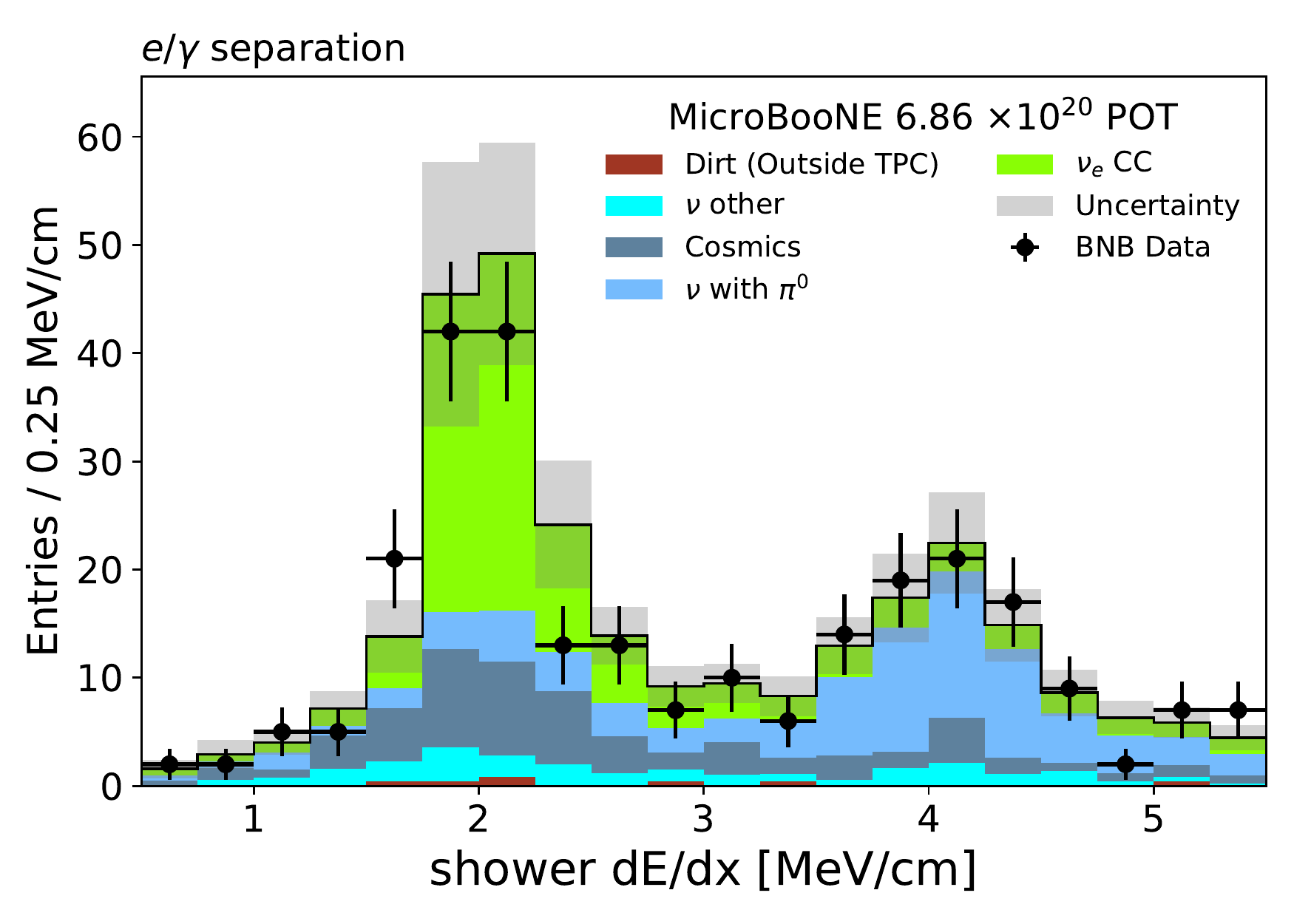}
\caption{\label{fig:dedx_1eXp} Energy deposited per unit length (\dedx) for electron-photon separation. The figure shows \dedx measured in the [0,4] cm range from the shower start point for a combination of events with and without protons. Data from the signal region ($E_{\nu} < 0.65$ GeV) is excluded from this validation plot. The contributions to the stacked histogram are comprised of charged-current intrinsic \nue interactions with any number of final state hadrons in green, \numu and neutral-current \nue interactions that produce one or more $\pi^0$s in the final-state in light blue, and all other $\nu$ interactions in cyan. Dirt backgrounds are in red, and cosmic backgrounds in grayish-blue. This categorization is used in all \nue selection figures.}
\end{figure}

Two key features are used to achieve electron-photon separation: the calorimetric measurement of \dedx at the start of the shower and the displacement of the electromagnetic shower’s start position from the primary vertex in neutrino interactions with hadronic activity. 
To evaluate \dedx, reconstructed showers are fit using a Kalman filter~\cite{Fruhwirth:1987fm} based procedure to identify the main shower trunk and reject hits that are transversely or longitudinally displaced. Values of \dedx measured in the first few centimeters of the electromagnetic shower, before it starts to cascade, are used to compute a median \dedx characteristic of the shower's energy loss~\cite{argoneutdedx}. Information from all three wire planes is used to optimize the ability to perform electron-photon separation independently of particle orientation.  Multiple ranges at the shower start point are used to evaluate \dedx to account for the potential impact of protons at the vertex and early branching of the electromagnetic shower and provide additional separation power. 
The \dedx variable is shown in Fig.~\ref{fig:dedx_1eXp}. Good separation between electron and photon showers is observed and contributes to the $\pi^0$ background rejection achieved by this analysis. 
In this and other data/simulation comparison plots shown in the article data points are shown with associated statistical uncertainty, computed as $\sqrt{N}$, while systematic uncertainties on the prediction are shown as a shaded gray band.

\subsection{\numu measurement}
\label{sec:numu}
\begin{figure}
\includegraphics[width=0.5\textwidth]{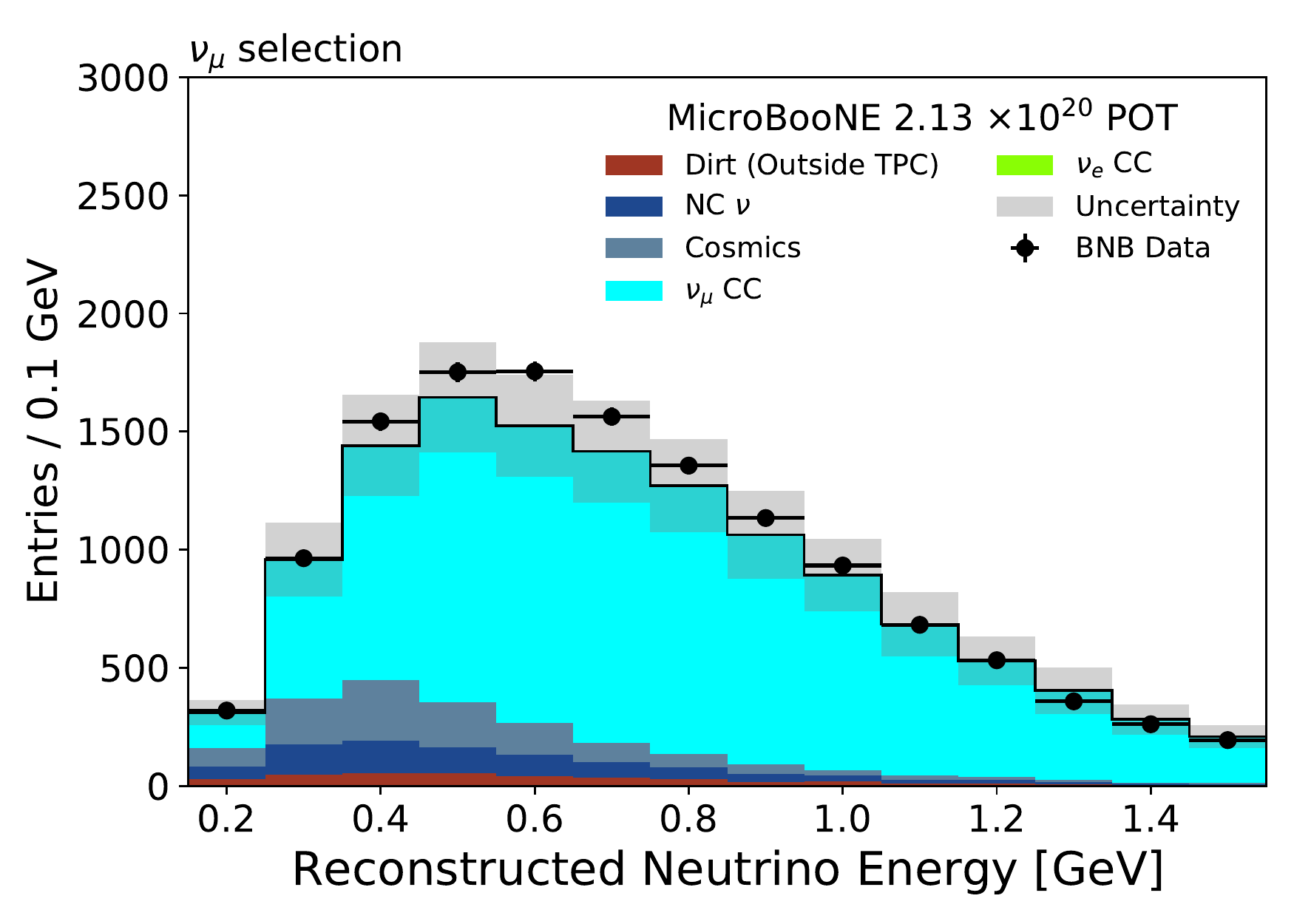}
\caption{\label{fig:numu} Selected muon neutrino events compared to simulation split by true interaction categories. The predicted stacked distribution is comprised of charged-current \numu interactions in cyan, neutral-current backgrounds in dark blue, \nue interactions in green, and cosmic backgrounds in grayish-blue. The shaded band is the systematic uncertainty.}
\end{figure}

The vast majority of neutrinos reaching the MicroBooNE detector are muon neutrinos. They come from the same flux of parent hadrons and interact on the same target argon in the detector as the electron neutrinos. This makes the measurement of high-statistics $\nu_{\mu}$ interactions a valuable handle with which to validate and constrain intrinsic \nue flux and cross section systematic uncertainties.
This is done with a \numu CC inclusive selection that allows any number of final state hadrons and prioritizes performance at low energy.  
A muon neutrino candidate is identified by the presence of a muon candidate inside the TPC fiducial volume. The muon is required to be contained, which preserves good efficiency for low-energy $\nu_{\mu}$ interactions, while suppressing cosmic-ray muon backgrounds.  Cosmic rays are the primary background for $\nu_{\mu}$ CC events, and an additional 64\% of these are removed using the CRT. The $\nu_{\mu}$ constraint only uses $2.13\times10^{20}$ POT of data collected after December 2017, when the CRT was available.
Events surviving these selection requirements are required to have a track PID with a muon-like value. Consistency is required between two independent measurements of muon energy: the range-based energy estimation and one based on multiple Coulomb scattering as described in Ref.~\cite{MCS}. 
The reconstructed neutrino energy distribution for the final $\nu_{\mu}$ selection is shown in Fig.~\ref{fig:numu}.  The data sample contains 13,346 events with a predicted \numu CC purity of 77\%. The main  backgrounds are cosmic ray and neutral-current neutrino interactions. Data and simulation are found to agree within systematic uncertainties in reconstructed neutrino energy as well as in other muon neutrino kinematic variables. 
These include the muon energy and angle with respect to the beam, and were tested quantitatively accounting for all uncertainties and their correlations.

\subsection{\nue selections}
\label{sec:nue}
Electron neutrinos are measured with two separate selections targeting events with and without visible protons. These are referred to as the \npsel and \zpsel selections where $N\geq1$. In simulation, we define a proton as visible if it has a kinetic energy of at least 40 MeV.
Combined, these span the signal definition of electron neutrinos measured by the MiniBooNE experiment: events with a single electron, any number of protons, and no pions.  

The analysis targets contained \nue interactions occurring in the fiducial volume, defined  by a boundary of \unit[10]{cm} in the drift coordinate, \unit[15]{cm} in the vertical, and 10 and \unit[50]{cm} from the front and end of the TPC respectively. The selections rely on a common pre-selection which identifies an event as \nue-like. 
An event is defined as \nue-like if there is a contained reconstructed electromagnetic shower with at least 70 MeV of deposited energy. 
The reconstructed energy requirement removes Michel electrons from cosmic- or neutrino-induced muons.
Events are then further classified depending on the presence or absence of proton candidates.

The \npsel and \zpsel selection definitions split after pre-selection, but the strategy and inputs used for the following steps are the same for both.
Events are classified based on topological and calorimetric information such as the track PID score and \dedx as described earlier.  
Additional handles are used to separate \nue events from events with a $\pi^0$. These are the distance between the neutrino interaction vertex and the start point of the shower, known as the conversion distance, and a second shower search.
This analysis does not use kinematic quantities in the selection criteria to limit the model dependence of the results.
A set of selection requirements called the ``loose'' selection is defined using these variables to remove large portions of the backgrounds for higher statistics data--simulation comparisons in more \nue-like regions.
Next, these variables are used to train boosted decision trees (BDTs) for the two channels used in the analysis.

The main backgrounds for the \nue selections are cosmic-rays, neutrino interactions with $\pi^0$ production, and neutrino interactions (referred to as ``$\nu$ other") that produce charged pions or muons that eventually produce a Michel electron that is mis-identified as an electron produced by a \nue interaction. After the full selection dirt events outside the TPC fiducial volume are a negligible contribution.
 
\subsubsection{\npsel selection}
\label{sec:np}
\begin{figure}
\includegraphics[width=0.45\textwidth]{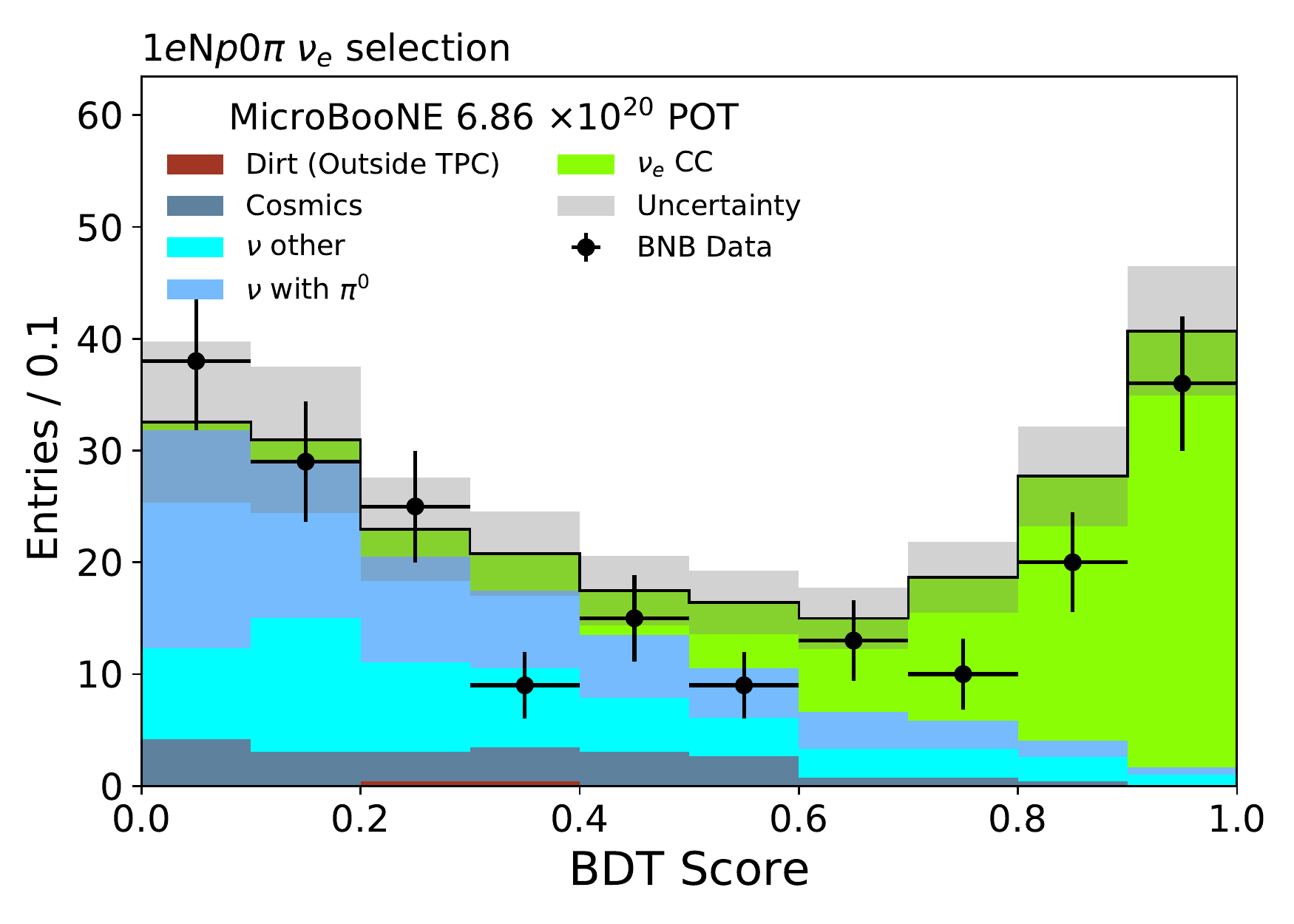}
\caption{\label{fig:BDT_NP} Response of the \npsel selection BDT designed to reject events with $\pi^{0}$s. Background events are predicted to peak at low BDT scores and electron neutrinos at high BDT scores. Events with BDT score above 0.67 are retained as part of the final selection. Gray bands denote the systematic uncertainty on the prediction.}
\end{figure}

The \npsel channel is most sensitive to the eLEE model as it is able to use tracks associated with the vertex in addition to the shower to select electron neutrino events and reject backgrounds. In this selection, two BDTs are trained with XGBoost~\cite{xgboost} to separate signal from background: one targets removal of backgrounds that contain a $\pi^0$, and the other backgrounds without $\pi^0$s. 
Samples of \nue events simulated with true neutrino energy below 0.8 GeV are used to define the signal when training the BDTs.
Simulated samples with \numu events from a variety of true interaction categories are used to train the BDT to identify backgrounds.
Sixteen topological and calorimetric variables are used to distinguish signal from background in these BDTs. The most important of these are the shower conversion distance, which separates $\nu_{e}$ from $\pi^0$ events, and the number of distinct branches in the shower, which separates mis-reconstructed \numu interactions from \nue-induced showers. The longest track-like particle in the interaction is required to be proton-like which further helps to suppress cosmic-ray backgrounds, \numu backgrounds, and \nue interactions with final-state charged pions. 
At pre-selection the purity of the \npsel selection is expected to be at the percent level. After the full selection is applied the \nue purity is expected to be 80\% with an efficiency of 15\% for true \npsel events defined based on the 40 MeV proton energy threshold. The response of the BDT targeting events with $\pi^{0}$'s is shown in Fig.~\ref{fig:BDT_NP} for the full data-set after the loose selection. The selected sample is obtained by rejecting events with BDT score less than 0.67 and 0.70 for the $\pi^0$ and non-$\pi^0$ BDTs respectively. Relative to pre-selection, cosmic background events are reduced by 99.98\% and background events with $\pi^0$s are reduced by 99.93\%. 
The predicted composition of the selected \npsel sample is shown in Table~\ref{tab:1eNp_predict}. The selected \nue CC events are predicted to be 95\% true \npsel events, with a $\sim$5\% contamination of events with pions.

\begin{table}[h]
\caption{\label{tab:1eNp_predict} Predicted composition of the \npsel selected events with unconstrained systematic uncertainties in the reconstructed neutrino energy range 0.01$\textrm{--}$2.39 GeV for $6.86\times 10^{20}$ POT.}
\begin{tabular}{p{0.15\textwidth} m{0.15\textwidth} m{0.15\textwidth}|}
\hline \hline
\multicolumn{2}{c}{\npsel Selection} \\
True Category & Predicted Events \\
\hline
\nue CC 0p0$\pi$ & $0.4\pm0.1$ \\
\nue CC Np0$\pi$ & $71.7\pm10.6$ \\
\nue CC XpN$\pi$ & $3.3\pm0.9$ \\
\hline
\nue CC total & $75.4\pm11.0$ \\
$\nu$ with $\pi^0$ & $5.1\pm1.4$ \\
$\nu$ other & $5.5\pm1.1$ \\
Cosmic-rays  & $0.8\pm0.5$ \\
\hline
Total & $86.8\pm11.5$ \\
\hline \hline
\end{tabular}
\end{table}


\subsubsection{\zpsel selection}
The \zpsel topology is sensitive to \nue events in the eLEE model, as well as potentially to single-electron events from a broader range of models. In addition, it complements the \npsel selection by mitigating migration effects that may arise from mis-reconstruction or mis-modeling of the multiplicity and kinematics of protons produced by neutrino interactions.

\begin{figure}[ht]
\begin{center}
\includegraphics[width=0.45\textwidth]{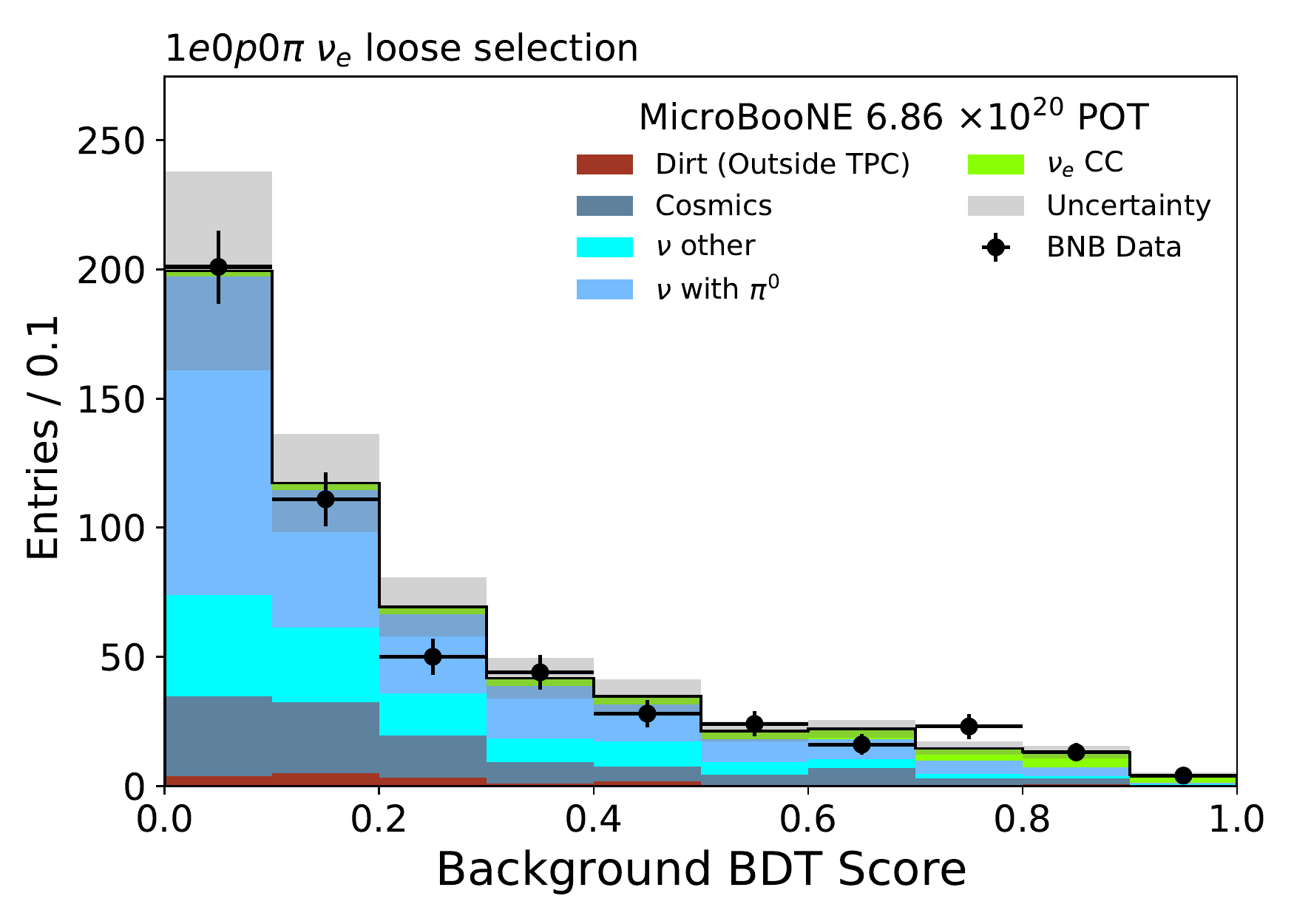}
\caption{\label{fig:BDT_ZP} \zpsel selection BDT response. Background events are predicted to peak at low BDT scores and electron neutrino events at high BDT scores. In the final selection, events with BDT score above 0.72 are retained. Gray bands denote the systematic uncertainty on the prediction.}
\end{center}
\end{figure}

A single BDT is trained to select true \zpsel events and true \npsel events in which protons are not reconstructed. The methods used are the same as those for the \npsel selection described in Sec.~\ref{sec:np}, except that only a single BDT is used to reject backgrounds.
The BDT leverages 28 topological and calorimetric variables, the most important of which are the measurements of \dedx which separate electrons from $\pi^{0}$s. The BDT response is shown in Fig.~\ref{fig:BDT_ZP} for the full data set after applying the loose selection. The final selection is made by requiring events have a BDT score greater than 0.72.

After pre-selection the \nue purity is estimated to be at the percent level.  After the full selection is applied the \nue purity is expected to be 43\% with an efficiency of 9\% for true \zpsel events.  
The selected \nue events are predicted to be 70\% true \zpsel events and 30\% true \npsel events.
Relative to pre-selection cosmic background events are reduced by 99.8\% and the backgrounds from events with $\pi^{0}$s are reduced by 99.7\%. Even with this level of $\pi^0$ background suppression, the overall $\pi^0$ contribution to the predicted event rate is, at low energies, comparable to that of electron neutrinos. This is due to the relatively low rate of \zpsel interactions as well as residual reconstruction limitations.
The predicted number of events after the BDT selection is shown in Table~\ref{tab:zp_sel_predict}.

\begin{table}[h]
\caption{\label{tab:zp_sel_predict} Predicted composition of the \zpsel selected events with unconstrained systematic uncertainties in the reconstructed neutrino energy range 0.01$\textrm{--}$2.39 GeV for $6.86\times 10^{20}$ POT.}
\begin{tabular}{p{0.15\textwidth} m{0.15\textwidth} m{0.15\textwidth}}
\hline \hline
\multicolumn{2}{c}{\zpsel Selection} \\
True Category & Predicted Events \\
\hline
\nue CC 0p0$\pi$ & $8.7\pm3.0$ \\
\nue CC Np0$\pi$ & $3.8\pm0.7$ \\
\nue CC XpN$\pi$ & $0.3\pm0.1$ \\
\hline
\nue CC total & $12.8\pm3.4$ \\
$\nu$ with $\pi^0$ & $8.6\pm1.9$ \\
$\nu$ other & $3.1\pm1.1$ \\
Cosmic-rays  & $5.7\pm1.5$ \\
\hline
Total & $30.1\pm4.3$ \\
\hline \hline
\end{tabular}
\end{table}

\section{Sidebands and Blind-Analysis Strategy}
\label{sec:sidebands}

This measurement of the \nue rate in the BNB and the corresponding exploration of the \nue nature of the MiniBooNE excess was designed as a blind analysis, without access to the \nue component of the BNB flux. This choice minimizes the risk of bias but also requires careful validation.  The flux, cross section, and detector models used in the \nue selections are validated using numerous data sidebands,  which include samples dominated by \numu and $\pi^0$ backgrounds, as well as the NuMI \cite{numibeam} neutrino beam data.
In addition, a small amount of BNB data, less than 10\% of the total data set, was fully open during analysis development.  Each of these sidebands are described in the following sections, followed by a description of the unblinding procedure. All sidebands are orthogonal with respect to the signal selection.

\subsection{Background-enriched sidebands}

\begin{figure}
\includegraphics[width=0.5\textwidth]{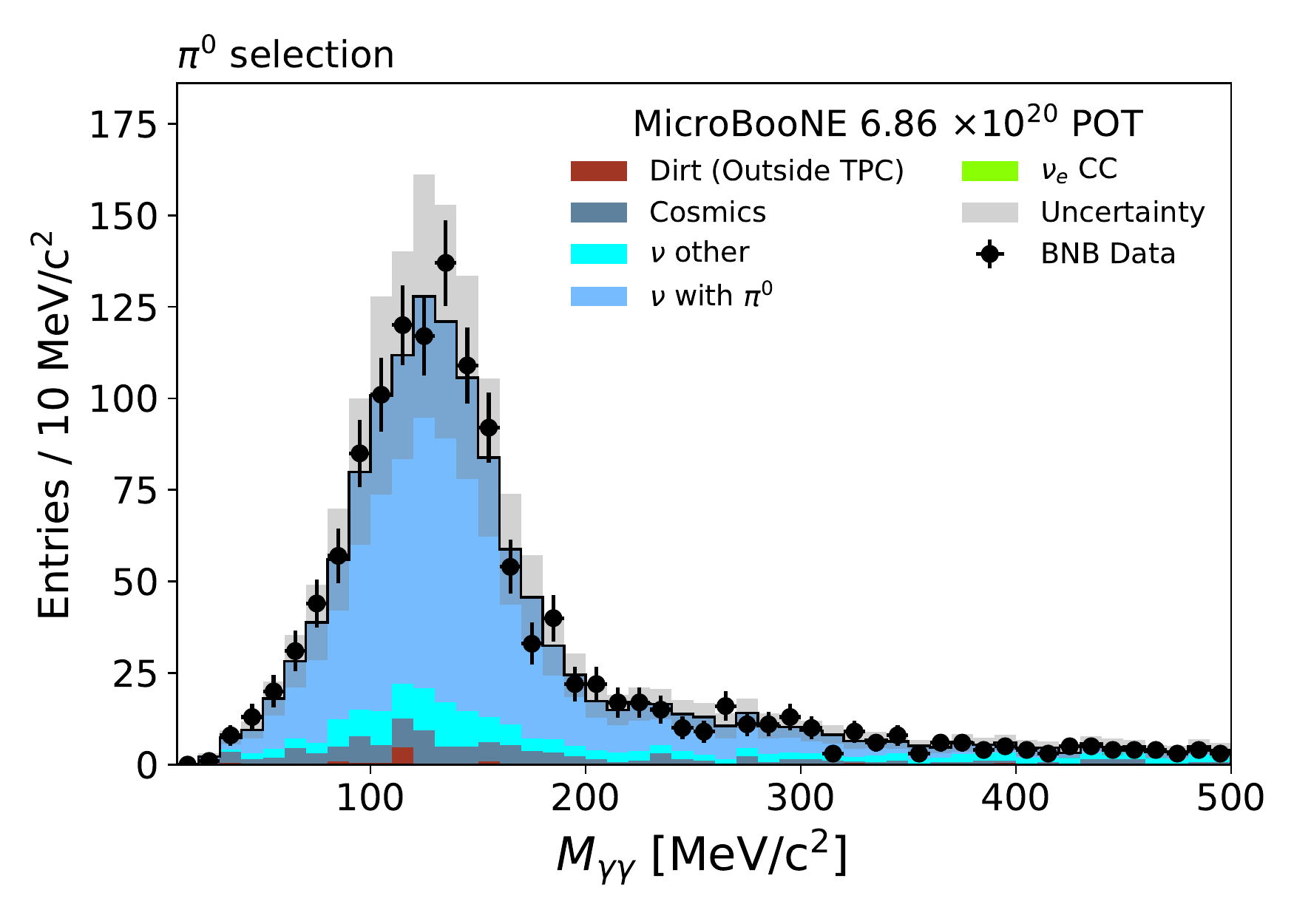}
\caption{\label{fig:pi0mass} Area-normalized comparison of data and simulation for the di-photon mass from $\pi^0$ candidates. The data and simulation peaks agree to within 1\%, and fall within 5\% of the accepted $\pi^0$ mass of 135 MeV/$c^2$, demonstrating good energy-scale calibration for EM showers.}
\end{figure}

\begin{figure*}[ht!]
\subfloat[The transverse development angle for events that would pass the \zpsel loose selection but have more than one shower. Most of the events contain $\pi^{0}$s, and good agreement between data and simulation indicates that this background is well modeled.\label{sfig:2shr}]{%
  \includegraphics[width=0.45\textwidth]{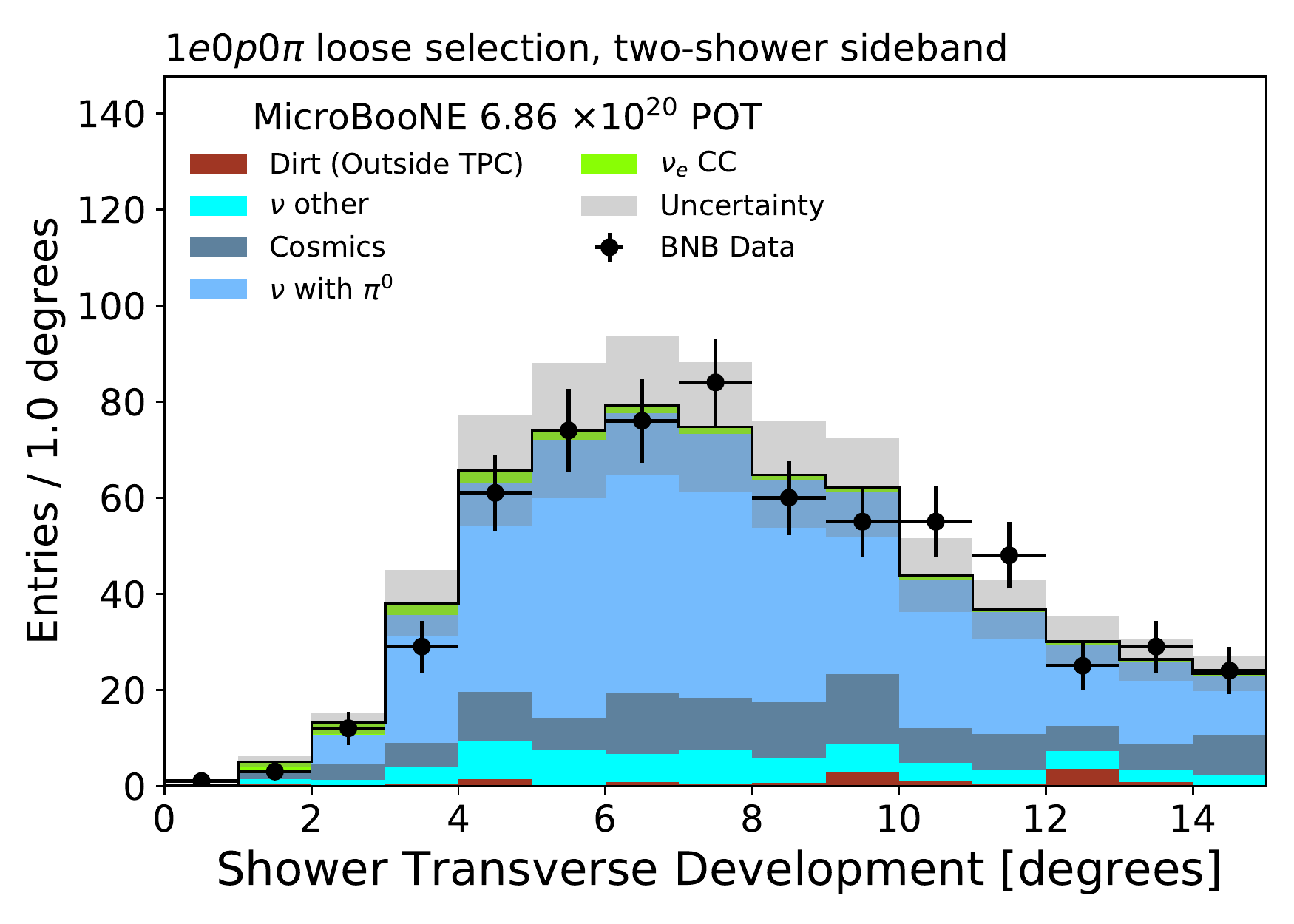}%
}\hfill
\subfloat[The shower conversion distance at pre-selection in the \npsel low-BDT sideband. Background events with $\pi^{0}$s are predicted to typically have a longer conversion distance than those without $\pi^{0}$s.\label{sfig:lowBDT}]{%
  \includegraphics[width=0.45\textwidth]{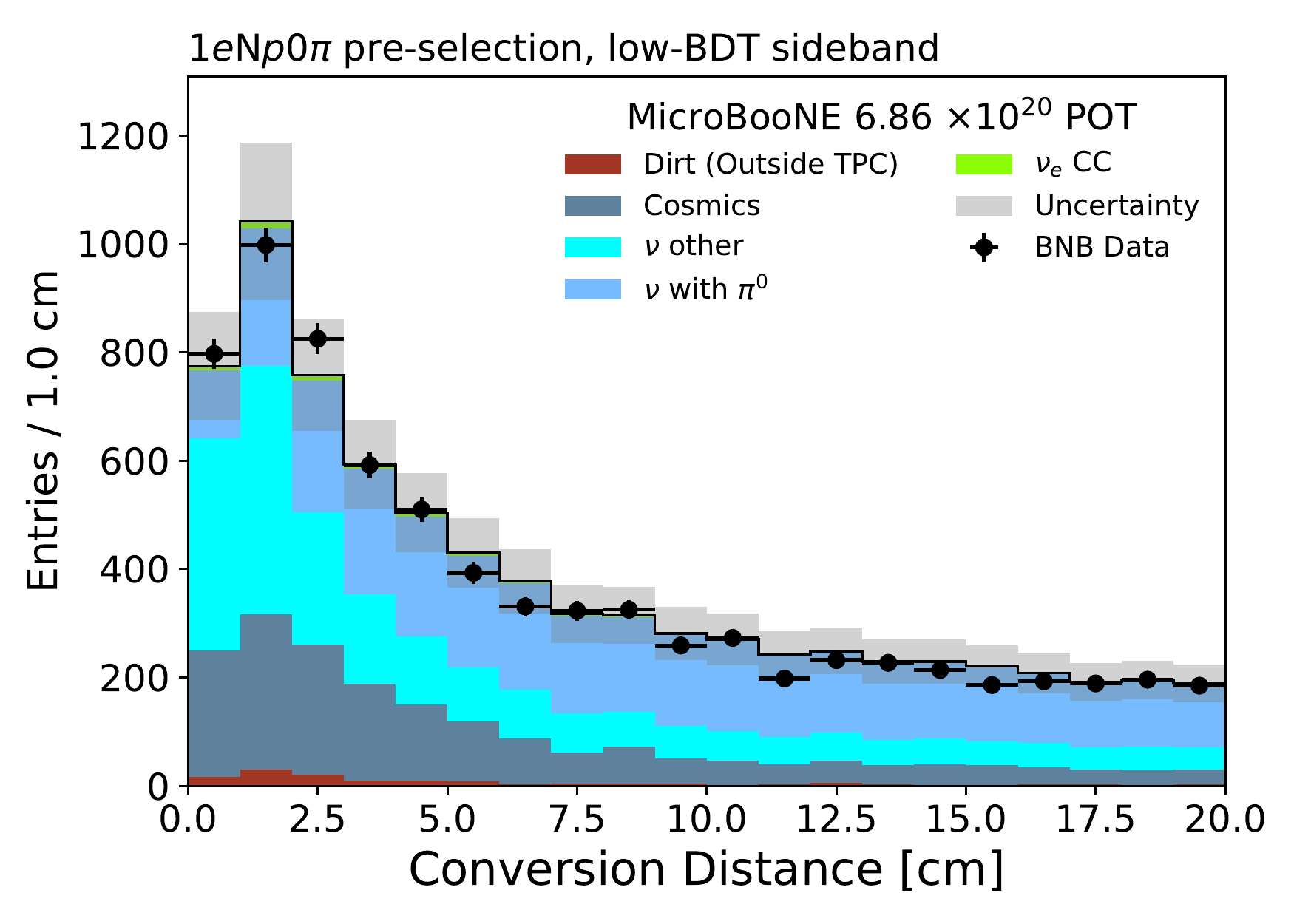}
}
\caption{\label{fig:2plus} Example distributions of BDT input variables in sideband regions.}
\end{figure*}

Multiple sideband samples were developed to validate the background modeling.
Neutral pions are particularly useful as they are both a well-understood standard candle to validate the shower energy reconstruction and an important background for the \nue measurement~\cite{uBpi0reco}.  The area-normalized data to simulation comparison of the reconstructed invariant mass, $M_{\gamma\gamma}$, from $\pi^0 \rightarrow \gamma\gamma$ decays in a high-statistics $\pi^{0}$ sample is shown in Fig.~\ref{fig:pi0mass}. It demonstrates good reconstruction performance for EM showers and well modeled energy-scale calibration.
An additional sideband dominated by neutrino interactions in which a $\pi^{0}$ is produced was developed by applying the \nue selection but requiring that there be at least two reconstructed showers instead of one.  This sample is $\pi^{0}$ rich and its similarity to the \nue selections makes this sideband ideal for validating the predicted $\pi^{0}$ background to the \nue measurement.
Overall, the prediction was found to be consistent with the observation in $\pi^0$ rich selections, with a trend showing more predicted $\pi^0$ events at higher energies compared to the observation. This trend is accounted for by the $\mathcal{O}$(20\%) systematic uncertainty associated with pion production in the neutrino interaction model. All inputs to the selection BDTs were checked in this sideband at each selection stage. As an example, Fig.~\ref{sfig:2shr} shows events in this sideband that would pass the \zpsel loose selection but have more than one shower.  The variable plotted is the shower transverse development angle which parameterizes the shower’s extension into the plane orthogonal to its principal axis.

Sidebands addressing all known major sources of background events were obtained by inverting the selection requirement on the BDT scores used by the electron neutrino selections. The result is the definition of separate sidebands for the \npsel and \zpsel selections. 
The corresponding datasets were opened progressively: first a ``low BDT" sideband, and then an ``intermediate BDT" sideband;  all input variables were checked in these sidebands.  As an example, Fig.~\ref{sfig:lowBDT} shows the distribution of the conversion distance for the reconstructed shower at  pre-selection stage for the \npsel low-BDT sideband. Events with a distance greater than 6 cm between the vertex and the shower are rejected by the loose selection, and the variable is further provided as input to the BDT.

Consistency between data and simulation is assessed through quantitative goodness-of-fit tests which show excellent agreement in all sidebands and selection stages, validating the background modeling with high statistics samples.

\subsection{NuMI beam data}
\begin{figure*}[ht!]
\subfloat[\npsel\label{sfig:numi_np}]{%
  \includegraphics[width=0.45\textwidth]{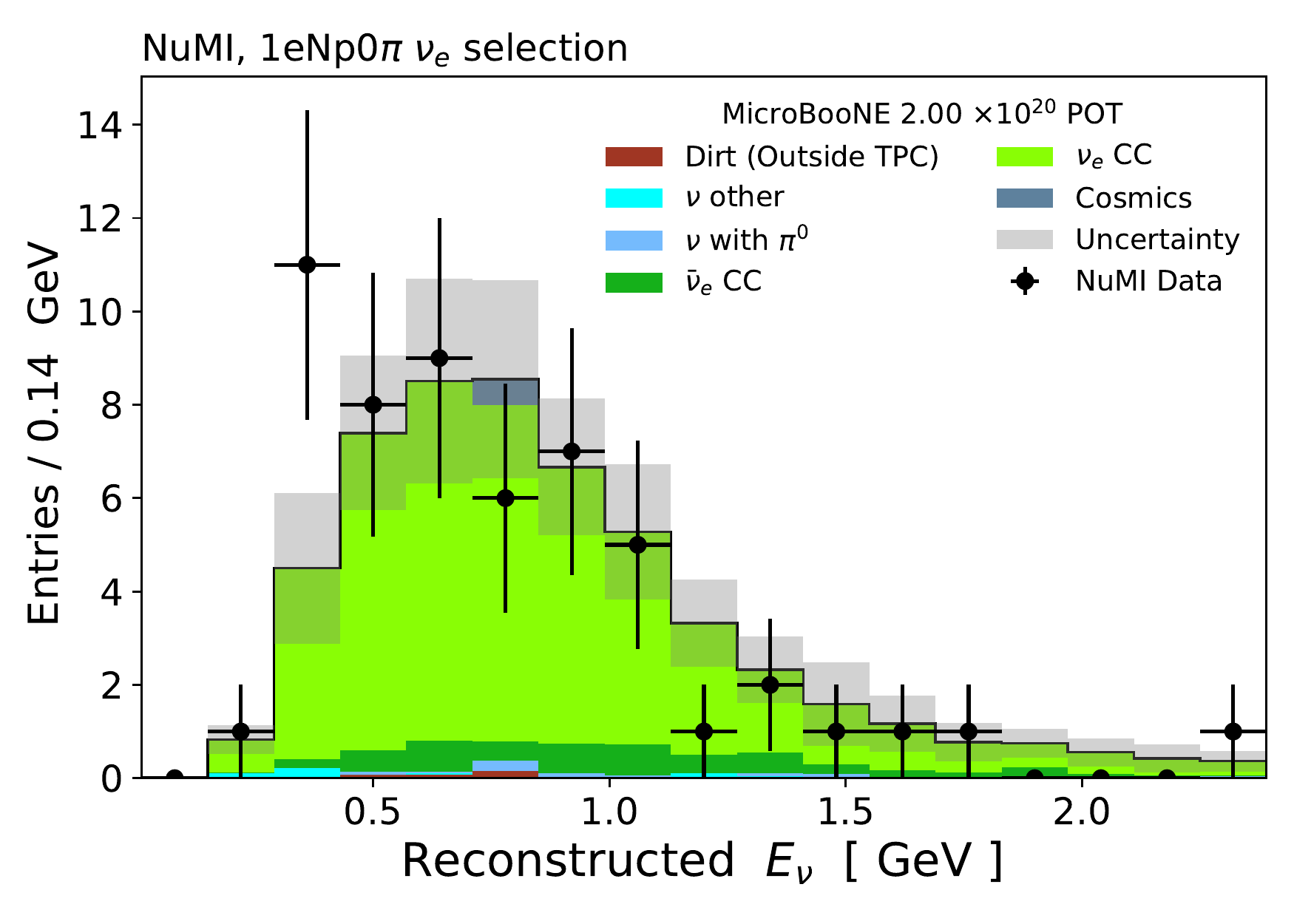}
}\hfill
\subfloat[\zpsel\label{sfig:numi_zp}]{%
  \includegraphics[width=0.45\textwidth]{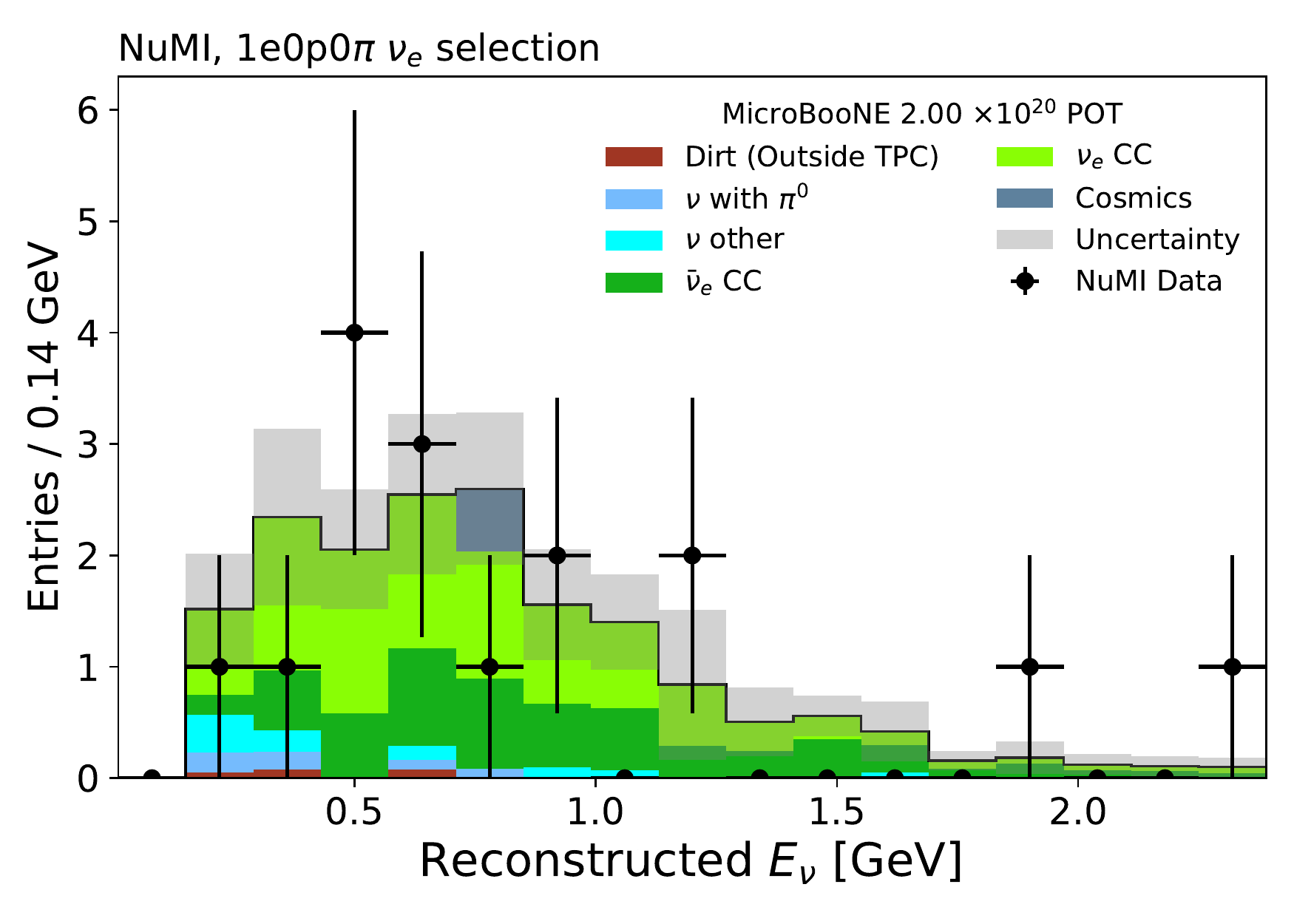}
}
\caption{\label{fig:numi} Reconstructed neutrino energy for events in \npsel~\protect\subref{sfig:numi_np} and \zpsel~\protect\subref{sfig:numi_zp} selections from the NuMI beam. In the \zpsel sample both the relatively large $\bar{\nu}_{e}$ content, due to similar \nue and $\bar{\nu}_{e}$ fluxes, as well as the low non-\nue background content, due to the relatively large \nue+ $\bar{\nu}_{e}$ fraction in the beam, are distinctive features in NuMI that make the spectrum different from the analogous predictions in BNB.}
\end{figure*}

Studying \nue interactions in data is crucial in order to validate the analysis selection performance and the \nue modeling. Measurements of \nue interactions on argon using the NuMI beamline~\cite{numibeam} have been performed by the ArgoNeuT~\cite{argoneutnue} collaboration as well as by MicroBooNE~\cite{ubnuminue,ubnuminueMCC9}. This makes the NuMI beam data collected at MicroBooNE a well understood sample which is particularly valuable for this validation.
Electron neutrinos from NuMI are produced almost entirely from the decay of unfocused kaons at the target, unlike those produced in the BNB which come approximately equally from pion and kaon decays.  They reach the MicroBooNE detector at about 27 degrees off the TPC axis and share a similar energy range and peak, around 1 GeV, with the electron neutrinos intrinsic to the BNB, but with a narrower distribution. 
Results from applying the \npsel and \zpsel selections to NuMI data from MicroBooNE's first year of operations are shown in Fig.~\ref{fig:numi}. Both the \npsel and \zpsel channels are predicted to have a \nue+ $\bar{\nu}_e$ purity of 87\%. The relatively high \nue + $\bar{\nu}_e$ content of the NuMI beam, 5\% relative to 0.5\% in the BNB, contributes to low predicted background levels in NuMI compared to those predicted in the BNB and makes the NuMI beam insensitive to new electron neutrino signatures in this analysis. 
The comparable \nue and $\bar{\nu}_e$ contributions to the NuMI flux, combined with the smaller rate of final-state protons in $\bar{\nu}_e$ interactions, leads to a $\sim$40\% $\bar{\nu}_e$ component in the \zpsel prediction.
In the \zpsel channel we observe 16 events with 16.9 predicted, and 54 with 53 predicted in the \npsel channel.
The good level of agreement between observed and predicted \nue events on a beamline that has been used for multiple \nue cross section measurements on argon provides a strong validation of the selection's ability to identify electron neutrinos in data.

\subsection{Progressive data opening for BNB events}

A small dataset was fully open during the development of this analysis: $4 \times 10^{19}$ POT from the first period of data taking and $1 \times 10^{19}$ POT from the third period of data taking. Together these correspond to less than 10\% of the full $6.86 \times 10^{20}$ POT used for the results presented in this article. This small open dataset was first used to develop and test the MicroBooNE LArTPC event reconstruction. It was also used to validate agreement between data and simulation during analysis development, where all inputs to the BDTs and other variables of physics interest were cross-checked at each selection stage.
After the selections were frozen and validated in sideband data the analysis was tested with several fake data sets. These were created with and without an injected electron-like signal as well as with simulation modifications. Results with the frozen analysis on these fake data sets were found to be consistent with the true injected signal.
Unblinding was performed in stages moving progressively from background-enriched sidebands and high-energy \nue regions towards the low-energy region in the BNB where the eLEE signal is predicted according to the model described in \ref{sec:eleemodel}.
This was done in three energy regions.
The high-energy region was defined as $E^{\nu_e}_{\textrm{reco}} > 0.85$ (0.90) GeV for the \npsel (\zpsel) selection,  the medium-energy region as $E^{\nu_e}_{\textrm{reco}} > 0.65$ GeV for both selections, and the low-energy region as $E^{\nu_e}_{\textrm{reco}} > 0.15$ GeV  for both selections.
Selection criteria, including BDTs, were frozen before opening the first high-energy \nue sideband. 
Following the opening of the medium and high energy regions, goodness-of-fit $p$-values of 0.277, 0.206, and 0.216 were measured respectively for \npsel events, \zpsel events, and the two combined. These are obtained comparing the observed data to the prediction from simulation after applying the \numu constraint procedure described in the next section.
The level of consistency with the underlying prediction supported the decision to move forward with unblinding the full energy range.

\section{Systematics and Application of the \numu Constraint}
\label{sec:constraint}

We separate the sources of systematic uncertainty into three main categories: flux, cross section, and detector response uncertainties.
Details about variations for each of these categories are presented in Sec.~\ref{sec:simulations}. Uncertainties associated with the statistics of the simulation samples used in the analysis are also included. Uncertainties related to the flux, particle propagation, and, partially, cross section are assessed through multi-universe simulations which are generated by varying the underlying model parameters within their range of uncertainty. Detector response and several cross section model uncertainties are assessed through single variations of the underlying simulation model, referred to as uni-sims. Systematic uncertainties are included in the analysis through the formalism of the covariance matrix:
\begin{eqnarray}
    \label{eq:errorsyst}
    C^\textrm{Syst} &=& C^\textrm{Flux} + C^\textrm{XSec} + C^\textrm{Detector} + C^\textrm{MCstat} \\
    \label{eq:multiverse}
    C_{ij} &=& \frac{1}{N}\sum_{k=1}^{N} \left(n_i^{k} - n_i^{\textrm{CV}}\right)\left(n_j^{k} - n_j^\textrm{CV}\right),
\end{eqnarray}
where $C$ indicates a covariance matrix, $i,j$ are indices over histogram bins, $n^{\textrm{CV}}$ is the nominal (central value) bin content, $n^{k}_{i}$ is the content of the $i$th bin in the alternative universe $k$, and $N$ is the total number of alternative universes.
Uni-sim variations are symmetrized in the covariance matrix approach adopted in the analysis.

Uncertainties are constrained by leveraging the correlations between \numu and \nue events, which share common flux parentage in their decay chain in the beamline and significant overlap in the cross sections that govern their interaction rate and final-state kinematics. Through correlations for shared sources of modeling uncertainty, the high-statistics measurement of \numu (see Sec.~\ref{sec:numu}) is used to update the \nue prediction and constrain the total systematic uncertainty. When considering all uncertainties (cross-section uncertainties only), the level of correlation between \numu and \nue events in the signal region is 60\% (80\%). Throughout this analysis, the \numu  flux is assumed to be unoscillated.

The \numu constraint is implemented by relying on the covariance between \numu and \nue bin contents, the predicted bin content in the different channels, and the observed \numu data, and leveraging the properties of block matrices~\cite{alma991015286869705251}. Given the bin-to-bin covariance matrices for the \numu channel ($C^{\mu\mu}$), \nue channel ($C^{ee}$), and the covariance between the two channels ($C^{e\mu}$), the predicted bin content in each bin $i$, $m_i^{e}$ and $m_i^{\mu}$ for \nue and \numu respectively, and the \numu observed bin contents $n_i^{\mu}$, the constrained \nue prediction is expressed as:
\begin{equation}
\label{eq:constraint:cv}
    m^{e\;\textrm{constrained}} = m^e + C^{e\mu} \left( C^{\mu\mu}\right)^{-1} \left( n^{\mu} - m^{\mu} \right),
\end{equation}
and the constrained covariance matrix as:
\begin{equation}
\label{eq:constraint:syst}
    C^{ee\;\textrm{constrained}} =  C^{ee} - C^{e\mu}\left( C^{\mu\mu}\right)^{-1}C^{\mu e}.
\end{equation}

\begin{figure*}[ht]
\begin{center}
\includegraphics[width=0.95\textwidth]{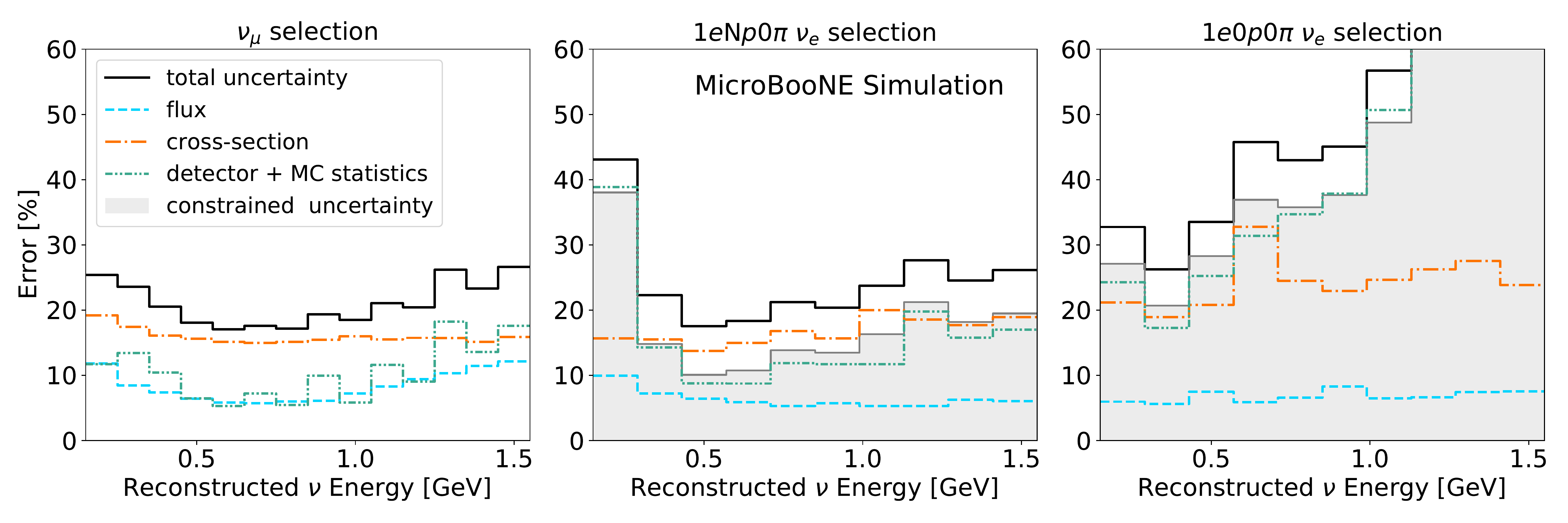}
\caption{\label{fig:systematics}Summary of the impact of systematic uncertainties presented in Sec.~\ref{sec:simulations} for all selected events in the three channels used in the analysis, shown in the 0.15--1.55 GeV energy range that is used for quantitative results. The percent systematic uncertainty by channel is shown in the top panel. The individual unconstrained contributions coming from flux, cross section, and detector plus simulation statistics are shown in blue, orange, and green respectively. Detector uncertainties account for both Geant4 re-interaction and detector response modeling uncertainties. The total unconstrained systematic uncertainty is in black and, for the \npsel and \zpsel selections, a grey area indicates the total constrained uncertainty. The \numu uncertainties are not changed by the constraint. In the high energy region of the \zpsel energy spectrum, where we have very few events, the uncertainties grow up to $\mathcal{O}$(100\%).}
\end{center}
\end{figure*}

The fractional systematic uncertainty is presented in Fig.~\ref{fig:systematics} for the three channels included in this analysis (\numu, \npsel, and \zpsel). 
The \numu selection has no requirement on hadron multiplicity, so the data it selects can constrain both the \npsel and \zpsel prediction. 
This selection aims to maximize the reduction of flux uncertainties, particularly at low reconstructed neutrino energy.
Overall, the constraint reduces the systematic uncertainties in the electron neutrino selections by 10$\textrm{--}$40\% relative to the pre-constraint uncertainties. In the analysis, the \numu constraint is performed on distributions of the \numu and \nue reconstructed neutrino energy and applied to all quantitative results presented in the next section.

\section{Results}
\label{sec:results}

We first present results from the \nue selections to test the agreement between the observation and neutrino interaction model prediction and then the tests of the eLEE model.
All statistical tests in this analysis are performed over the range 0.15-1.55 GeV in reconstructed neutrino energy with ten 0.14 GeV bins.
The test-statistic used is a $\chi^2$ defined as
\begin{eqnarray}
    \label{eq:chisq}
    \chi^2 &=& \sum_{i,j=1}^{N}\left(n_i - m_i\right) C^{-1}_{ij}\left( n_j - m_j \right) \\
    C_{ij} &=& C^{\textrm{stat CNP}}_{ij} + C^{\textrm{syst}}_{ij}, 
    \label{eq:errmatrix}
\end{eqnarray}
with $n_i$ the observed number of events in bin $i$, $m_i$ the predicted number of events in bin $i$ for the model being tested, and $C_{ij}$ the covariance matrix, defined in Eq.~\ref{eq:errmatrix}.
Statistical uncertainties are the largest in this analysis and are included through the error matrix $C^{\textrm{stat CNP}}_{ij} = 3/\left(1/n_i +2/m_i  \right)\delta_{ij}$ that is constructed using the combined Neyman-Pearson $\chi^2$ definition of Ref.~\cite{cnp}. The systematic error matrix $C^{\textrm{syst}}_{ij}$ is defined in Eq.~\ref{eq:errorsyst}.
Toy experiments are generated incorporating systematic variations with a Gaussian sampling of the constrained covariance matrix and then Poisson fluctuating the result to account for statistical variations. For each toy experiment the test-statistic is evaluated, and the distribution is compared to the test-statistic of the data to extract a p-value for the observation.
Alternative statistical procedures were also used to validate the results shown in the next sections, and led to similar conclusions as the ones presented. 

\subsection{Modeling of electron neutrinos}
\label{sec:results:gof}
\begin{figure*}[ht!]
\subfloat[\npsel\label{sfig:bnb_np}]{%
  \includegraphics[width=0.45\textwidth]{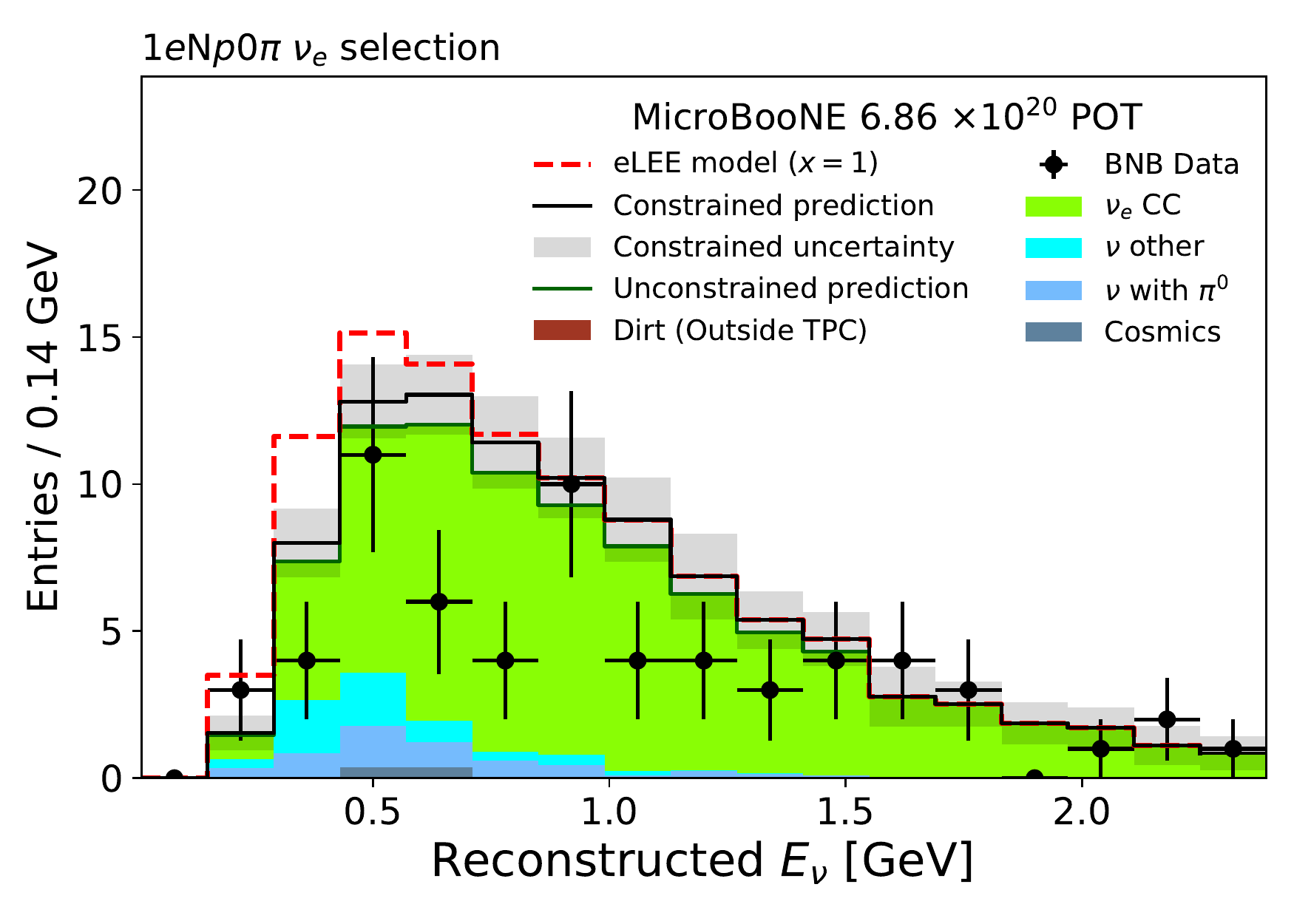}
}\hfill
\subfloat[\zpsel\label{sfig:bnb_zp}]{%
  \includegraphics[width=0.45\textwidth]{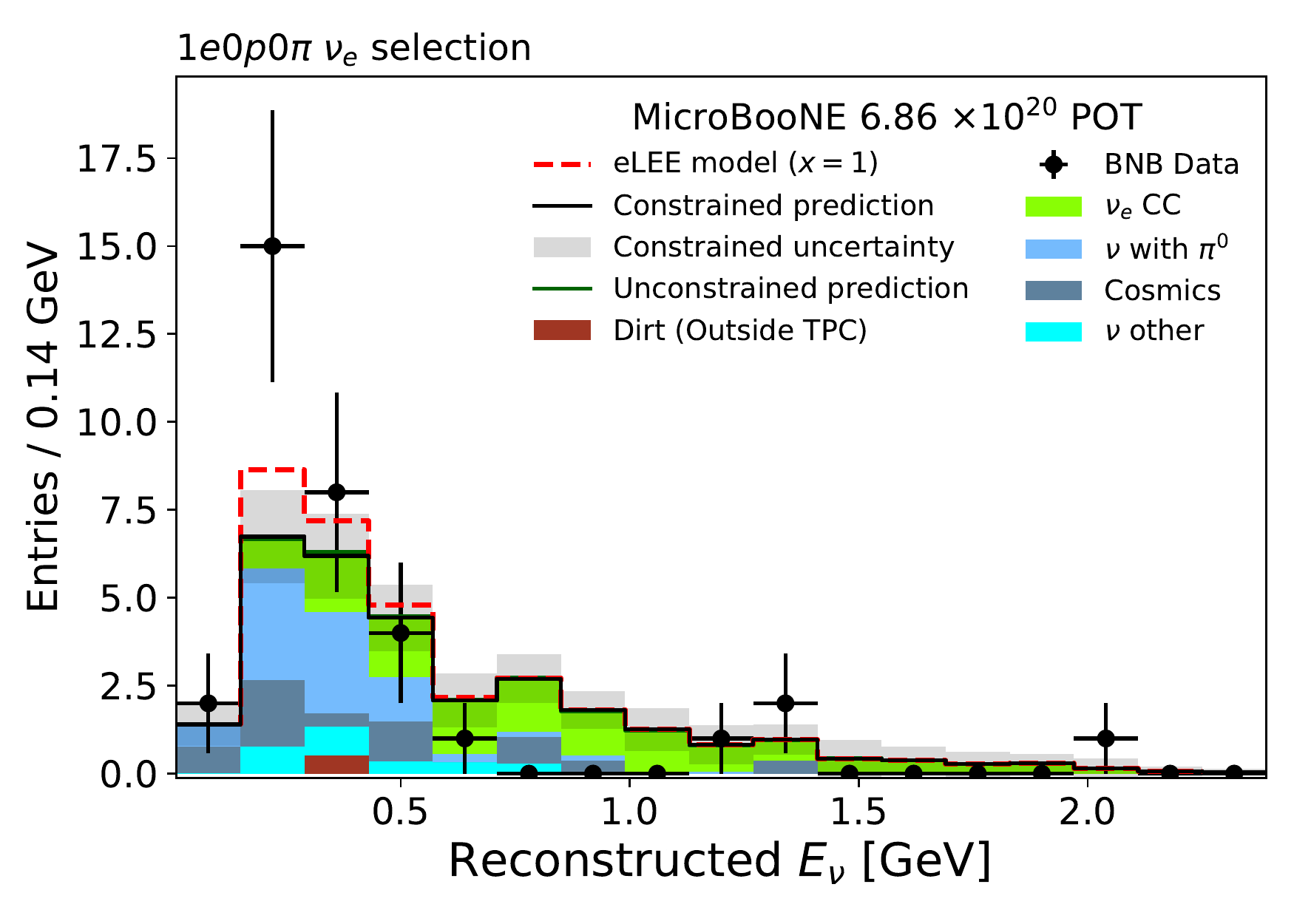}
}
\caption{\label{fig:ereco} Reconstructed neutrino energy for the selected \npsel~\protect\subref{sfig:bnb_np} and \zpsel~\protect\subref{sfig:bnb_zp} events. The pre-constraint number of predicted events is shown broken down by true interaction topology.  The constrained prediction using the muon neutrino data is also shown both with (red) and without (black) the eLEE model included. Systematic uncertainties on the constrained prediction are shown as a shaded band. While not shown in the figure, systematic uncertainties on the eLEE model are considered in the analysis. Quantitative results are calculated in 10 bins from 0.15 to 1.55 GeV, shown here starting in the second bin. The lower bound of the first bin is 0.01 GeV.}
\end{figure*}

The observed \npsel and \zpsel event rates are plotted as a function of reconstructed energy in Fig.~\ref{fig:ereco}, where data is compared to the prediction after the \numu constraint. Given the agreement observed with the \numu selection (Fig.~\ref{fig:numu}), the effect of the constraint procedure on the \nue prediction is relatively small (less than $10\%$). 

The first statistical test performed is a goodness of fit to quantify how well the intrinsic \nue model matches the data in reconstructed neutrino energy. 
The results are presented in Table~\ref{tab:gof}. 
The data are consistent with the intrinsic \nue model with $p$-values of 0.182, 0.126, and 0.098 for \npsel events, \zpsel events, and the two combined, respectively.  

\begin{table}[ht!]
\begin{center}
\caption{\label{tab:gof}Summary of $\chi^2$ and $p$-value results for the goodness-of-fit tests for the intrinsic \nue model. The $p$-values are computed with frequentist studies based on toy experiments. }
\begin{tabular}{{c c c c}}
\hline \hline
Channel & $\chi^2$ & $\chi^2$/dof & $p$-value\\
\hline 
\npsel & 15.2 & 1.52 & 0.182 \\
\zpsel & 16.3 & 1.63 & 0.126 \\
\npsel + \zpsel & 31.50 & 1.58 & 0.098 \\
\hline \hline
\end{tabular}
\end{center}
\end{table}

\begin{figure*}[ht!]
\subfloat[Electron angle relative to beam direction. \label{sfig:bnb_np_etheta}]{%
  \includegraphics[width=0.45\textwidth]{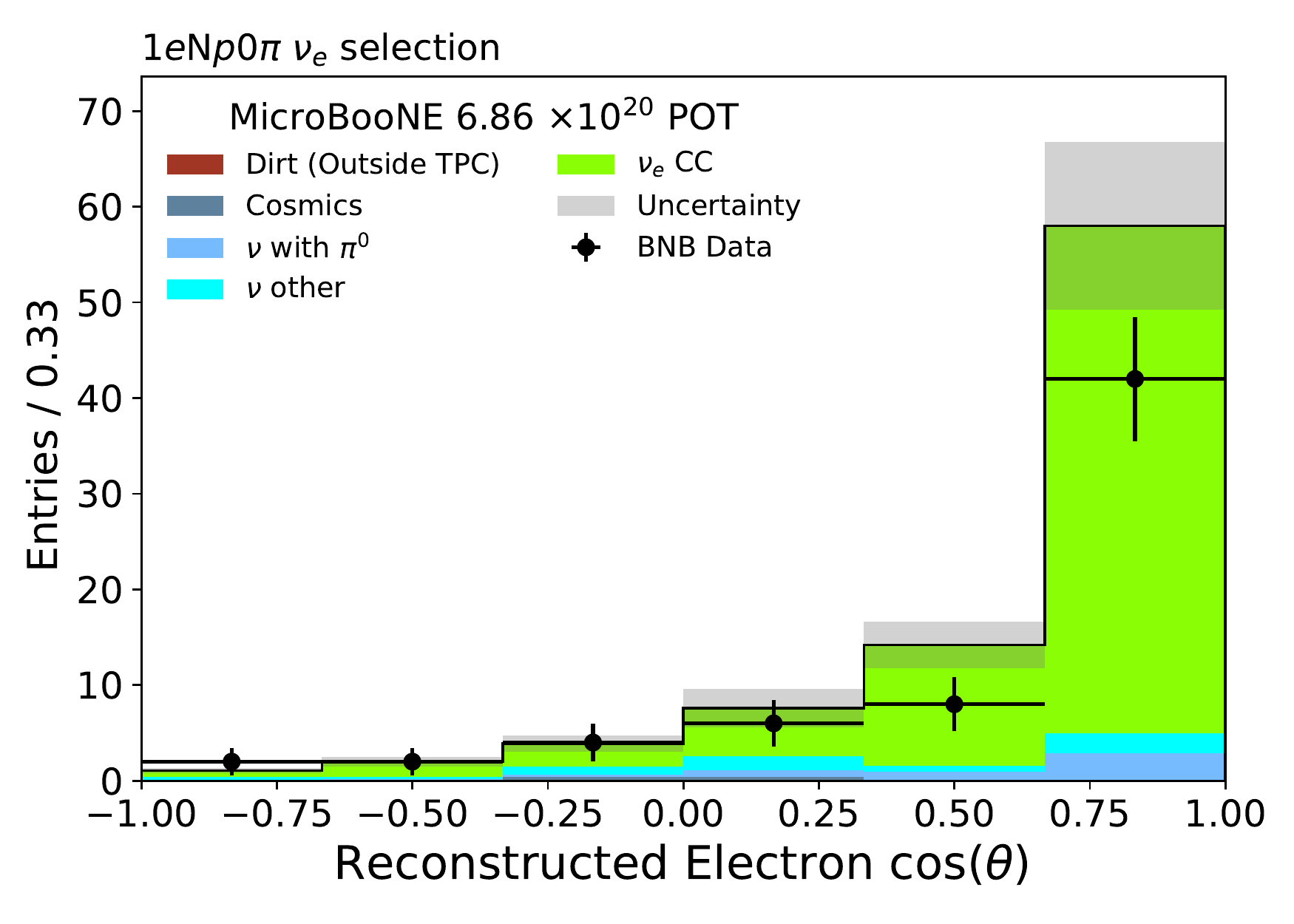}
}\hfill
\subfloat[Proton kinetic energy. \label{sfig:bnb_np_pke}]{%
  \includegraphics[width=0.45\textwidth]{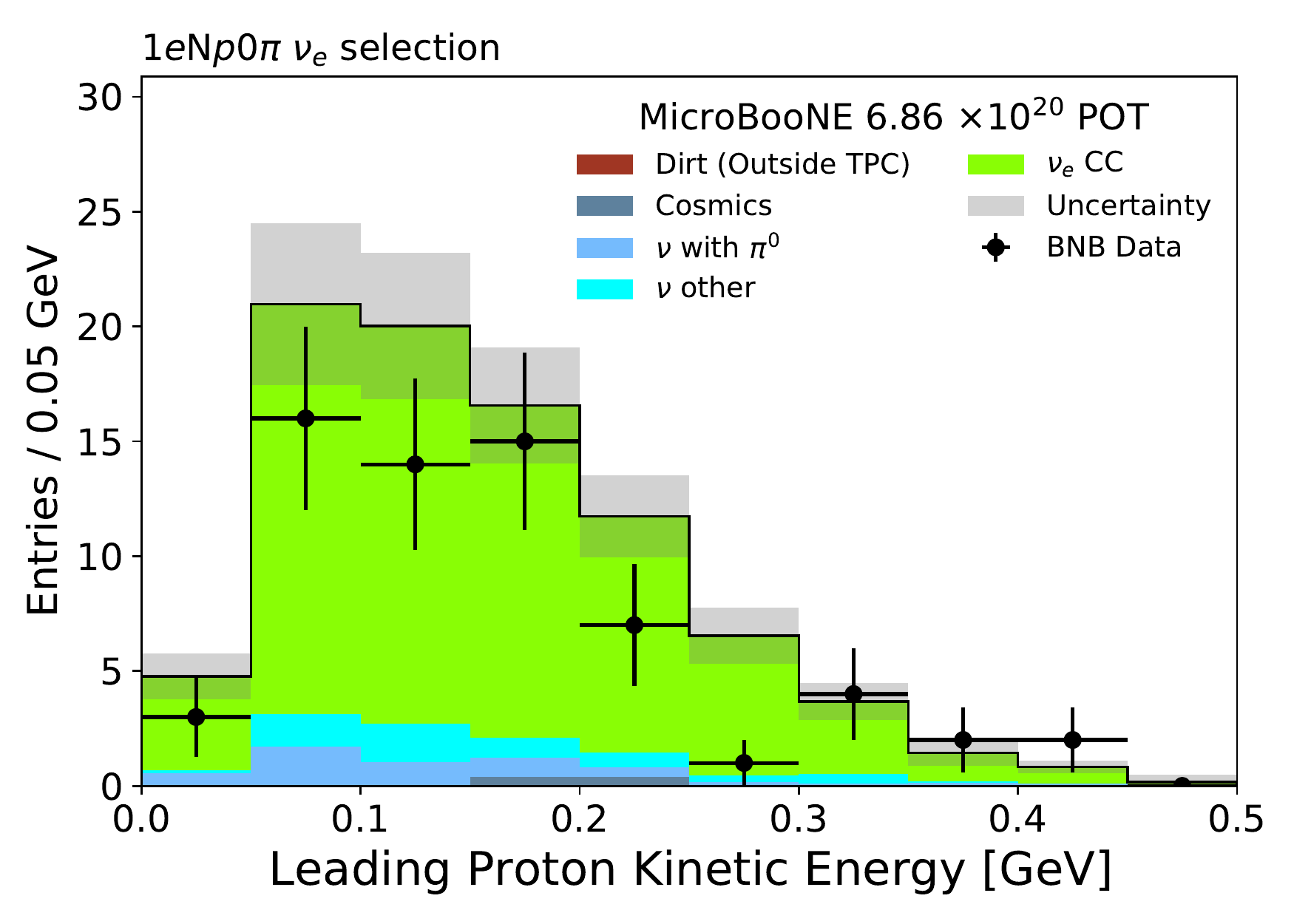}
}
\caption{\label{fig:1eNp:kinematics} Selected kinematic distributions for events that pass the \npsel selection. Expected events and uncertainties are shown as predicted by the nominal simulation.}
\end{figure*}

Electron neutrino events can be further characterized in terms of their kinematics. Figure~\ref{fig:1eNp:kinematics} shows the angle ($\theta$) of the reconstructed electron candidate relative to the beam direction and the kinetic energy of the leading proton for \npsel events.
Considering only normalization, in the 0.15-1.55 GeV range, 53 events are observed by the \npsel channel with a constrained prediction of $78.9 \pm 11.6$ events, corresponding to a deficit of 1.7$\sigma$.
The deficit is most pronounced at intermediate energies and in the forward direction. As demonstrated by the $p$-value obtained by the goodness of fit test, the estimated combined statistical and systematic uncertainties accommodate the observed difference.

\begin{figure*}[ht!]
\subfloat[All selected events. \label{sfig:bnb_zp_costheta}]{%
  \includegraphics[width=0.45\textwidth]{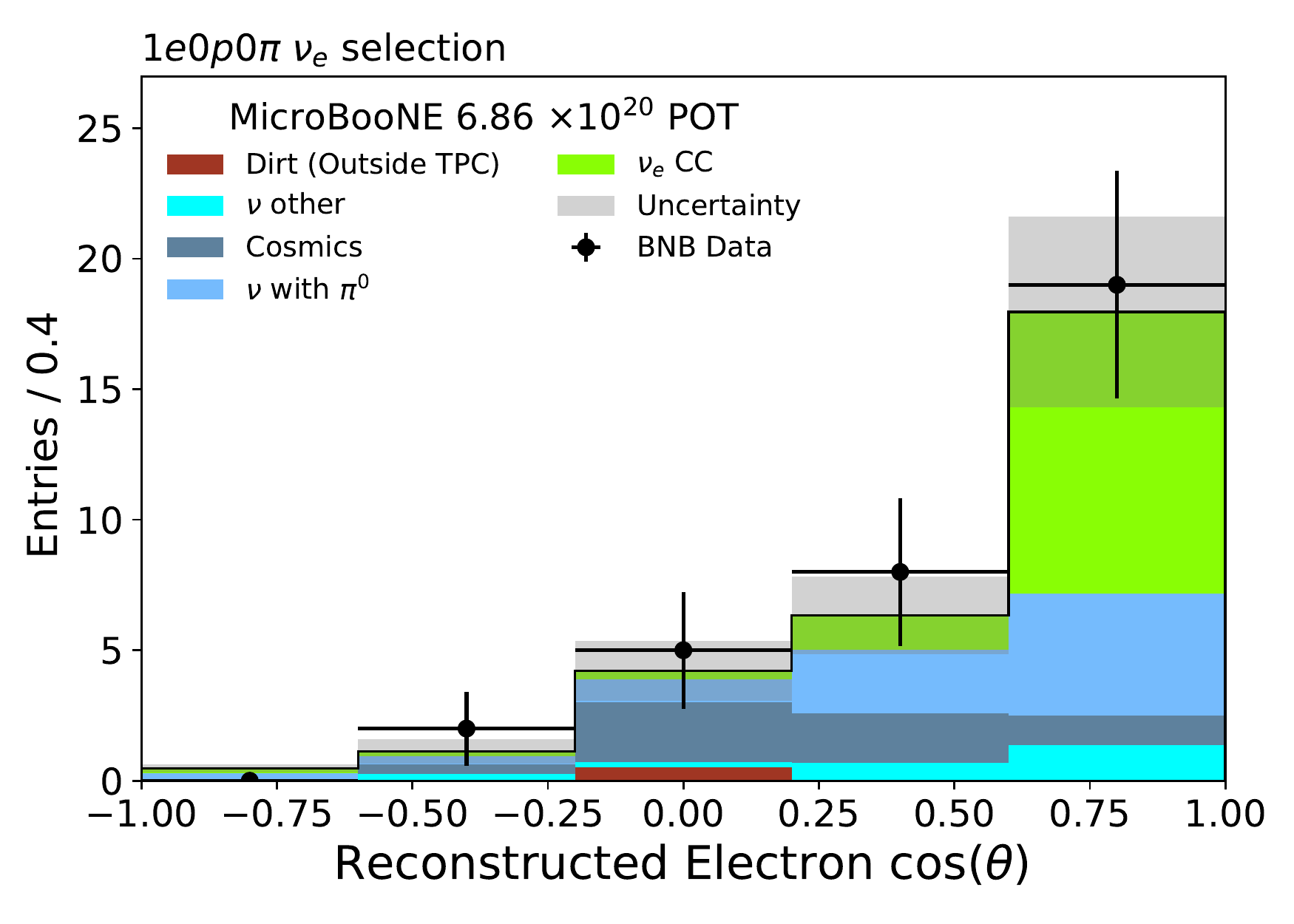}
}\hfill
\subfloat[Low energy selected events from 0.15$\textrm{--}$0.43 GeV. \label{sfig:bnb_zp_costheta_lowe}]{%
  \includegraphics[width=0.45\textwidth]{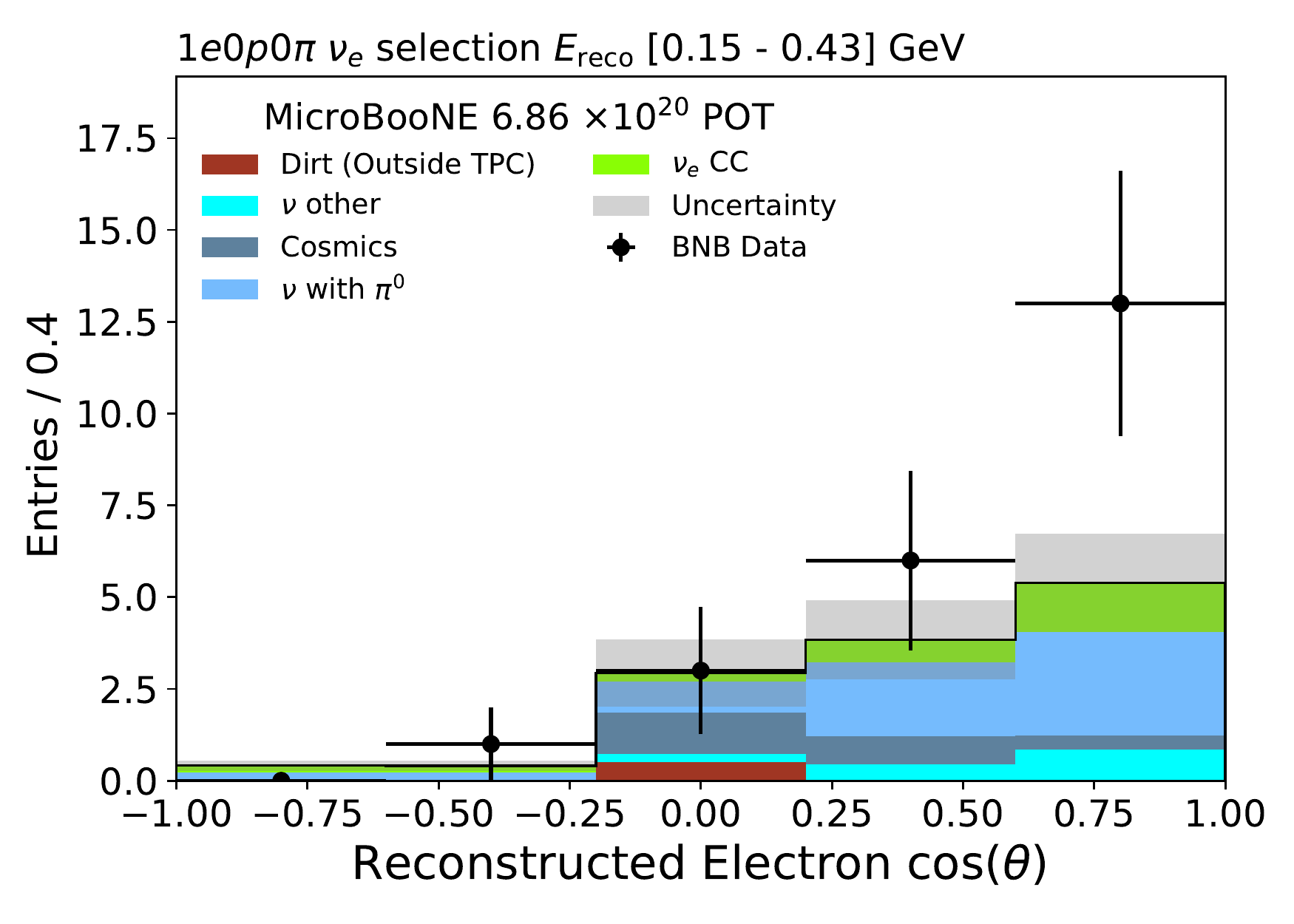}
}
\caption{\label{fig:1e0p_angle} Selected \zpsel events as a function of electron angle with respect to the beam. Expected events and uncertainties are shown, without the \numu constraint applied.}
\end{figure*}

In the \zpsel channel we observe good overall normalization agreement between data and simulation, with 31 events observed compared to a constrained prediction of $27.8 \pm 4.4$, 
but the simulation under-predicts the data in the energy bins corresponding to $0.15 < E_{\text{reco}} < 0.43$ GeV.
In multiple $\pi^0$ enriched sideband regions, however, the data is consistent with the prediction within statistical and systematic uncertainties as reported in Sec.~\ref{sec:sidebands}.
In the $0.15 < E_{\text{reco}} < 0.43$ GeV energy bins of the NuMI sideband, the high electron neutrino purity and satisfactory agreement between data and simulation validate the interaction model for low energy electron neutrinos.
We find that data events in this range are dominated by single-shower events with a \dedx profile consistent with a minimum ionizing particle, as expected for both signal and most surviving background events from simulation.
Figure~\ref{fig:1e0p_angle} shows the angular distribution of \zpsel events, both over the full energy range and in the region corresponding to $0.15 \textrm{--} 0.43$ GeV.
Integrated over the full energy range, the angular distribution shows good agreement with simulation. 
In the low energy bins, where the simulation under-predicts the observed data, events primarily populate the region with $\cos\theta>0$.
More data will be needed to further study these events.
The observation of good shape agreement between the data and the prediction in the leading proton kinetic energy distribution shown in Fig.~\ref{sfig:bnb_np_pke}, as well as a visual scan of selected events, suggests that the migration between the \npsel and \zpsel selections is minimal.

\subsection{Test of the eLEE model}
\label{sec:results:elee}

\begin{figure*}[ht!]
\subfloat[\npsel\label{sfig:statsnpsh}]{%
  \includegraphics[width=6cm]{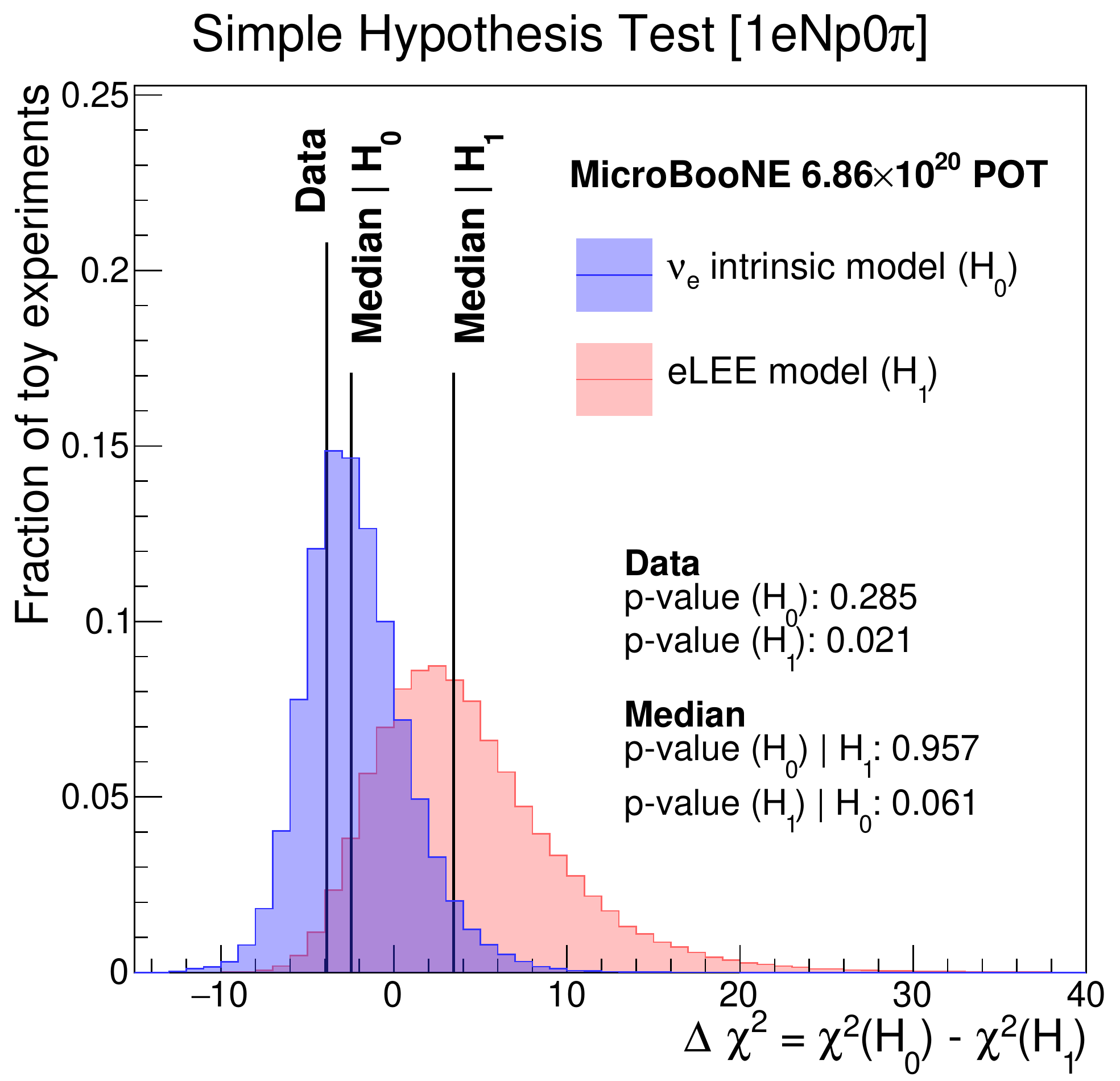}}\hspace{-0.2    cm}
\subfloat[\zpsel\label{sfig:statszpsh}]{%
  \includegraphics[width=6cm]{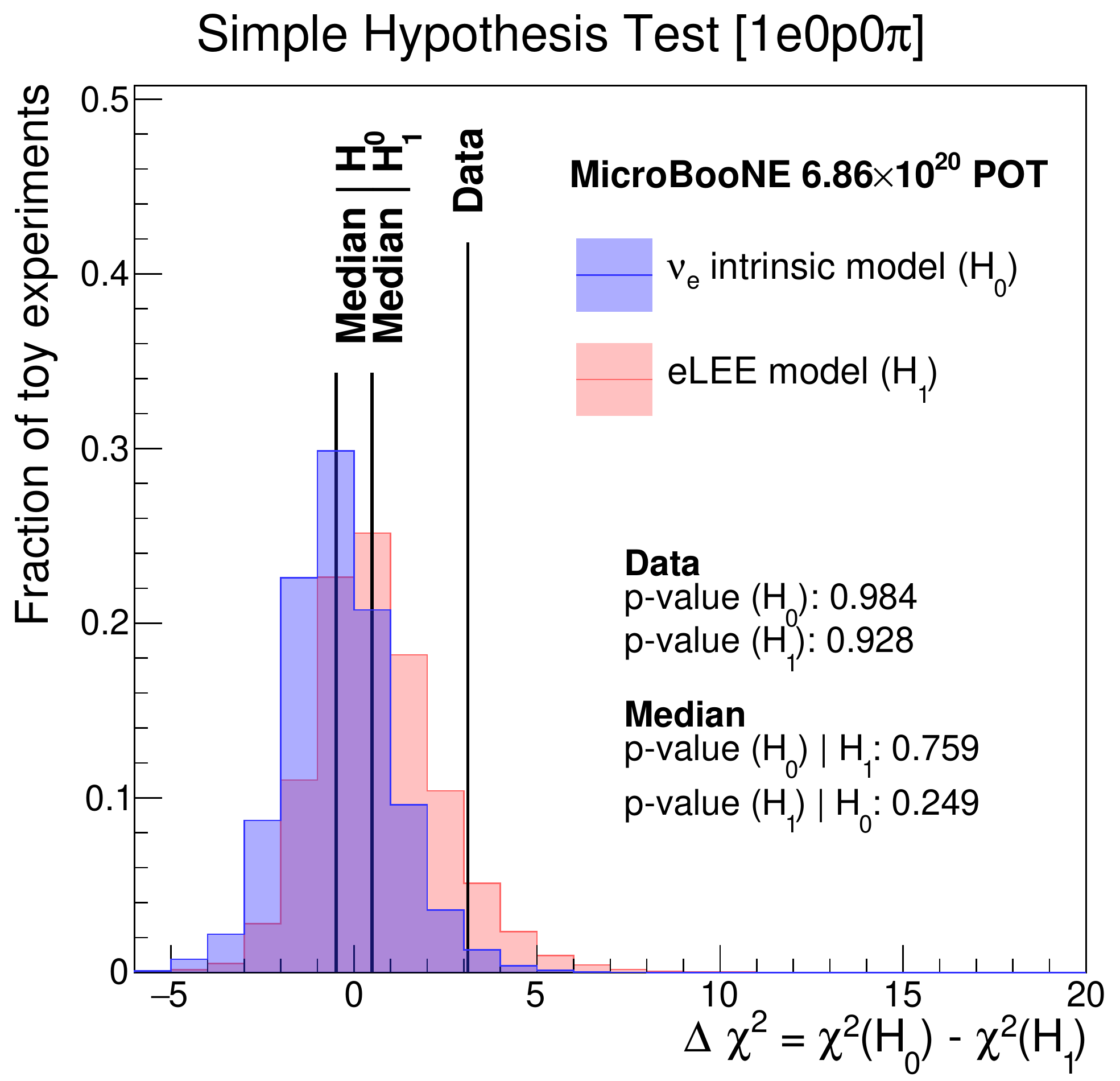}}\hspace{-0.2cm}
\subfloat[\npsel+\zpsel\label{sfig:statsnpzpsh}]{%
  \includegraphics[width=6cm]{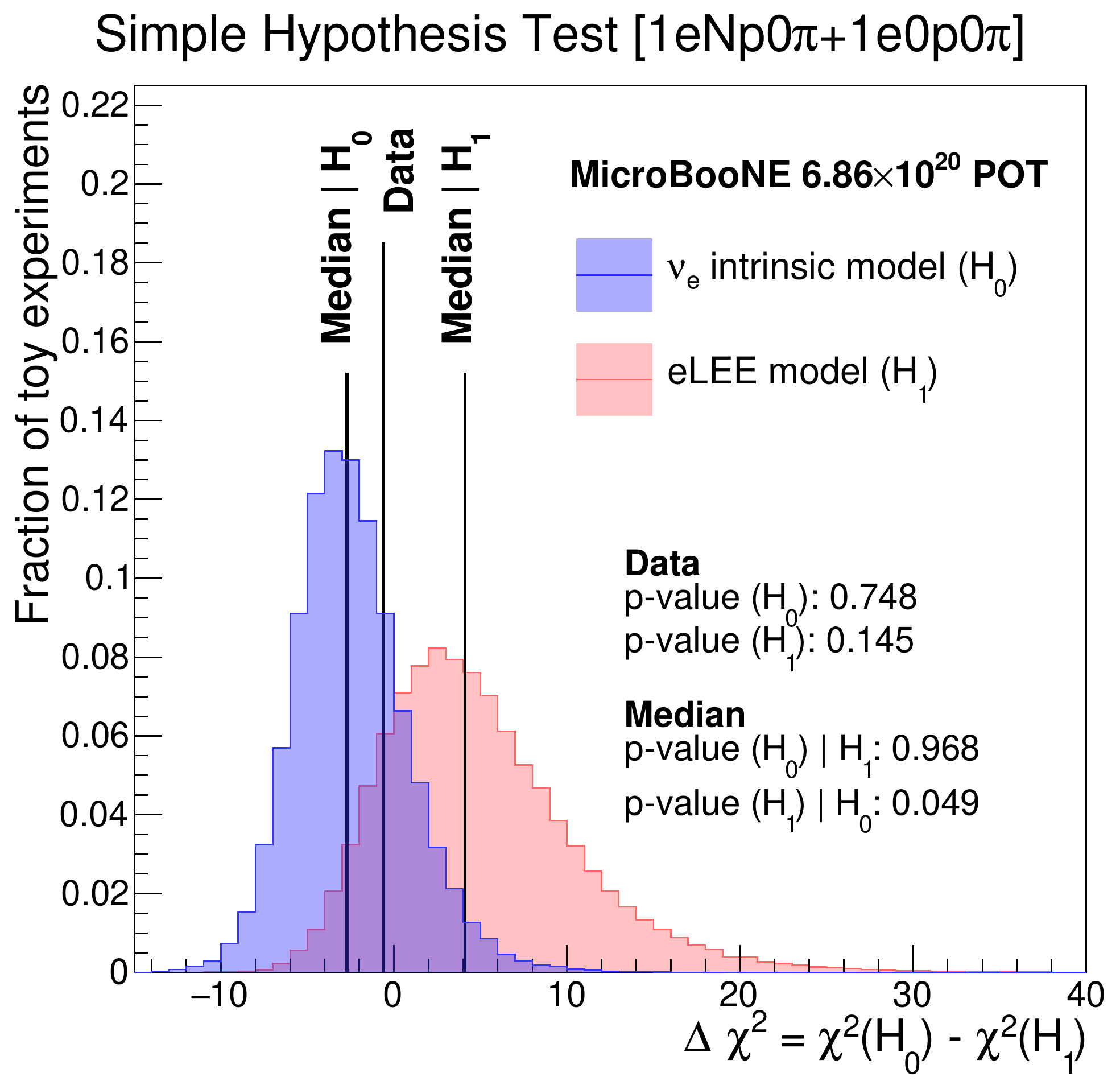}}
\caption{\label{fig:statsSH} 
Results for the simple hypothesis test in the \npsel~\protect\subref{sfig:statsnpsh}, \zpsel~\protect\subref{sfig:statszpsh}, and combined~\protect\subref{sfig:statsnpzpsh} channels.
The $\Delta \chi^2$ between the intrinsic \nue model and the eLEE hypotheses is plotted for toy experiments generated with these hypotheses. The $p$-values indicate the fraction of toy experiments with $\Delta \chi^2$ smaller than the observation. The median $p$-value for toy experiments produced assuming the intrinsic \nue model is also reported.}
\end{figure*}

The statistical tests performed to assess the probability that the eLEE model introduced in Sec.~\ref{sec:simulations} is present in the data are described in this section.
The first is a simple hypothesis test in which two  hypotheses  are  tested  in  order  to  assess  the probability of rejecting one hypothesis assuming the other is true. The hypotheses tested are the intrinsic \nue prediction ($H_0$) and the intrinsic \nue prediction plus the eLEE model contribution ($H_1$). The result reported from this test is the $p$-value based on the $\Delta \chi^2$ between $H_0$ and $H_1$ defined as:  
\begin{equation}
\label{eq:deltachisq}
\Delta\chi^2 = \chi^2\left(H_0\right) - \chi^2\left(H_1\right).
\end{equation}
The expected sensitivity and the data results are presented in Fig.~\ref{fig:statsSH} and summarized in Table~\ref{tab:simple}.
In the \npsel channel, the data are consistent with $H_0$ with a $p$-value of 0.285, which corresponds to 28.5\% of the toy experiments that assume $H_0$ is true showing a $\Delta\chi^2$ smaller than the observed value. 
When the inverse test is performed, with toy experiments assuming $H_1$ is true, we find a $p$-value of 0.021 thus implying that the \npsel channel alone excludes the $H_1$ hypothesis at the 97.9\% confidence level (CL).
In the \zpsel channel, as shown in Fig.~\ref{sfig:statszpsh}, the observed $\Delta\chi^2$ falls in the tail of the expected distribution from both hypotheses. The observation indicates a preference for the $H_1$ over the $H_0$ hypothesis with a fraction of toys in the tail of 0.072 for $H_1$ and of 0.016 for $H_0$.
While the combined results are expected to be driven by the larger sensitivity of the \npsel channel to the model tested, the preference for $H_1$ in the \zpsel channel leads to an intermediate result between the two hypotheses. 

\begin{table*}[t!]
\begin{center}
\caption{\label{tab:simple} Summary of the simple hypothesis tests. Reported $p$-value$(H_0)$ ($p$-value$(H_1)$) results reflect the probability for the $H_0$ ($H_1$) hypothesis to give $\Delta\chi^2 = \chi^2\left(H_0\right) - \chi^2\left(H_1\right)$ smaller than the observed value. The observed value of $\chi^2(H_0)$ is reported in Table~\ref{tab:gof}. The median sensitivity in terms of these $p$-values is also reported under the assumption that the eLEE model $H_1$ (no-signal scenario $H_0$) is true. The fraction of toy experiments generated under the $H_0$ hypothesis with $\Delta \chi^2$ larger than the median value obtained for the eLEE model $H_1$ is $1-p$-value$(H_0)$ so the combined \npsel + \zpsel median sensitivity to reject $H_0$ if $H_1$ is true is 0.032.}
\begin{tabular}{{cccccc}}
\hline \hline
 & obs. & $\Delta\chi^2 < \textrm{obs.}$ & $\Delta\chi^2 < \textrm{obs.}$ & Sensitivity & Sensitivity \\
Channel & $\Delta\chi^2$ & $p$-value$(H_0)$ & $p$-value$(H_1)$ & $p$-value$(H_0)$ \(|\)$H_1$  & $p$-value$(H_1)$ \(|\)$H_0$ \\
\hline
\npsel & -3.89 & 0.285 & 0.021 & 0.957 & 0.061 \\ 
\zpsel &  3.11 & 0.984 & 0.928 & 0.759 & 0.249 \\
\npsel+\zpsel & -0.58 & 0.748 & 0.145 & 0.968 & 0.049 \\
\hline \hline
\end{tabular}
\end{center}
\end{table*}

\begin{figure*}[ht!]
\subfloat[\npsel\label{sfig:statsnpfc}]{%
  \includegraphics[width=6.cm]{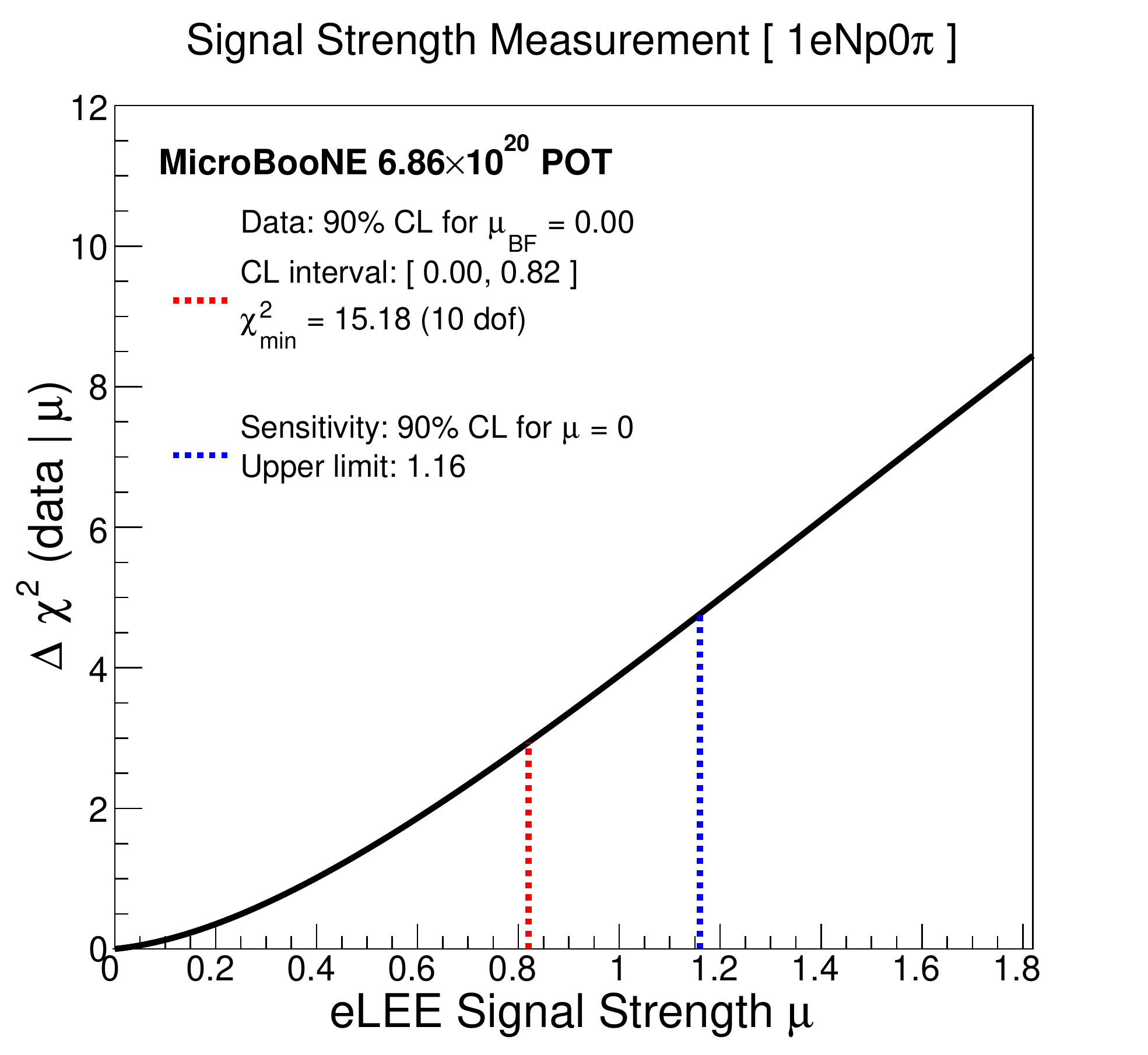}
}\hspace{-0.3cm}
\subfloat[\zpsel\label{sfig:statszpfc}]{%
  \includegraphics[width=6.cm]{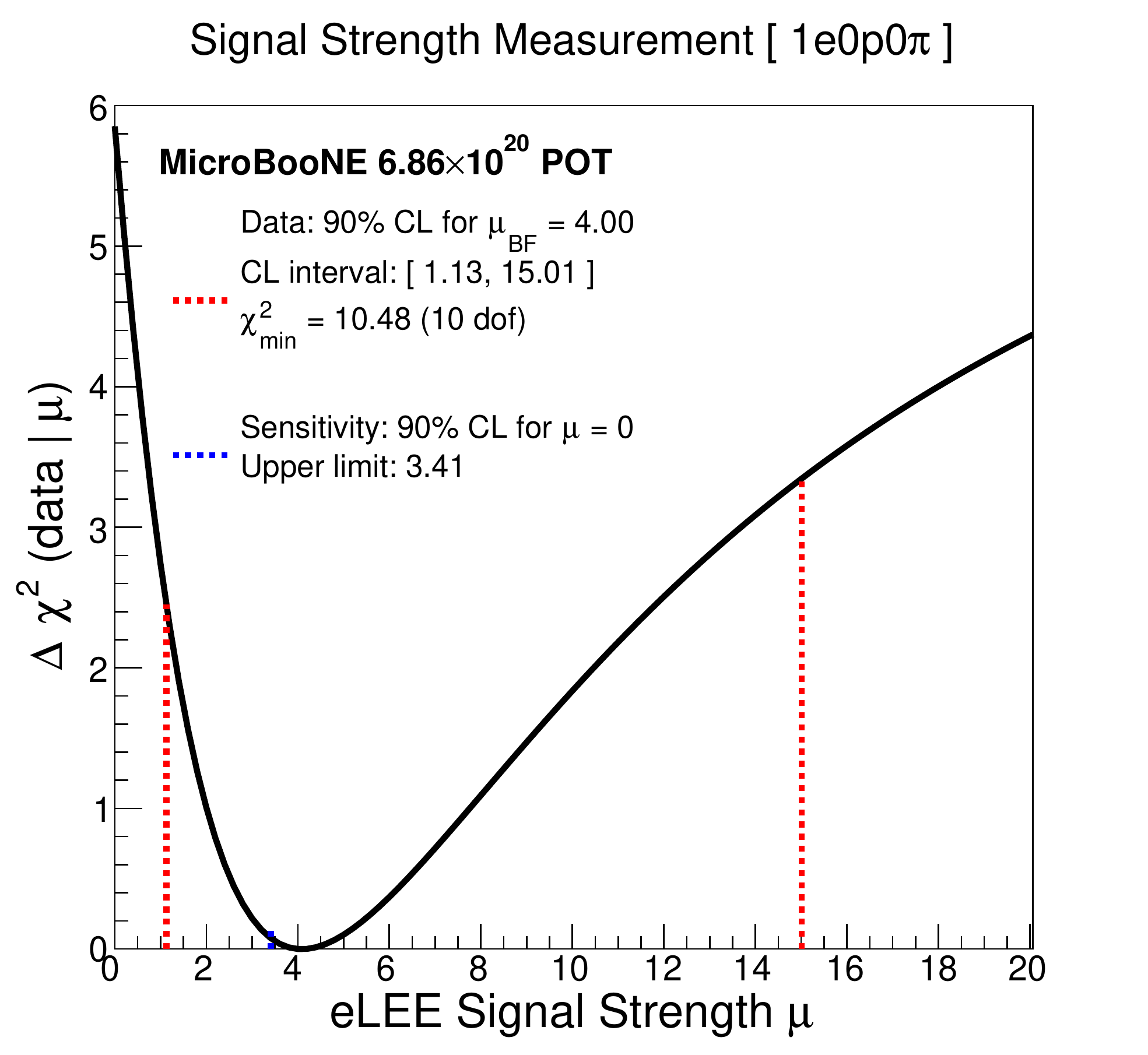}
}\hspace{-0.3cm}
\subfloat[\npsel+\zpsel\label{sfig:statsnpzpfc}]{%
  \includegraphics[width=6.cm]{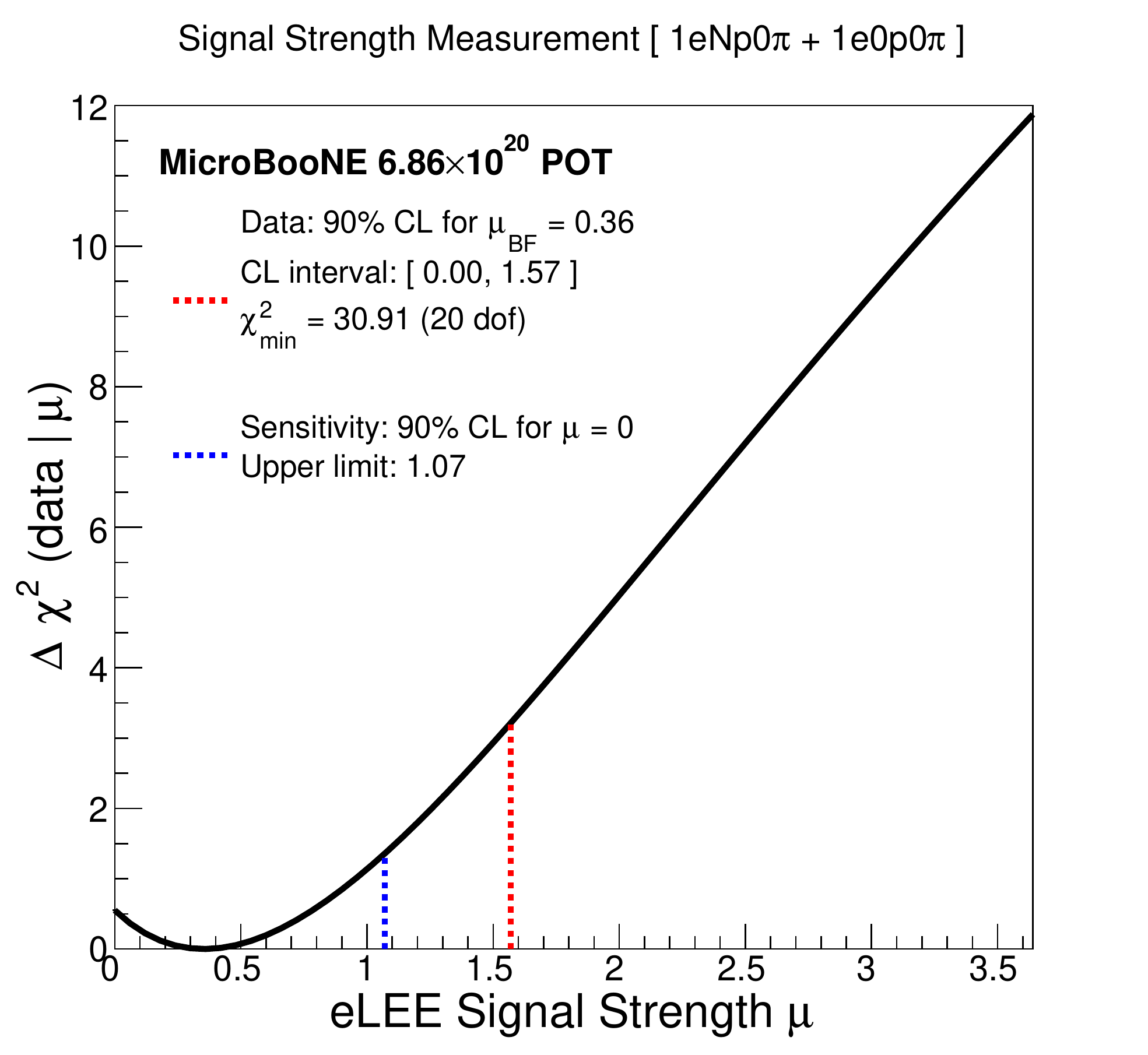}
}
\caption{\label{fig:statsFC} 
Results for the signal strength test in the \npsel~\protect\subref{sfig:statsnpfc}, \zpsel~\protect\subref{sfig:statszpfc}, and combined~\protect\subref{sfig:statsnpzpfc} channels.
The $\Delta \chi^2$ as a function of the signal strength is evaluated with respect to the best-fit signal strength value. The observed confidence interval at 90\% confidence level is indicated with a vertical lines, as well as the expected upper limit in case of no signal.}
\end{figure*}

The Feldman-Cousins procedure~\cite{feldmancousins} is used to  test the signal-strength $\mu$, where $\mu$ is a flat scaling parameter of the eLEE model, and is intended to provide further quantitative insight into a possible signal enhancement at low energy.
Toy experiments are generated for different values of true signal-strength $\mu$.
In this test the metric for defining the ordering rule based on the likelihood ratio $R\left(x|\mu\right)$ is approximated as
\begin{equation}
    R\left(x|\mu\right) \sim \Delta\chi^2 \left(x|\mu \right) = \chi^2 (x, \mu) - \chi^2 (x, \mu_{\textrm{BF}})
\end{equation}
where $\mu_{\textrm{BF}}$ is the value of $\mu$ that maximizes the likelihood ratio for a given toy-experiment $x$. 
Given the observed data, $\chi^2\left(\textrm{data}, \mu \right)$ values are computed for all values of $\mu$ and the best-fit value is identified. 
Confidence intervals are extracted based on the fraction of the toy experiments that give a larger $\Delta\chi^2\left(x|\mu \right)$ than $\Delta\chi^2\left(\textrm{data}|\mu \right)$.
Results are shown in Fig.~\ref{fig:statsFC}. Intervals at the 90\% CL are reported in Table~\ref{tab:sensitivity90} where the best-fit value $\mu_{\textrm{BF}}$ and the expected sensitivity are also reported.
In the \npsel channel we find that $\mu_{\textrm{BF}}$ is 0, and values of $\mu > 0.82$ are excluded at the 90\% CL. Due to the low sensitivity to the eLEE model in the \zpsel channel we find that the 90\% confidence interval covers a wide range of $\mu$ values, from 1.1 to 15.0. The combined measurement excludes $\mu > 1.57$ at the 90\% CL.

\begin{table*}[t]
\caption{\label{tab:sensitivity90} Best-fit eLEE model signal strength ($\mu$) and 90\% confidence intervals. The sensitivity is quantified by reporting the expected upper limits assuming $\mu=0$.}
\begin{tabular}{cccc}
\hline \hline
 & Data & Data & Sensitivity \\
Channel & $\mu_{\text{BF}}$ & 90\% CL interval on $\mu$ & 90\% upper limit on $\mu$ \\
\hline
\npsel & 0.00 & [0.00 , 0.82] & 1.16 \\
\zpsel & 4.00 & [1.13 , 15.01] & 3.41 \\
\npsel + \zpsel & 0.36 & [0.00 , 1.57] & 1.07 \\
\hline \hline
\end{tabular}
\end{table*}

Overall, the data are consistent with the intrinsic \nue model, as shown in Sec.~\ref{sec:results:gof}, but an enhancement of the event topologies measured in the \zpsel channel cannot be ruled out. The data in the separate \npsel and \zpsel channels suggest that a simple energy-dependent scaling of the \nue beam content as defined in the eLEE model tested is not favored.

\section{Conclusions}
This article presents a measurement of charged current \nue interactions without final-state pions in the MicroBooNE detector from the Fermilab BNB.
This analysis incorporates numerous sidebands to validate the modeling of the detector response as well as the neutrino flux and interaction model. Good compatibility between data and simulation is found in all validation data sets, including the \numu selection used to constrain the flux and cross section uncertainties of the intrinsic electron neutrino interactions. Electron neutrino interactions are observed with high purity leveraging the power of the MicroBooNE LArTPC detector and are found to be consistent with the \nue prediction through a goodness of fit test at the 10--20\% level.  These events are further characterized using their measured kinematic properties of electron angle and kinetic energy of the leading proton.

Comparison to a signal model based on the median MiniBooNE eLEE observation is also studied.
When the presence of the eLEE model is tested against the intrinsic electron neutrino interaction model, data in the two channels combined (\zpsel+\npsel) does not indicate a strong preference between the two hypotheses.
The eLEE signal model is disfavored by the \npsel channel at the 97.9\% CL.
The eLEE model is further parametrized in terms of the signal strength $\mu$, and confidence intervals for this parameter are extracted with the Feldman-Cousins procedure.
The \zpsel selection, which is overall less sensitive to the eLEE model, observes more events than predicted in the lowest energy region and we find that the 90\% confidence interval covers a wide range of $\mu$, with a lower bound of 1.1.
The \npsel selection, which drives the analysis sensitivity due to its higher statistics and purity, indicates a preference for no excess of low-energy electron neutrinos with respect to the intrinsic beam content prediction resulting in an upper 90\% CL limit on the signal strength of 0.82. 
More data and tests of additional models will provide further insight into these results.

This analysis is part of a broad effort by the MicroBooNE collaboration to measure low-energy electromagnetic interactions.
It will be followed by additional analyses, including those which use the full MicroBooNE dataset, roughly twice the size of that used in this result as well as advances in low energy shower reconstruction and analysis tools. Finally, the full SBN program~\cite{SBN}, with the introduction of a near detector and a third detector at a longer baseline, will further expand this investigation.

\section*{Acknowledgements}

This document was prepared by the MicroBooNE collaboration using the resources of the Fermi National Accelerator Laboratory (Fermilab), a U.S. Department of Energy, Office of Science, HEP User Facility. Fermilab is managed by Fermi Research Alliance, LLC (FRA), acting under Contract No. DE-AC02-07CH11359.  MicroBooNE is supported by the following: the U.S. Department of Energy, Office of Science, Offices of High Energy Physics and Nuclear Physics; the U.S. National Science Foundation; the Swiss National Science Foundation; the Science and Technology Facilities Council (STFC), part of the United Kingdom Research and Innovation; the Royal Society (United Kingdom); and The European Union’s Horizon 2020 Marie Sklodowska-Curie Actions. Additional support for the laser calibration system and cosmic ray tagger was provided by the Albert Einstein Center for Fundamental Physics, Bern, Switzerland. We also acknowledge the contributions of technical and scientific staff to the design, construction, and operation of the MicroBooNE detector as well as the contributions of past collaborators to the development of MicroBooNE analyses, without whom this work would not have been possible.

\bibliographystyle{apsrev4-1}
\bibliography{biblio}

\begin{thebibliography}{68}%
\makeatletter
\providecommand \@ifxundefined [1]{%
 \@ifx{#1\undefined}
}%
\providecommand \@ifnum [1]{%
 \ifnum #1\expandafter \@firstoftwo
 \else \expandafter \@secondoftwo
 \fi
}%
\providecommand \@ifx [1]{%
 \ifx #1\expandafter \@firstoftwo
 \else \expandafter \@secondoftwo
 \fi
}%
\providecommand \natexlab [1]{#1}%
\providecommand \enquote  [1]{``#1''}%
\providecommand \bibnamefont  [1]{#1}%
\providecommand \bibfnamefont [1]{#1}%
\providecommand \citenamefont [1]{#1}%
\providecommand \href@noop [0]{\@secondoftwo}%
\providecommand \href [0]{\begingroup \@sanitize@url \@href}%
\providecommand \@href[1]{\@@startlink{#1}\@@href}%
\providecommand \@@href[1]{\endgroup#1\@@endlink}%
\providecommand \@sanitize@url [0]{\catcode `\\12\catcode `\$12\catcode
  `\&12\catcode `\#12\catcode `\^12\catcode `\_12\catcode `\%12\relax}%
\providecommand \@@startlink[1]{}%
\providecommand \@@endlink[0]{}%
\providecommand \url  [0]{\begingroup\@sanitize@url \@url }%
\providecommand \@url [1]{\endgroup\@href {#1}{\urlprefix }}%
\providecommand \urlprefix  [0]{URL }%
\providecommand \Eprint [0]{\href }%
\providecommand \doibase [0]{http://dx.doi.org/}%
\providecommand \selectlanguage [0]{\@gobble}%
\providecommand \bibinfo  [0]{\@secondoftwo}%
\providecommand \bibfield  [0]{\@secondoftwo}%
\providecommand \translation [1]{[#1]}%
\providecommand \BibitemOpen [0]{}%
\providecommand \bibitemStop [0]{}%
\providecommand \bibitemNoStop [0]{.\EOS\space}%
\providecommand \EOS [0]{\spacefactor3000\relax}%
\providecommand \BibitemShut  [1]{\csname bibitem#1\endcsname}%
\let\auto@bib@innerbib\@empty
\bibitem [{\citenamefont {Zyla}\ \emph {et~al.}(2020)\citenamefont {Zyla} \emph
  {et~al.}}]{PDG}%
  \BibitemOpen
  \bibfield  {author} {\bibinfo {author} {\bibfnamefont {P.~A.}\ \bibnamefont
  {Zyla}} \emph {et~al.} (\bibinfo {collaboration} {Particle Data Group}),\
  }\href {\doibase 10.1093/ptep/ptaa104} {\bibfield  {journal} {\bibinfo
  {journal} {PTEP}\ }\textbf {\bibinfo {volume} {2020}},\ \bibinfo {pages}
  {083C01} (\bibinfo {year} {2020})}\BibitemShut {NoStop}%
\bibitem [{\citenamefont {Giunti}\ and\ \citenamefont
  {Laveder}(2011)}]{gallium}%
  \BibitemOpen
  \bibfield  {author} {\bibinfo {author} {\bibfnamefont {C.}~\bibnamefont
  {Giunti}}\ and\ \bibinfo {author} {\bibfnamefont {M.}~\bibnamefont
  {Laveder}},\ }\href {\doibase 10.1103/PhysRevC.83.065504} {\bibfield
  {journal} {\bibinfo  {journal} {Phys. Rev. C}\ }\textbf {\bibinfo {volume}
  {83}},\ \bibinfo {pages} {065504} (\bibinfo {year} {2011})}\BibitemShut
  {NoStop}%
\bibitem [{\citenamefont {Mention}\ \emph {et~al.}(2011)\citenamefont
  {Mention}, \citenamefont {Fechner}, \citenamefont {Lasserre}, \citenamefont
  {Mueller}, \citenamefont {Lhuillier}, \citenamefont {Cribier},\ and\
  \citenamefont {Letourneau}}]{reactor}%
  \BibitemOpen
  \bibfield  {author} {\bibinfo {author} {\bibfnamefont {G.}~\bibnamefont
  {Mention}}, \bibinfo {author} {\bibfnamefont {M.}~\bibnamefont {Fechner}},
  \bibinfo {author} {\bibfnamefont {T.}~\bibnamefont {Lasserre}}, \bibinfo
  {author} {\bibfnamefont {T.~A.}\ \bibnamefont {Mueller}}, \bibinfo {author}
  {\bibfnamefont {D.}~\bibnamefont {Lhuillier}}, \bibinfo {author}
  {\bibfnamefont {M.}~\bibnamefont {Cribier}}, \ and\ \bibinfo {author}
  {\bibfnamefont {A.}~\bibnamefont {Letourneau}},\ }\href {\doibase
  10.1103/PhysRevD.83.073006} {\bibfield  {journal} {\bibinfo  {journal} {Phys.
  Rev. D}\ }\textbf {\bibinfo {volume} {83}},\ \bibinfo {pages} {073006}
  (\bibinfo {year} {2011})}\BibitemShut {NoStop}%
\bibitem [{\citenamefont {Aguilar}\ \emph {et~al.}(2001)\citenamefont {Aguilar}
  \emph {et~al.}}]{LSND}%
  \BibitemOpen
  \bibfield  {author} {\bibinfo {author} {\bibfnamefont {A.}~\bibnamefont
  {Aguilar}} \emph {et~al.} (\bibinfo {collaboration} {LSND}),\ }\href
  {\doibase 10.1103/PhysRevD.64.112007} {\bibfield  {journal} {\bibinfo
  {journal} {Phys. Rev. D}\ }\textbf {\bibinfo {volume} {64}},\ \bibinfo
  {pages} {112007} (\bibinfo {year} {2001})}\BibitemShut {NoStop}%
\bibitem [{\citenamefont {Aguilar-Arevalo}\ \emph {et~al.}(2018)\citenamefont
  {Aguilar-Arevalo} \emph {et~al.}}]{MiniBooNELEE}%
  \BibitemOpen
  \bibfield  {author} {\bibinfo {author} {\bibfnamefont {A.~A.}\ \bibnamefont
  {Aguilar-Arevalo}} \emph {et~al.} (\bibinfo {collaboration} {MiniBooNE}),\
  }\href {\doibase 10.1103/PhysRevLett.121.221801} {\bibfield  {journal}
  {\bibinfo  {journal} {Phys. Rev. Lett.}\ }\textbf {\bibinfo {volume} {121}},\
  \bibinfo {pages} {221801} (\bibinfo {year} {2018})}\BibitemShut {NoStop}%
\bibitem [{\citenamefont {Barinov}\ \emph {et~al.}()\citenamefont {Barinov}
  \emph {et~al.}}]{BEST}%
  \BibitemOpen
  \bibfield  {author} {\bibinfo {author} {\bibfnamefont {V.~V.}\ \bibnamefont
  {Barinov}} \emph {et~al.} (\bibinfo {collaboration} {BEST}),\ }\href@noop {}
  {\ }\Eprint {http://arxiv.org/abs/2109.11482} {arXiv:2109.11482 [nucl-ex]}
  \BibitemShut {NoStop}%
\bibitem [{\citenamefont {Abazajian}\ \emph {et~al.}()\citenamefont {Abazajian}
  \emph {et~al.}}]{sterileWP}%
  \BibitemOpen
  \bibfield  {author} {\bibinfo {author} {\bibfnamefont {K.~N.}\ \bibnamefont
  {Abazajian}} \emph {et~al.},\ }\href@noop {} {\ }\Eprint
  {http://arxiv.org/abs/1204.5379} {arXiv:1204.5379 [hep-ph]} \BibitemShut
  {NoStop}%
\bibitem [{\citenamefont {Aartsen}\ \emph {et~al.}(2020)\citenamefont {Aartsen}
  \emph {et~al.}}]{Aartsen:2020iky}%
  \BibitemOpen
  \bibfield  {author} {\bibinfo {author} {\bibfnamefont {M.~G.}\ \bibnamefont
  {Aartsen}} \emph {et~al.} (\bibinfo {collaboration} {IceCube}),\ }\href
  {\doibase 10.1103/PhysRevLett.125.141801} {\bibfield  {journal} {\bibinfo
  {journal} {Phys. Rev. Lett.}\ }\textbf {\bibinfo {volume} {125}},\ \bibinfo
  {pages} {141801} (\bibinfo {year} {2020})}\BibitemShut {NoStop}%
\bibitem [{\citenamefont {Adamson}\ \emph {et~al.}(2020)\citenamefont {Adamson}
  \emph {et~al.}}]{disappearance}%
  \BibitemOpen
  \bibfield  {author} {\bibinfo {author} {\bibfnamefont {P.}~\bibnamefont
  {Adamson}} \emph {et~al.} (\bibinfo {collaboration} {Daya Bay and
  $\mathrm{MINOS}+$}),\ }\href {\doibase 10.1103/PhysRevLett.125.071801}
  {\bibfield  {journal} {\bibinfo  {journal} {Phys. Rev. Lett.}\ }\textbf
  {\bibinfo {volume} {125}},\ \bibinfo {pages} {071801} (\bibinfo {year}
  {2020})}\BibitemShut {NoStop}%
\bibitem [{\citenamefont {Vergani}\ \emph {et~al.}()\citenamefont {Vergani},
  \citenamefont {Kamp}, \citenamefont {Diaz}, \citenamefont {Arg{\"u}elles},
  \citenamefont {Conrad}, \citenamefont {Shaevitz},\ and\ \citenamefont
  {Uchida}}]{Vergani2021}%
  \BibitemOpen
  \bibfield  {author} {\bibinfo {author} {\bibfnamefont {S.}~\bibnamefont
  {Vergani}}, \bibinfo {author} {\bibfnamefont {N.~W.}\ \bibnamefont {Kamp}},
  \bibinfo {author} {\bibfnamefont {A.}~\bibnamefont {Diaz}}, \bibinfo {author}
  {\bibfnamefont {C.~A.}\ \bibnamefont {Arg{\"u}elles}}, \bibinfo {author}
  {\bibfnamefont {J.~M.}\ \bibnamefont {Conrad}}, \bibinfo {author}
  {\bibfnamefont {M.~H.}\ \bibnamefont {Shaevitz}}, \ and\ \bibinfo {author}
  {\bibfnamefont {M.~A.}\ \bibnamefont {Uchida}},\ }\href@noop {} {\ }\Eprint
  {http://arxiv.org/abs/2105.06470} {arXiv:2105.06470 [hep-ph]} \BibitemShut
  {NoStop}%
\bibitem [{\citenamefont {Fischer}\ \emph {et~al.}(2020)\citenamefont
  {Fischer}, \citenamefont {Hern\'andez-Cabezudo},\ and\ \citenamefont
  {Schwetz}}]{Fischer2021}%
  \BibitemOpen
  \bibfield  {author} {\bibinfo {author} {\bibfnamefont {O.}~\bibnamefont
  {Fischer}}, \bibinfo {author} {\bibfnamefont {A.}~\bibnamefont
  {Hern\'andez-Cabezudo}}, \ and\ \bibinfo {author} {\bibfnamefont
  {T.}~\bibnamefont {Schwetz}},\ }\href {\doibase 10.1103/PhysRevD.101.075045}
  {\bibfield  {journal} {\bibinfo  {journal} {Phys. Rev. D}\ }\textbf {\bibinfo
  {volume} {101}},\ \bibinfo {pages} {075045} (\bibinfo {year}
  {2020})}\BibitemShut {NoStop}%
\bibitem [{\citenamefont {Alvarez-Ruso}\ and\ \citenamefont
  {Saul-Sala}(2017)}]{Alvarez-Ruso:2017hdm}%
  \BibitemOpen
  \bibfield  {author} {\bibinfo {author} {\bibfnamefont {L.}~\bibnamefont
  {Alvarez-Ruso}}\ and\ \bibinfo {author} {\bibfnamefont {E.}~\bibnamefont
  {Saul-Sala}},\ }in\ \href@noop {} {\emph {\bibinfo {booktitle} {{Prospects in
  Neutrino Physics}}}}\ (\bibinfo {year} {2017})\ \Eprint
  {http://arxiv.org/abs/1705.00353} {arXiv:1705.00353 [hep-ph]} \BibitemShut
  {NoStop}%
\bibitem [{\citenamefont {Abdullahi}\ \emph {et~al.}(2021)\citenamefont
  {Abdullahi}, \citenamefont {Hostert},\ and\ \citenamefont
  {Pascoli}}]{Abdullahi2020}%
  \BibitemOpen
  \bibfield  {author} {\bibinfo {author} {\bibfnamefont {A.}~\bibnamefont
  {Abdullahi}}, \bibinfo {author} {\bibfnamefont {M.}~\bibnamefont {Hostert}},
  \ and\ \bibinfo {author} {\bibfnamefont {S.}~\bibnamefont {Pascoli}},\ }\href
  {\doibase 10.1016/j.physletb.2021.136531} {\bibfield  {journal} {\bibinfo
  {journal} {Phys. Lett. B}\ }\textbf {\bibinfo {volume} {820}},\ \bibinfo
  {pages} {136531} (\bibinfo {year} {2021})}\BibitemShut {NoStop}%
\bibitem [{\citenamefont {Bertuzzo}\ \emph {et~al.}(2018)\citenamefont
  {Bertuzzo}, \citenamefont {Jana}, \citenamefont {Machado},\ and\
  \citenamefont {Zukanovich~Funchal}}]{Bertuzzo2018}%
  \BibitemOpen
  \bibfield  {author} {\bibinfo {author} {\bibfnamefont {E.}~\bibnamefont
  {Bertuzzo}}, \bibinfo {author} {\bibfnamefont {S.}~\bibnamefont {Jana}},
  \bibinfo {author} {\bibfnamefont {P.~A.~N.}\ \bibnamefont {Machado}}, \ and\
  \bibinfo {author} {\bibfnamefont {R.}~\bibnamefont {Zukanovich~Funchal}},\
  }\href {\doibase 10.1103/PhysRevLett.121.241801} {\bibfield  {journal}
  {\bibinfo  {journal} {Phys. Rev. Lett.}\ }\textbf {\bibinfo {volume} {121}},\
  \bibinfo {pages} {241801} (\bibinfo {year} {2018})}\BibitemShut {NoStop}%
\bibitem [{\citenamefont {Ballett}\ \emph {et~al.}(2019)\citenamefont
  {Ballett}, \citenamefont {Pascoli},\ and\ \citenamefont
  {Ross-Lonergan}}]{Ballett2019}%
  \BibitemOpen
  \bibfield  {author} {\bibinfo {author} {\bibfnamefont {P.}~\bibnamefont
  {Ballett}}, \bibinfo {author} {\bibfnamefont {S.}~\bibnamefont {Pascoli}}, \
  and\ \bibinfo {author} {\bibfnamefont {M.}~\bibnamefont {Ross-Lonergan}},\
  }\href {\doibase 10.1103/PhysRevD.99.071701} {\bibfield  {journal} {\bibinfo
  {journal} {Phys. Rev. D}\ }\textbf {\bibinfo {volume} {99}},\ \bibinfo
  {pages} {071701} (\bibinfo {year} {2019})}\BibitemShut {NoStop}%
\bibitem [{\citenamefont {Gninenko}(2011)}]{Gninenko2011}%
  \BibitemOpen
  \bibfield  {author} {\bibinfo {author} {\bibfnamefont {S.~N.}\ \bibnamefont
  {Gninenko}},\ }\href@noop {} {\bibfield  {journal} {\bibinfo  {journal}
  {Phys. Rev. D}\ }\textbf {\bibinfo {volume} {83}},\ \bibinfo {pages} {015015}
  (\bibinfo {year} {2011})}\BibitemShut {NoStop}%
\bibitem [{\citenamefont {Abdallah}\ \emph {et~al.}(2021)\citenamefont
  {Abdallah}, \citenamefont {Gandhi},\ and\ \citenamefont
  {Roy}}]{AbdallahEtAl}%
  \BibitemOpen
  \bibfield  {author} {\bibinfo {author} {\bibfnamefont {W.}~\bibnamefont
  {Abdallah}}, \bibinfo {author} {\bibfnamefont {R.}~\bibnamefont {Gandhi}}, \
  and\ \bibinfo {author} {\bibfnamefont {S.}~\bibnamefont {Roy}},\ }\href
  {\doibase 10.1103/PhysRevD.104.055028} {\bibfield  {journal} {\bibinfo
  {journal} {Phys. Rev. D}\ }\textbf {\bibinfo {volume} {104}},\ \bibinfo
  {pages} {055028} (\bibinfo {year} {2021})}\BibitemShut {NoStop}%
\bibitem [{\citenamefont {Asaadi}\ \emph {et~al.}(2018)\citenamefont {Asaadi},
  \citenamefont {Church}, \citenamefont {Guenette}, \citenamefont {Jones},\
  and\ \citenamefont {Szelc}}]{AsaadiEtAl}%
  \BibitemOpen
  \bibfield  {author} {\bibinfo {author} {\bibfnamefont {J.}~\bibnamefont
  {Asaadi}}, \bibinfo {author} {\bibfnamefont {E.}~\bibnamefont {Church}},
  \bibinfo {author} {\bibfnamefont {R.}~\bibnamefont {Guenette}}, \bibinfo
  {author} {\bibfnamefont {B.~J.~P.}\ \bibnamefont {Jones}}, \ and\ \bibinfo
  {author} {\bibfnamefont {A.~M.}\ \bibnamefont {Szelc}},\ }\href {\doibase
  10.1103/PhysRevD.97.075021} {\bibfield  {journal} {\bibinfo  {journal} {Phys.
  Rev. D}\ }\textbf {\bibinfo {volume} {97}},\ \bibinfo {pages} {075021}
  (\bibinfo {year} {2018})}\BibitemShut {NoStop}%
\bibitem [{\citenamefont {Dutta}\ \emph {et~al.}(2020)\citenamefont {Dutta},
  \citenamefont {Ghosh},\ and\ \citenamefont {Li}}]{DuttaEtAl}%
  \BibitemOpen
  \bibfield  {author} {\bibinfo {author} {\bibfnamefont {B.}~\bibnamefont
  {Dutta}}, \bibinfo {author} {\bibfnamefont {S.}~\bibnamefont {Ghosh}}, \ and\
  \bibinfo {author} {\bibfnamefont {T.}~\bibnamefont {Li}},\ }\href {\doibase
  10.1103/PhysRevD.102.055017} {\bibfield  {journal} {\bibinfo  {journal}
  {Phys. Rev. D}\ }\textbf {\bibinfo {volume} {102}},\ \bibinfo {pages}
  {055017} (\bibinfo {year} {2020})}\BibitemShut {NoStop}%
\bibitem [{\citenamefont {Giunti}\ \emph {et~al.}(2020)\citenamefont {Giunti},
  \citenamefont {Ioannisian},\ and\ \citenamefont {Ranucci}}]{deltarad}%
  \BibitemOpen
  \bibfield  {author} {\bibinfo {author} {\bibfnamefont {C.}~\bibnamefont
  {Giunti}}, \bibinfo {author} {\bibfnamefont {A.}~\bibnamefont {Ioannisian}},
  \ and\ \bibinfo {author} {\bibfnamefont {G.}~\bibnamefont {Ranucci}},\ }\href
  {\doibase 10.1007/JHEP11(2020)146} {\bibfield  {journal} {\bibinfo  {journal}
  {JHEP}\ }\textbf {\bibinfo {volume} {11}},\ \bibinfo {pages} {146} (\bibinfo
  {year} {2020})},\ \bibinfo {note} {[Erratum: JHEP 02, 078
  (2021)]}\BibitemShut {NoStop}%
\bibitem [{\citenamefont {Acciarri}\ \emph
  {et~al.}(2017{\natexlab{a}})\citenamefont {Acciarri} \emph
  {et~al.}}]{ubdetector}%
  \BibitemOpen
  \bibfield  {author} {\bibinfo {author} {\bibfnamefont {R.}~\bibnamefont
  {Acciarri}} \emph {et~al.} (\bibinfo {collaboration} {MicroBooNE}),\ }\href
  {\doibase 10.1088/1748-0221/12/02/P02017} {\bibfield  {journal} {\bibinfo
  {journal} {JINST}\ }\textbf {\bibinfo {volume} {12}},\ \bibinfo {pages}
  {P02017} (\bibinfo {year} {2017}{\natexlab{a}})}\BibitemShut {NoStop}%
\bibitem [{\citenamefont {Abratenko}\ \emph
  {et~al.}({\natexlab{a}})\citenamefont {Abratenko} \emph {et~al.}}]{eLEEPRL}%
  \BibitemOpen
  \bibfield  {author} {\bibinfo {author} {\bibfnamefont {P.}~\bibnamefont
  {Abratenko}} \emph {et~al.} (\bibinfo {collaboration} {MicroBooNE}),\
  }\Eprint {http://arxiv.org/abs/2110.14054} {arXiv:2110.14054 [hep-ex]}
  \BibitemShut {NoStop}%
\bibitem [{\citenamefont {Abratenko}\ \emph
  {et~al.}({\natexlab{b}})\citenamefont {Abratenko} \emph {et~al.}}]{DLPRD}%
  \BibitemOpen
  \bibfield  {author} {\bibinfo {author} {\bibfnamefont {P.}~\bibnamefont
  {Abratenko}} \emph {et~al.} (\bibinfo {collaboration} {MicroBooNE}),\
  }\Eprint {http://arxiv.org/abs/2110.14080} {arXiv:2110.14080 [hep-ex]}
  \BibitemShut {NoStop}%
\bibitem [{\citenamefont {Abratenko}\ \emph
  {et~al.}({\natexlab{c}})\citenamefont {Abratenko} \emph {et~al.}}]{WCPRD}%
  \BibitemOpen
  \bibfield  {author} {\bibinfo {author} {\bibfnamefont {P.}~\bibnamefont
  {Abratenko}} \emph {et~al.} (\bibinfo {collaboration} {MicroBooNE}),\
  }\Eprint {http://arxiv.org/abs/2110.13978} {arXiv:2110.13978 [hep-ex]}
  \BibitemShut {NoStop}%
\bibitem [{\citenamefont {Abratenko}\ \emph
  {et~al.}({\natexlab{d}})\citenamefont {Abratenko} \emph {et~al.}}]{gLEE}%
  \BibitemOpen
  \bibfield  {author} {\bibinfo {author} {\bibfnamefont {P.}~\bibnamefont
  {Abratenko}} \emph {et~al.} (\bibinfo {collaboration} {MicroBooNE}),\
  }\Eprint {http://arxiv.org/abs/2110.00409} {arXiv:2110.00409 [hep-ex]}
  \BibitemShut {NoStop}%
\bibitem [{\citenamefont {Aguilar-Arevalo}\ \emph {et~al.}(2007)\citenamefont
  {Aguilar-Arevalo} \emph {et~al.}}]{MiniBooNE2007}%
  \BibitemOpen
  \bibfield  {author} {\bibinfo {author} {\bibfnamefont {A.~A.}\ \bibnamefont
  {Aguilar-Arevalo}} \emph {et~al.} (\bibinfo {collaboration} {MiniBooNE}),\
  }\href {\doibase 10.1103/PhysRevLett.98.231801} {\bibfield  {journal}
  {\bibinfo  {journal} {Phys. Rev. Lett.}\ }\textbf {\bibinfo {volume} {98}},\
  \bibinfo {pages} {231801} (\bibinfo {year} {2007})}\BibitemShut {NoStop}%
\bibitem [{\citenamefont {Chang}\ \emph {et~al.}(2021)\citenamefont {Chang},
  \citenamefont {Chen}, \citenamefont {Ho},\ and\ \citenamefont
  {Tseng}}]{Chang:2021myh}%
  \BibitemOpen
  \bibfield  {author} {\bibinfo {author} {\bibfnamefont {C.-H.~V.}\
  \bibnamefont {Chang}}, \bibinfo {author} {\bibfnamefont {C.-R.}\ \bibnamefont
  {Chen}}, \bibinfo {author} {\bibfnamefont {S.-Y.}\ \bibnamefont {Ho}}, \ and\
  \bibinfo {author} {\bibfnamefont {S.-Y.}\ \bibnamefont {Tseng}},\ }\href
  {\doibase 10.1103/PhysRevD.104.015030} {\bibfield  {journal} {\bibinfo
  {journal} {Phys. Rev. D}\ }\textbf {\bibinfo {volume} {104}},\ \bibinfo
  {pages} {015030} (\bibinfo {year} {2021})}\BibitemShut {NoStop}%
\bibitem [{\citenamefont {Stancu}(2001)}]{BNB}%
  \BibitemOpen
  \bibfield  {author} {\bibinfo {author} {\bibfnamefont {I.}~\bibnamefont
  {Stancu}},\ }\href {\doibase 10.2172/1212167} {\bibfield  {journal} {\bibinfo
   {journal} {FERMILAB-DESIGN-2001-03}\ } (\bibinfo {year} {2001}),\
  10.2172/1212167}\BibitemShut {NoStop}%
\bibitem [{\citenamefont {Aguilar-Arevalo}\ \emph {et~al.}(2009)\citenamefont
  {Aguilar-Arevalo} \emph {et~al.}}]{mbflux}%
  \BibitemOpen
  \bibfield  {author} {\bibinfo {author} {\bibfnamefont {A.~A.}\ \bibnamefont
  {Aguilar-Arevalo}} \emph {et~al.} (\bibinfo {collaboration} {MiniBooNE}),\
  }\href {\doibase 10.1103/PhysRevD.79.072002} {\bibfield  {journal} {\bibinfo
  {journal} {Phys. Rev. D}\ }\textbf {\bibinfo {volume} {79}},\ \bibinfo
  {pages} {072002} (\bibinfo {year} {2009})}\BibitemShut {NoStop}%
\bibitem [{\citenamefont {Adams}\ \emph
  {et~al.}(2018{\natexlab{a}})\citenamefont {Adams} \emph {et~al.}}]{ubsig1}%
  \BibitemOpen
  \bibfield  {author} {\bibinfo {author} {\bibfnamefont {C.}~\bibnamefont
  {Adams}} \emph {et~al.} (\bibinfo {collaboration} {MicroBooNE}),\ }\href
  {\doibase 10.1088/1748-0221/13/07/P07006} {\bibfield  {journal} {\bibinfo
  {journal} {JINST}\ }\textbf {\bibinfo {volume} {13}},\ \bibinfo {pages}
  {P07006} (\bibinfo {year} {2018}{\natexlab{a}})}\BibitemShut {NoStop}%
\bibitem [{\citenamefont {Adams}\ \emph
  {et~al.}(2019{\natexlab{a}})\citenamefont {Adams} \emph {et~al.}}]{ubCRT}%
  \BibitemOpen
  \bibfield  {author} {\bibinfo {author} {\bibfnamefont {C.}~\bibnamefont
  {Adams}} \emph {et~al.} (\bibinfo {collaboration} {MicroBooNE}),\ }\href
  {\doibase 10.1088/1748-0221/14/04/P04004} {\bibfield  {journal} {\bibinfo
  {journal} {JINST}\ }\textbf {\bibinfo {volume} {14}},\ \bibinfo {pages}
  {P04004} (\bibinfo {year} {2019}{\natexlab{a}})}\BibitemShut {NoStop}%
\bibitem [{\citenamefont {Snider}\ and\ \citenamefont
  {Petrillo}(2017)}]{LArSoft}%
  \BibitemOpen
  \bibfield  {author} {\bibinfo {author} {\bibfnamefont {E.~L.}\ \bibnamefont
  {Snider}}\ and\ \bibinfo {author} {\bibfnamefont {G.}~\bibnamefont
  {Petrillo}},\ }\href {\doibase 10.1088/1742-6596/898/4/042057} {\bibfield
  {journal} {\bibinfo  {journal} {J. Phys. Conf. Ser.}\ }\textbf {\bibinfo
  {volume} {898}},\ \bibinfo {pages} {042057} (\bibinfo {year}
  {2017})}\BibitemShut {NoStop}%
\bibitem [{\citenamefont {Aguilar-Arevalo}\ \emph {et~al.}(2021)\citenamefont
  {Aguilar-Arevalo} \emph {et~al.}}]{MiniBooNE2020}%
  \BibitemOpen
  \bibfield  {author} {\bibinfo {author} {\bibfnamefont {A.~A.}\ \bibnamefont
  {Aguilar-Arevalo}} \emph {et~al.} (\bibinfo {collaboration} {MiniBooNE}),\
  }\href {\doibase 10.1103/PhysRevD.103.052002} {\bibfield  {journal} {\bibinfo
   {journal} {Phys. Rev. D}\ }\textbf {\bibinfo {volume} {103}},\ \bibinfo
  {pages} {052002} (\bibinfo {year} {2021})}\BibitemShut {NoStop}%
\bibitem [{\citenamefont {Abratenko}\ \emph
  {et~al.}(2021{\natexlab{a}})\citenamefont {Abratenko} \emph
  {et~al.}}]{ubnuminue}%
  \BibitemOpen
  \bibfield  {author} {\bibinfo {author} {\bibfnamefont {P.}~\bibnamefont
  {Abratenko}} \emph {et~al.} (\bibinfo {collaboration} {MicroBooNE}),\ }\href
  {\doibase 10.1103/PhysRevD.104.052002} {\bibfield  {journal} {\bibinfo
  {journal} {Phys. Rev. D}\ }\textbf {\bibinfo {volume} {104}},\ \bibinfo
  {pages} {052002} (\bibinfo {year} {2021}{\natexlab{a}})}\BibitemShut
  {NoStop}%
\bibitem [{\citenamefont {Abratenko}\ \emph
  {et~al.}({\natexlab{e}})\citenamefont {Abratenko} \emph
  {et~al.}}]{ubnuminueMCC9}%
  \BibitemOpen
  \bibfield  {author} {\bibinfo {author} {\bibfnamefont {P.}~\bibnamefont
  {Abratenko}} \emph {et~al.} (\bibinfo {collaboration} {MicroBooNE}),\
  }\Eprint {http://arxiv.org/abs/2109.06832} {arXiv:2109.06832 [hep-ex]}
  \BibitemShut {NoStop}%
\bibitem [{\citenamefont {Abratenko}\ \emph
  {et~al.}(2020{\natexlab{a}})\citenamefont {Abratenko} \emph
  {et~al.}}]{ubnumunp}%
  \BibitemOpen
  \bibfield  {author} {\bibinfo {author} {\bibfnamefont {P.}~\bibnamefont
  {Abratenko}} \emph {et~al.} (\bibinfo {collaboration} {MicroBooNE}),\ }\href
  {\doibase 10.1103/PhysRevD.102.112013} {\bibfield  {journal} {\bibinfo
  {journal} {Phys. Rev. D}\ }\textbf {\bibinfo {volume} {102}},\ \bibinfo
  {pages} {112013} (\bibinfo {year} {2020}{\natexlab{a}})}\BibitemShut
  {NoStop}%
\bibitem [{\citenamefont {Abratenko}\ \emph
  {et~al.}(2020{\natexlab{b}})\citenamefont {Abratenko} \emph
  {et~al.}}]{ubccqexsec}%
  \BibitemOpen
  \bibfield  {author} {\bibinfo {author} {\bibfnamefont {P.}~\bibnamefont
  {Abratenko}} \emph {et~al.} (\bibinfo {collaboration} {MicroBooNE}),\ }\href
  {\doibase 10.1103/PhysRevLett.125.201803} {\bibfield  {journal} {\bibinfo
  {journal} {Phys. Rev. Lett.}\ }\textbf {\bibinfo {volume} {125}},\ \bibinfo
  {pages} {201803} (\bibinfo {year} {2020}{\natexlab{b}})}\BibitemShut
  {NoStop}%
\bibitem [{\citenamefont {Abratenko}\ \emph {et~al.}(2019)\citenamefont
  {Abratenko} \emph {et~al.}}]{ubccincl}%
  \BibitemOpen
  \bibfield  {author} {\bibinfo {author} {\bibfnamefont {P.}~\bibnamefont
  {Abratenko}} \emph {et~al.} (\bibinfo {collaboration} {MicroBooNE}),\ }\href
  {\doibase 10.1103/PhysRevLett.123.131801} {\bibfield  {journal} {\bibinfo
  {journal} {Phys. Rev. Lett.}\ }\textbf {\bibinfo {volume} {123}},\ \bibinfo
  {pages} {131801} (\bibinfo {year} {2019})}\BibitemShut {NoStop}%
\bibitem [{\citenamefont {Adams}\ \emph
  {et~al.}(2019{\natexlab{b}})\citenamefont {Adams} \emph {et~al.}}]{ubccpi0}%
  \BibitemOpen
  \bibfield  {author} {\bibinfo {author} {\bibfnamefont {C.}~\bibnamefont
  {Adams}} \emph {et~al.} (\bibinfo {collaboration} {MicroBooNE}),\ }\href
  {\doibase 10.1103/PhysRevD.99.091102} {\bibfield  {journal} {\bibinfo
  {journal} {Phys. Rev. D}\ }\textbf {\bibinfo {volume} {99}},\ \bibinfo
  {pages} {091102} (\bibinfo {year} {2019}{\natexlab{b}})}\BibitemShut
  {NoStop}%
\bibitem [{\citenamefont {Aguilar-Arevalo}\ \emph {et~al.}(2013)\citenamefont
  {Aguilar-Arevalo} \emph {et~al.}}]{mbflux2}%
  \BibitemOpen
  \bibfield  {author} {\bibinfo {author} {\bibfnamefont {A.~A.}\ \bibnamefont
  {Aguilar-Arevalo}} \emph {et~al.} (\bibinfo {collaboration} {MiniBooNE}),\
  }\href {\doibase 10.1103/PhysRevLett.110.161801} {\bibfield  {journal}
  {\bibinfo  {journal} {Phys. Rev. Lett.}\ }\textbf {\bibinfo {volume} {110}},\
  \bibinfo {pages} {161801} (\bibinfo {year} {2013})}\BibitemShut {NoStop}%
\bibitem [{\citenamefont {Andreopoulos}\ \emph {et~al.}(2010)\citenamefont
  {Andreopoulos} \emph {et~al.}}]{GENIE}%
  \BibitemOpen
  \bibfield  {author} {\bibinfo {author} {\bibfnamefont {C.}~\bibnamefont
  {Andreopoulos}} \emph {et~al.},\ }\href {\doibase 10.1016/j.nima.2009.12.009}
  {\bibfield  {journal} {\bibinfo  {journal} {Nucl. Instrum. Meth. A}\ }\textbf
  {\bibinfo {volume} {614}},\ \bibinfo {pages} {87} (\bibinfo {year}
  {2010})}\BibitemShut {NoStop}%
\bibitem [{\citenamefont {Abratenko}\ \emph
  {et~al.}({\natexlab{f}})\citenamefont {Abratenko} \emph {et~al.}}]{ubtune}%
  \BibitemOpen
  \bibfield  {author} {\bibinfo {author} {\bibfnamefont {P.}~\bibnamefont
  {Abratenko}} \emph {et~al.} (\bibinfo {collaboration} {MicroBooNE}),\
  }\Eprint {http://arxiv.org/abs/2110.14028} {arXiv:2110.14028 [hep-ex]}
  \BibitemShut {NoStop}%
\bibitem [{\citenamefont {Agostinelli}\ \emph {et~al.}(2003)\citenamefont
  {Agostinelli} \emph {et~al.}}]{geant4}%
  \BibitemOpen
  \bibfield  {author} {\bibinfo {author} {\bibfnamefont {S.}~\bibnamefont
  {Agostinelli}} \emph {et~al.} (\bibinfo {collaboration} {GEANT4}),\ }\href
  {\doibase 10.1016/S0168-9002(03)01368-8} {\bibfield  {journal} {\bibinfo
  {journal} {Nucl. Instrum. Meth. A}\ }\textbf {\bibinfo {volume} {506}},\
  \bibinfo {pages} {250} (\bibinfo {year} {2003})}\BibitemShut {NoStop}%
\bibitem [{\citenamefont {Adams}\ \emph
  {et~al.}(2018{\natexlab{b}})\citenamefont {Adams} \emph {et~al.}}]{ubsig2}%
  \BibitemOpen
  \bibfield  {author} {\bibinfo {author} {\bibfnamefont {C.}~\bibnamefont
  {Adams}} \emph {et~al.} (\bibinfo {collaboration} {MicroBooNE}),\ }\href
  {\doibase 10.1088/1748-0221/13/07/P07007} {\bibfield  {journal} {\bibinfo
  {journal} {JINST}\ }\textbf {\bibinfo {volume} {13}},\ \bibinfo {pages}
  {P07007} (\bibinfo {year} {2018}{\natexlab{b}})}\BibitemShut {NoStop}%
\bibitem [{\citenamefont {Adams}\ \emph
  {et~al.}(2020{\natexlab{a}})\citenamefont {Adams} \emph {et~al.}}]{ubefield}%
  \BibitemOpen
  \bibfield  {author} {\bibinfo {author} {\bibfnamefont {C.}~\bibnamefont
  {Adams}} \emph {et~al.} (\bibinfo {collaboration} {MicroBooNE}),\ }\href
  {\doibase 10.1088/1748-0221/15/07/P07010} {\bibfield  {journal} {\bibinfo
  {journal} {JINST}\ }\textbf {\bibinfo {volume} {15}},\ \bibinfo {pages}
  {P07010} (\bibinfo {year} {2020}{\natexlab{a}})}\BibitemShut {NoStop}%
\bibitem [{\citenamefont {Abratenko}\ \emph
  {et~al.}(2020{\natexlab{c}})\citenamefont {Abratenko} \emph
  {et~al.}}]{efieldcosmics}%
  \BibitemOpen
  \bibfield  {author} {\bibinfo {author} {\bibfnamefont {P.}~\bibnamefont
  {Abratenko}} \emph {et~al.} (\bibinfo {collaboration} {MicroBooNE}),\ }\href
  {\doibase 10.1088/1748-0221/15/12/P12037} {\bibfield  {journal} {\bibinfo
  {journal} {JINST}\ }\textbf {\bibinfo {volume} {15}},\ \bibinfo {pages}
  {P12037} (\bibinfo {year} {2020}{\natexlab{c}})}\BibitemShut {NoStop}%
\bibitem [{\citenamefont {Acciarri}\ \emph {et~al.}(2013)\citenamefont
  {Acciarri} \emph {et~al.}}]{argoneutrecomb}%
  \BibitemOpen
  \bibfield  {author} {\bibinfo {author} {\bibfnamefont {R.}~\bibnamefont
  {Acciarri}} \emph {et~al.} (\bibinfo {collaboration} {ArgoNeuT}),\ }\href
  {\doibase 10.1088/1748-0221/8/08/P08005} {\bibfield  {journal} {\bibinfo
  {journal} {JINST}\ }\textbf {\bibinfo {volume} {8}},\ \bibinfo {pages}
  {P08005} (\bibinfo {year} {2013})}\BibitemShut {NoStop}%
\bibitem [{\citenamefont {Abratenko}\ \emph
  {et~al.}({\natexlab{g}})\citenamefont {Abratenko} \emph {et~al.}}]{DetSyst}%
  \BibitemOpen
  \bibfield  {author} {\bibinfo {author} {\bibfnamefont {P.}~\bibnamefont
  {Abratenko}} \emph {et~al.} (\bibinfo {collaboration} {MicroBooNE}),\
  }\bibinfo {note}
  {\url{https://microboone.fnal.gov/wp-content/uploads/MICROBOONE-NOTE-1075-PUB.pdf}}\BibitemShut
  {NoStop}%
\bibitem [{\citenamefont {Calcutt}\ \emph {et~al.}(2021)\citenamefont
  {Calcutt}, \citenamefont {Thorpe}, \citenamefont {Mahn},\ and\ \citenamefont
  {Fields}}]{G4Reweight}%
  \BibitemOpen
  \bibfield  {author} {\bibinfo {author} {\bibfnamefont {J.}~\bibnamefont
  {Calcutt}}, \bibinfo {author} {\bibfnamefont {C.}~\bibnamefont {Thorpe}},
  \bibinfo {author} {\bibfnamefont {K.}~\bibnamefont {Mahn}}, \ and\ \bibinfo
  {author} {\bibfnamefont {L.}~\bibnamefont {Fields}},\ }\href {\doibase
  10.1088/1748-0221/16/08/P08042} {\bibfield  {journal} {\bibinfo  {journal}
  {JINST}\ }\textbf {\bibinfo {volume} {16}},\ \bibinfo {pages} {P08042}
  (\bibinfo {year} {2021})}\BibitemShut {NoStop}%
\bibitem [{\citenamefont {Casper}(2002)}]{nuance}%
  \BibitemOpen
  \bibfield  {author} {\bibinfo {author} {\bibfnamefont {D.}~\bibnamefont
  {Casper}},\ }\href {\doibase 10.1016/S0920-5632(02)01756-5} {\bibfield
  {journal} {\bibinfo  {journal} {Nucl. Phys. B Proc. Suppl.}\ }\textbf
  {\bibinfo {volume} {112}},\ \bibinfo {pages} {161} (\bibinfo {year}
  {2002})}\BibitemShut {NoStop}%
\bibitem [{\citenamefont {Acciarri}\ \emph
  {et~al.}(2017{\natexlab{b}})\citenamefont {Acciarri} \emph
  {et~al.}}]{ubnoise}%
  \BibitemOpen
  \bibfield  {author} {\bibinfo {author} {\bibfnamefont {R.}~\bibnamefont
  {Acciarri}} \emph {et~al.} (\bibinfo {collaboration} {MicroBooNE}),\ }\href
  {\doibase 10.1088/1748-0221/12/08/P08003} {\bibfield  {journal} {\bibinfo
  {journal} {JINST}\ }\textbf {\bibinfo {volume} {12}},\ \bibinfo {pages}
  {P08003} (\bibinfo {year} {2017}{\natexlab{b}})}\BibitemShut {NoStop}%
\bibitem [{\citenamefont {Acciarri}\ \emph {et~al.}(2018)\citenamefont
  {Acciarri} \emph {et~al.}}]{pandora}%
  \BibitemOpen
  \bibfield  {author} {\bibinfo {author} {\bibfnamefont {R.}~\bibnamefont
  {Acciarri}} \emph {et~al.} (\bibinfo {collaboration} {MicroBooNE}),\ }\href
  {\doibase 10.1140/epjc/s10052-017-5481-6} {\bibfield  {journal} {\bibinfo
  {journal} {Eur. Phys. J. C}\ }\textbf {\bibinfo {volume} {78}},\ \bibinfo
  {pages} {82} (\bibinfo {year} {2018})}\BibitemShut {NoStop}%
\bibitem [{\citenamefont {Adams}\ \emph
  {et~al.}(2019{\natexlab{c}})\citenamefont {Adams} \emph
  {et~al.}}]{ubtrkmult}%
  \BibitemOpen
  \bibfield  {author} {\bibinfo {author} {\bibfnamefont {C.}~\bibnamefont
  {Adams}} \emph {et~al.} (\bibinfo {collaboration} {MicroBooNE}),\ }\href
  {\doibase 10.1140/epjc/s10052-019-6742-3} {\bibfield  {journal} {\bibinfo
  {journal} {Eur. Phys. J. C}\ }\textbf {\bibinfo {volume} {79}},\ \bibinfo
  {pages} {248} (\bibinfo {year} {2019}{\natexlab{c}})}\BibitemShut {NoStop}%
\bibitem [{\citenamefont {Abratenko}\ \emph
  {et~al.}(2021{\natexlab{b}})\citenamefont {Abratenko} \emph
  {et~al.}}]{ubhportal}%
  \BibitemOpen
  \bibfield  {author} {\bibinfo {author} {\bibfnamefont {P.}~\bibnamefont
  {Abratenko}} \emph {et~al.} (\bibinfo {collaboration} {MicroBooNE}),\ }\href
  {\doibase 10.1103/PhysRevLett.127.151803} {\bibfield  {journal} {\bibinfo
  {journal} {Phys. Rev. Lett.}\ }\textbf {\bibinfo {volume} {127}},\ \bibinfo
  {pages} {151803} (\bibinfo {year} {2021}{\natexlab{b}})}\BibitemShut
  {NoStop}%
\bibitem [{\citenamefont {Abratenko}\ \emph
  {et~al.}(2020{\natexlab{d}})\citenamefont {Abratenko} \emph
  {et~al.}}]{ubhsl}%
  \BibitemOpen
  \bibfield  {author} {\bibinfo {author} {\bibfnamefont {P.}~\bibnamefont
  {Abratenko}} \emph {et~al.} (\bibinfo {collaboration} {MicroBooNE}),\ }\href
  {\doibase 10.1103/PhysRevD.101.052001} {\bibfield  {journal} {\bibinfo
  {journal} {Phys. Rev. D}\ }\textbf {\bibinfo {volume} {101}},\ \bibinfo
  {pages} {052001} (\bibinfo {year} {2020}{\natexlab{d}})}\BibitemShut
  {NoStop}%
\bibitem [{\citenamefont {Adams}\ \emph
  {et~al.}(2020{\natexlab{b}})\citenamefont {Adams} \emph {et~al.}}]{ubcalib}%
  \BibitemOpen
  \bibfield  {author} {\bibinfo {author} {\bibfnamefont {C.}~\bibnamefont
  {Adams}} \emph {et~al.} (\bibinfo {collaboration} {MicroBooNE}),\ }\href
  {\doibase 10.1088/1748-0221/15/03/P03022} {\bibfield  {journal} {\bibinfo
  {journal} {JINST}\ }\textbf {\bibinfo {volume} {15}},\ \bibinfo {pages}
  {P03022} (\bibinfo {year} {2020}{\natexlab{b}})}\BibitemShut {NoStop}%
\bibitem [{\citenamefont {Adams}\ \emph
  {et~al.}(2020{\natexlab{c}})\citenamefont {Adams} \emph
  {et~al.}}]{uBpi0reco}%
  \BibitemOpen
  \bibfield  {author} {\bibinfo {author} {\bibfnamefont {C.}~\bibnamefont
  {Adams}} \emph {et~al.} (\bibinfo {collaboration} {MicroBooNE}),\ }\href
  {\doibase 10.1088/1748-0221/15/02/P02007} {\bibfield  {journal} {\bibinfo
  {journal} {JINST}\ }\textbf {\bibinfo {volume} {15}},\ \bibinfo {pages}
  {P02007} (\bibinfo {year} {2020}{\natexlab{c}})}\BibitemShut {NoStop}%
\bibitem [{\citenamefont {Abratenko}\ \emph
  {et~al.}({\natexlab{h}})\citenamefont {Abratenko} \emph {et~al.}}]{llrpid}%
  \BibitemOpen
  \bibfield  {author} {\bibinfo {author} {\bibfnamefont {P.}~\bibnamefont
  {Abratenko}} \emph {et~al.} (\bibinfo {collaboration} {MicroBooNE}),\
  }\Eprint {http://arxiv.org/abs/2109.02460} {arXiv:2109.02460
  [physics.ins-det]} \BibitemShut {NoStop}%
\bibitem [{\citenamefont {Fruhwirth}(1987)}]{Fruhwirth:1987fm}%
  \BibitemOpen
  \bibfield  {author} {\bibinfo {author} {\bibfnamefont {R.}~\bibnamefont
  {Fruhwirth}},\ }\href {\doibase 10.1016/0168-9002(87)90887-4} {\bibfield
  {journal} {\bibinfo  {journal} {Nucl. Instrum. Meth. A}\ }\textbf {\bibinfo
  {volume} {262}},\ \bibinfo {pages} {444} (\bibinfo {year}
  {1987})}\BibitemShut {NoStop}%
\bibitem [{\citenamefont {Acciarri}\ \emph
  {et~al.}(2017{\natexlab{c}})\citenamefont {Acciarri} \emph
  {et~al.}}]{argoneutdedx}%
  \BibitemOpen
  \bibfield  {author} {\bibinfo {author} {\bibfnamefont {R.}~\bibnamefont
  {Acciarri}} \emph {et~al.} (\bibinfo {collaboration} {ArgoNeuT}),\ }\href
  {\doibase 10.1103/PhysRevD.95.072005} {\bibfield  {journal} {\bibinfo
  {journal} {Phys. Rev. D}\ }\textbf {\bibinfo {volume} {95}},\ \bibinfo
  {pages} {072005} (\bibinfo {year} {2017}{\natexlab{c}})}\BibitemShut
  {NoStop}%
\bibitem [{\citenamefont {Abratenko}\ \emph {et~al.}(2017)\citenamefont
  {Abratenko} \emph {et~al.}}]{MCS}%
  \BibitemOpen
  \bibfield  {author} {\bibinfo {author} {\bibfnamefont {P.}~\bibnamefont
  {Abratenko}} \emph {et~al.} (\bibinfo {collaboration} {MicroBooNE}),\ }\href
  {\doibase 10.1088/1748-0221/12/10/P10010} {\bibfield  {journal} {\bibinfo
  {journal} {JINST}\ }\textbf {\bibinfo {volume} {12}},\ \bibinfo {pages}
  {P10010} (\bibinfo {year} {2017})}\BibitemShut {NoStop}%
\bibitem [{\citenamefont {Chen}\ and\ \citenamefont {Guestrin}()}]{xgboost}%
  \BibitemOpen
  \bibfield  {author} {\bibinfo {author} {\bibfnamefont {T.}~\bibnamefont
  {Chen}}\ and\ \bibinfo {author} {\bibfnamefont {C.}~\bibnamefont
  {Guestrin}},\ }\href@noop {} {\ }\Eprint {http://arxiv.org/abs/1603.02754}
  {arXiv:1603.02754 [cs.LG]} \BibitemShut {NoStop}%
\bibitem [{\citenamefont {Adamson}\ \emph {et~al.}(2016)\citenamefont {Adamson}
  \emph {et~al.}}]{numibeam}%
  \BibitemOpen
  \bibfield  {author} {\bibinfo {author} {\bibfnamefont {P.}~\bibnamefont
  {Adamson}} \emph {et~al.},\ }\href {\doibase 10.1016/j.nima.2015.08.063}
  {\bibfield  {journal} {\bibinfo  {journal} {Nucl. Instrum. Meth. A}\ }\textbf
  {\bibinfo {volume} {806}},\ \bibinfo {pages} {279} (\bibinfo {year}
  {2016})}\BibitemShut {NoStop}%
\bibitem [{\citenamefont {Acciarri}\ \emph {et~al.}(2020)\citenamefont
  {Acciarri} \emph {et~al.}}]{argoneutnue}%
  \BibitemOpen
  \bibfield  {author} {\bibinfo {author} {\bibfnamefont {R.}~\bibnamefont
  {Acciarri}} \emph {et~al.} (\bibinfo {collaboration} {ArgoNeuT}),\ }\href
  {\doibase 10.1103/PhysRevD.102.011101} {\bibfield  {journal} {\bibinfo
  {journal} {Phys. Rev. D}\ }\textbf {\bibinfo {volume} {102}},\ \bibinfo
  {pages} {011101} (\bibinfo {year} {2020})}\BibitemShut {NoStop}%
\bibitem [{\citenamefont {Eaton}(1983)}]{alma991015286869705251}%
  \BibitemOpen
  \bibfield  {author} {\bibinfo {author} {\bibfnamefont {M.~L.}\ \bibnamefont
  {Eaton}},\ }\href@noop {} {\emph {\bibinfo {title} {{Multivariate statistics:
  a vector space approach}}}}\ (\bibinfo  {publisher} {Wiley},\ \bibinfo
  {address} {New York},\ \bibinfo {year} {1983})\ pp.\ \bibinfo {pages}
  {116--117}\BibitemShut {NoStop}%
\bibitem [{\citenamefont {Ji}\ \emph {et~al.}(2020)\citenamefont {Ji},
  \citenamefont {Gu}, \citenamefont {Qian}, \citenamefont {Wei},\ and\
  \citenamefont {Zhang}}]{cnp}%
  \BibitemOpen
  \bibfield  {author} {\bibinfo {author} {\bibfnamefont {X.}~\bibnamefont
  {Ji}}, \bibinfo {author} {\bibfnamefont {W.}~\bibnamefont {Gu}}, \bibinfo
  {author} {\bibfnamefont {X.}~\bibnamefont {Qian}}, \bibinfo {author}
  {\bibfnamefont {H.}~\bibnamefont {Wei}}, \ and\ \bibinfo {author}
  {\bibfnamefont {C.}~\bibnamefont {Zhang}},\ }\href {\doibase
  10.1016/j.nima.2020.163677} {\bibfield  {journal} {\bibinfo  {journal} {Nucl.
  Instrum. Meth. A}\ }\textbf {\bibinfo {volume} {961}},\ \bibinfo {pages}
  {163677} (\bibinfo {year} {2020})}\BibitemShut {NoStop}%
\bibitem [{\citenamefont {Feldman}\ and\ \citenamefont
  {Cousins}(1998)}]{feldmancousins}%
  \BibitemOpen
  \bibfield  {author} {\bibinfo {author} {\bibfnamefont {G.~J.}\ \bibnamefont
  {Feldman}}\ and\ \bibinfo {author} {\bibfnamefont {R.~D.}\ \bibnamefont
  {Cousins}},\ }\href {\doibase 10.1103/PhysRevD.57.3873} {\bibfield  {journal}
  {\bibinfo  {journal} {Phys. Rev. D}\ }\textbf {\bibinfo {volume} {57}},\
  \bibinfo {pages} {3873} (\bibinfo {year} {1998})}\BibitemShut {NoStop}%
\bibitem [{\citenamefont {Antonello}\ \emph {et~al.}()\citenamefont {Antonello}
  \emph {et~al.}}]{SBN}%
  \BibitemOpen
  \bibfield  {author} {\bibinfo {author} {\bibfnamefont {M.}~\bibnamefont
  {Antonello}} \emph {et~al.} (\bibinfo {collaboration} {MicroBooNE, LAr1-ND,
  ICARUS-WA104}),\ }\href@noop {} {\ }\Eprint {http://arxiv.org/abs/1503.01520}
  {arXiv:1503.01520 [physics.ins-det]} \BibitemShut {NoStop}%
\end{thebibliography}%

\end{document}